\documentclass[12pt,final,letterpaper,onecolumn,romanappendices]{IEEEtran}
\usepackage{cite}
\usepackage{epstopdf}
\usepackage{epsfig}
\usepackage{amsthm}
\usepackage{amsmath,amssymb,amsfonts,amstext,amsbsy,amsopn,dsfont}
\usepackage{cases}
\usepackage{bigints}
\usepackage{sublabel}
\usepackage{array}

\newtheorem{theorem}{\textbf{Theorem}}

\newtheorem{definition}{\textbf{Definition}}
\newtheorem{corollary}{\textbf{Corollary}}

\newcommand{\defn}{\triangleq}
\newcommand{\dif}{\textmd{d}}

\newcommand{\sinr}{\mathsf{SIR}}

\begin{document}

\title{Generalized SIR Analysis for Stochastic Heterogeneous Wireless Networks: \\Theory and Applications}

\author{Chun-Hung Liu, Zhu-Kuan Yang and Di-Chun Liang
\thanks{C.-H. Liu, Z.-K. Yang and D.-C. Liang are with the Department of Electrical and Computer Engineering, National Chiao Tung University, Hsinchu Taiwan. The contact author is Dr. Liu  (Email: chungliu@nctu.edu.tw). Manuscript date: \today.}}

\maketitle

\begin{abstract}
This paper provides an analytically tractable framework of investigating the statistical properties of the signal-to-interference power ratio (SIR) with a general distribution in a heterogeneous wireless ad hoc network in which there are $K$ different types of transmitters (TXs) communicating with their unique intended receiver (RX). The TXs of each type form an independent homogeneous Poisson point process. In the first part of this paper, we introduce a novel approach to deriving the Laplace transform of the reciprocal of the SIR and use it to characterize the distribution of the SIR.  Our main findings show that the closed-form expression of the distribution of the SIR can be obtained whenever the receive signal power has an Erlang distribution, and an almost closed-form expression can be found if the power-law pathloss model has a pathloss exponent of four. In the second part of this paper, we aim to apply the derived distribution of the SIR in finding the two important performance metrics: the success probability and ergodic link capacity. For each type of the RXs, the success probability with (without) interference cancellation and that with (without) the proposed stochastic power control are found in a compact form. With the aid of the derived Shannon transform identity, the ergodic link capacities of $K$-type RXs are derived with low complexity, and they can be applied to many transmitting scenarios, such as multi-antenna communication and stochastic power control. Finally, we analyze the spatial throughput capacity of the heterogeneous network defined based on the derived $K$ success probabilities and ergodic link capacities and show the existence of its maximum.
\end{abstract}


\section{Introduction}
Consider a large-scale heterogeneous wireless ad hoc network in which there are $K$ different types of transmitters (TXs) and the TXs of each specific type form an independent Poisson point process (PPP) with a certain intensity (density). Each TX has a unique intended receiver (RX) away from it by some (random) distance. Namely, such a heterogeneous wireless network can be viewed as consisting of $K$-type TX-RX pairs independently scattering on an infinitely large plane $\mathbb{R}^2$.  Usually, interference in such a wireless network significantly dominates the transmission performance that is effectively evaluated by the metric of the signal-to-interference power ratio (SIR) at RXs. By assuming all TXs in the network transmit narrow band signals and share the same spectrum bandwidth, the SIR of a typical type-$k$ RX located at the origin, called type-$k$ SIR, can be written as
\begin{align}\label{Eqn:IntroSINR}
\sinr_k\defn \frac{S_k}{I_k}=\frac{P_kH_kR^{-\alpha}_k}{I_k}>\theta,\,\,k\in\mathcal{K}\defn\{1,2,\ldots,K\},
\end{align}
where $I_k$ denotes the interference power of the typical type-$k$ RX from all interferers in the network, $S_k=P_kH_kR^{-\alpha}_k$ is the received (desired) signal power of the typical type-$k$ RX, $P_k$ is the transmit power, $H_k$ is the random channel (power) gain, $R_k$ is the (random) distance between the typical RX and its associated TX, $\alpha>2$ is the pathloss exponent, and $\theta$ is the minimum required SIR for successfully decoding. Note that $S_k$ and $I_k$ are both random variables whose distributions depend upon (random) transmit power, random channel gain as well as pathloss models between TX-RX pairs. The SIR pertaining to several important transmitting performance metrics, such as success/outage probability, ergodic link capacity, network capacity, etc. Understanding the statistical properties of the SIR not only helps us realize how the received random signal powers affect the distribution of the SIR, but also provides us a crucial clue indicating the interplay of the transmitting policies and behaviors among many different TXs. 

\subsection{Prior Work and Motivation}
Traditionally, the statistical properties of the SIR in a Poisson-distributed wireless network are only analytically accessible in very few special cases. Some prior works have already made a good progress on the analysis of the distribution of the SIR by presuming a specific channel gain model (typically see \cite{FBBBPM06,SWXYJGAGDV05,MHRKG09,MHJGAFBODMF10,CHLJGA12}). In reference \cite{FBBBPM06},  for example, the closed-form success probability, which is essentially the contemporary cumulative density function (CCDF) of the SIR in a single-type Poisson ad hoc network, was firstly found by assuming independent Rayleigh fading channels since the Rayleigh fading channel model gives rise to the  solvable Laplace transform of the interference by means of the probability generating function (PGF) of a homogeneous PPP \cite{DSWKJM13,FBBBL101,MHRKG09}. The outage probability, which is essentially the CDF of the  SIR, is studied  in \cite{SWXYJGAGDV05} without considering random channel gain models and only its bounds are obtained. Although the closed-form Laplace transform of the interference with a general random channel gain model is found in \cite{MHRKG09}, it can only be applied to find the CDF/CCDF of the SIR with an exponential-distributed received signal power. In \cite{CHLJGA12}, the bounds on the temporally averaged outage probability are studied specifically for Rayleigh fading channels due to the tractability in mathematical analysis. These prior works aim to study how channel gain randomness affects the success/outage probability so that they simply use constant transmit power and distance while doing analysis.  

Since the SIR significantly depends upon the randomness of the received signal and interference powers, some previous works focus on exploiting the SIR randomness by distributed channel-aware scheduling and power control in order to improve the success probability is \cite{SWJGANJ07,CHL13,CHLYCT13, NJSPWJGA08,XZMH12,CHLBRSC15}. However, the success/outage probability in these works is only characterized by some lower and upper bounds since it lacks of a tractable Laplace transform of the interference. These bounds may not be always tight in different ranges of the TX intensity even though they are claimed to be asymptotically tight. In addition to the success/outage probability, another important performance metric regarding to the SIR is the ergodic link capacity (rate) of a TX and its analytical results are barely completely and deeply investigated. In the literature, the ergodic link capacity is either characterized by its bounds or obtained by integrating the CCDF of the link capacity that can be written in terms of the success probability\cite{YGIBEZ13,CHLJGA11}. In other words, its accurate and simple expression is never discovered so that many prior works on network capacity (throughput) just simply use a minimum required constant link rate to define their capacity/throughput metrics \cite{FBBBPM06,SWXYJGAGDV05,MHRKG09,MHJGAFBODMF10,CHLJGA12,SWJGANJ07,RVKTTSWRH11,CHL13,CHLYCT13, NJSPWJGA08,XZMH12,CHLBRSC15}. Accordingly, the network capacity evaluated in the prior works may be far away from the real fundamental limit of the network capacity. 

In the literature, the distribution of the SIR is tractably derived only in the context where the received signal power contains an exponential random variable that creates a natural condition of applying the PGF of homogeneous PPPs to resolve the Laplace transform of the interference.  In other words, the distribution of the SIR is not significantly tractable if the receive signal power no longer possesses an exponential random variable. As a result, any transmitting policies adopted by TXs that make the received signal power loose/change its original exponential randomness cannot result in a tractable analysis in the distribution of the SIR. A straightforward example is to let TXs control their transmit power to compensate or cancel the Rayleigh fading gain in their channel and the success probability under this power control cannot be found in closed-form despite the fact that the original success probability without power control has a closed-form \cite{NJSPWJGA08}. To tractably study the statistical properties of the SIR with a general distribution, we need to find another way to deal with the interference generated in a Poisson field without utilizing the exponential randomness of the received signal power. Also, the heterogeneity of TXs is hardly modeled in prior works on Poisson ad hoc networks. Such a heterogeneity could exist in the future network of machine-to-machine (M2M) communication and internet of things (IoTs) and how it impacts the randomness of the SIR is still fairly unclear \cite{HSRRSIALTTRJAG15,SAOGAPMGTTJTJSMDYK15}. These aforementioned issues foster our motive to develop a generalized framework of analyzing the distribution of the SIR in a heterogeneous wireless network.

\subsection{Main Results and Contributions}
In this work, our first main contribution is to derive the integral identity of the Shannon transform as well as devise the novel theoretical framework of tractably analyzing the CDF of the type-$k$ SIR defined in \eqref{Eqn:IntroSINR}. The main idea behind this framework is to first find the explicitly result of the Laplace transform of the reciprocal of $\sinr_k$ with a general distribution since we can tractably deal with it by the PGL of homogeneous PPPs. Then substituting it into the exploited  fundamental identity between the CDF of  $\sinr_k$ and the Laplace transform of the reciprocal of $\sinr_k$. For the analytical framework regarding to the statistical properties of $\sinr_k$ with a general distribution, the following summarizes our main findings.
\begin{itemize}
\item The general expression of the CDF of $\sinr_k$ without and with interference cancellation is characterized\footnote{For the case of interference cancellation considered in this work, we assume each RX can cancel its first $L$ ``strongest" interferers, which is different from prior works that mainly assume RXs can cancel their ``nearest" interferers \cite{SWJGAXYGVJ07,XZMH14}. Considering the case of canceling the strongest interferers is more like the realistic situation since in practice the strongest interfering signal may not come from the nearest interfering TX due to channel fades.}, which can be practically evaluated by the numerical inverse Laplace transform. Its nearly closed-form result for pathloss exponent $\alpha=4$ is found, whereas its low-complexity and tight bounds for an arbitrary $\alpha>2$ are obtained as well.
\item We show that the closed-form CDF of the $\sinr_k$ exists if and only if received signal power $S_k$ has an Erlang distribution. Namely, any randomness in $S_k$ (from the random transmit power, channel gain and distance) that lets $S_k$ have an Erlang distribution can make the CDF of $\sinr_k$ have a closed-form result. 
\item The fractional moment of the $\sinr_k$ without interference cancellation is derived in closed-form and that with interference cancellation can be obtained in a neat integral expression.  
\end{itemize}

Due to the generality of the CDF and fractional moment of $\sinr_k$, they can be used to find the explicitly results of some important performance metrics in many transmitting and receiving scenarios. In this work, our second main contribution is to apply our developed analytical framework to tractably study the success probability and the ergodic link capacity that are the two paramount metrics of evaluating transmission performance.  For the success probability, the following are our main findings:
\begin{itemize}
\item The type-$k$ success probability without and with interference cancellation can be acquired directly from the CCDF of $\sinr_k$. Thus, its nearly closed-form expression also exists for $\alpha=4$ and its tight bounds for any $\alpha>2$ are also found. 
\item Since the success probability is found with a general-distributed SIR, it can be used to explicitly evaluate the success probabilities with specified random models involved in the SIR. This fact helps us theoretically show that random channel gains do not necessarily jeopardize/benefit the success probability (relative to constant channel gains) as the TX intensities change and it very likely improves the success probability in a dense network.  
\item Due to the generality of the derived success probability, the success probability with the proposed stochastic power control is tractably characterized by its bounds or found in a nearly closed-form expression depending on if $\alpha$ is equal to 4 or not. It also reveals how to design the power control scheme to improve the success probability by exploiting the randomness of the received signal power. 
\end{itemize}

The explicit expression of the ergodic link capacity in a Poisson network is hardly found in the literature when the SIR has a general distribution. In this work, we derive a low-complexity and general expression of the ergodic link capacity without and with interference cancellation by jointly using the derived integral identity of the Shannon transform and a novel integrating technique. According to the derived general results of the ergodic link capacity and the success probability, we define the spatial throughput capacity of the heterogeneous network that characterizes the area spectrum efficiency in an ergodic sense. We summarize some key observations of the ergodic link capacity and spatial throughput capacity as follows.     
\begin{itemize}
	\item Due to the generality of the derived ergodic link capacity, we can easily find the ergodic link capacity without and with channel fading so that we are able to conclude that channel fading does not always reduce or increase the ergodic link capacity (relative to no channel fading) as TX intensities change.
	\item The ergodic link capacity with the proposed stochastic power control is found and its fundamental upper and lower bounds are also characterized. These analytical results help us understand how to design the stochastic power control so that it can benefit the ergodic link capacity.    
	\item The spatial throughput capacity proposed in this work can characterize the network capacity with TX heterogeneity, and it is closer to the fundamental limit of the network capacity than other network metrics with a constant link rate in prior works. 
\end{itemize}
The salient trait of the derived CDF of $\sinr_k$ and its corresponding performance metrics, compared with related prior works, is to indicate how they are impacted by the interferences of other types. This provides a very useful clue in optimally deploying the network with some performance constraints. In addition, our main analytical results and findings are correctly validated by numerical simulations so that they can offer a quick and correct approach to evaluating the network performance with a new protocol and/or deployment design.

\section{System Model and Preliminaries} 
\subsection{Network Modeling and Performance Metrics}
Consider a large-scale and interference-limited heterogeneous wireless ad hoc network on the plane $\mathbb{R}^2$  in which there are $K$ different types of TXs and the TXs of each type form an independent homogeneous Poisson point process (PPP). Specifically, assume each TX has a unique intended RX and the set $\Phi_k$ consisting of the type-$k$ TXs with intensity $\lambda_k$ is expressed as
\begin{align}
\Phi_k\defn&\{(X_{k_i},H_{k_i},P_{k_i} ,R_{k_i}): X_{k_i}\in\mathbb{R}^2, P_{k_i}, H_{k_i}\in\mathbb{R}_+, R_{k_i}\in[1,\infty), i\in\mathbb{N}\},
\end{align}
where $k\in\mathcal{K}\defn\{1,2,\ldots,K\}$, $X_{k_i}$ denotes the $i$th nearest TX of type $k$ to the origin and its location, $H_{k_i}$ represents the random channel (power) gain from $X_{k_i}$ to the typical RX located at the origin induced by fading and/or shadowing effects, $P_{k_i}$ is the (random) transmit power of $X_{k_i}$, $R_{k_i}$ is the (random) distance between $X_{k_i}$ and its receiver. Throughout this paper, all random variables (RVs) with subscript ``$k_i$'' are independent for all $k\in\mathcal{K}$ and $i\in\mathbb{N}_+$ and they are i.i.d. for the same $k$. In addition, all channel gains $\{H_{k_i}\}$ have unit mean for all $k\in\mathcal{K}$ and $i\in\mathbb{N}_+$. The main variables and symbols used throughout this paper are listed in Table \ref{Tab:MathNotation}.

\begin{table}[!t]
\centering
\caption{Notation of Main Variables, Symbols and Functions}\label{Tab:MathNotation}
\begin{tabular}{|c|c|}
\hline
Symbol & Meaning\\ \hline
$\Phi_k$ ($\Phi$) &  Homogeneous PPP of the type-$k$ TXs ($\Phi\defn\bigcup^K_{k=1}\Phi_k$)\\
$\lambda_k$ & Intensity of the type-$k$ TXs\\
$\alpha>2$ & Pathloss exponent\\
$H_{k_i}$  & Channel gain of TX $X_{k_i}$ with unit mean\\
$P_{k_i}$ & Transmit power of TX $X_{k_i}$\\
$R_{k_i}$ & Transmit distance of TX $X_{k_i}$\\
$S_k (S^{pc}_k)$ & Type-$k$ received signal power (with power control)\\
$I_k (I^{pc}_k)$  & Interference at a type-$k$ RX (with power control) \\
$\sinr_k$ & Type-$k$ signal-to-interference power ratio\\
$\mathcal{L}\{\cdot\}(\mathcal{L}^{-1}\{\cdot\})$ & Laplace (Inverse Laplace) transform operator\\
$\mathcal{M}_Z(\tau)$ & $\mathbb{E}[e^{\tau Z}]$, Moment-generating function of RV $Z$ for $\tau>0$ \\
$a\gtrapprox b (a\lessapprox b)$ & $b$ is the tight lower (upper) bound on $a$\\
$p_k (p^{pc}_k)$ & Type-$k$ success probability (with power control)\\
$\theta$ & SIR decoding threshold\\
$c_k (c^{pc}_k)$ & Type-$k$ ergodic link capacity (with power control)\\
$C(\theta)$ & Throughput capacity with threshold $\theta$\\
$f_Z(\cdot) $ & pdf of random variable (RV) $Z$\\
$F_Z(\cdot) (F^c_Z(\cdot))$ & CDF (CCDF) of RV $Z$\\
$\widehat{Z}$ & $Z/\mathbb{E}[Z]$ (RV $Z$ normalized by its mean)\\
$\widetilde{\lambda}_k (\widetilde{\lambda})$ & $\lambda_k\mathbb{E}[H^{2/\alpha}_k]\mathbb{E}[P^{2/\alpha}_k]$ ($\widetilde{\lambda}\defn\sum_{k=1}^{K}\widetilde{\lambda}_k$)\\
$\gamma_k$ & Power control exponent\\
$L$ & Number of the canceled interferers \\
$D_L$ & Erlang RV with parameters $L$ and $\pi\widetilde{\lambda}$\\
$V_{k,L}$ & Variable $V_k$ with interference cancellation\\
$\Gamma(a)$ & $\int_{0}^{\infty} t^{a-1}e^{-t}\dif t$, Gamma function  \\
$\Gamma(a,b)$ & $\int_{b}^{\infty} t^{a-1}e^{-t}\dif t$, Upper incomplete Gamma function  \\
$\ell_z(y,x)$ & $z[1-xy^x\Gamma(-x,y)]$, $x\in(0,1)$ \\
$\mathrm{erf}(x)$ & $\frac{2}{\sqrt{\pi}}\int_{x}^{0}e^{-t^2}\dif t$, error function\\
$\mathrm{erfc}(x)$ & $1-\mathrm{erf}(x)$, complementary error function\\
\hline
\end{tabular}
\end{table}

Assume all TXs adopt the slotted Aloha protocol to access the channel shared in the network so that the type-$k$ typical RX receives the interference given by\footnote{We call this receiver located at the origin ``typical receiver'' since our following analysis will be based on the location of this typical RX and the statistical results obtained at this receiver are the same as those at all other RXs in the network based on the Slivnyak theorem \cite{DSWKJM13,FBBBL101,CHLJGA12}. }
\begin{align}\label{Eqn:Interference}
I_k\defn \sum_{X_{k_i}\in\Phi\setminus X_k} \frac{H_{k_i}P_{k_i}}{\|X_{k_i}\|^{\alpha}},
\end{align}
where $\Phi\defn\bigcup_{k=1}^K \Phi_k$, $\|X_i-X_j\|$ denotes the Euclidean distance between TXs $X_i$ and $X_j$, and $\alpha>2$ is the pathloss exponent. Accordingly, the type-$k$ SIR, as already defined in \eqref{Eqn:IntroSINR}, can be explicitly rewritten as follows
\begin{align}\label{Eqn:DefnSINR}
\sinr_k = \frac{S_k}{I_k}=\frac{P_kH_kR^{-\alpha}_k}{\sum_{X_{k_i}\in\Phi\setminus X_k}H_{k_i}P_{k_i}\|X_{k_i}\|^{-\alpha}}.
\end{align}
Assuming the minimum SIR threshold for  successful decoding the received signals at any RXs is $\theta$, the (transmitting) \textit{success probability} of a type-$k$ TX is defined as
\begin{align}\label{Eqn:DefnSuccessProb}
p_k(\theta)\defn \mathbb{P}\left[\sinr_k>\theta\right],
\end{align}
whereas $1-p_k$ is called the outage probability of a type-$k$ TX. Using the definitions of SIR and success probability, we define the type-$k$ ergodic link capacity as follows.
\begin{definition}[Ergodic Link Capacity]\label{Defn:LinkNetworkCapacity}
If the capacity-approaching code is used by all TXs, the ergodic link capacity (per unit bandwidth) of a type-$k$ TX-RX pair, called type-$k$ ergodic link capacity in the heterogeneous wireless ad hoc network, is defined as 
\begin{align}\label{Eqn:DefnErgodicRate}
c_k \defn \mathbb{E}\left[\log_2(1+\sinr_k)\right]\,\,(\text{bps/Hz}),\,\,k\in\mathcal{K}.
\end{align}

\end{definition}

In prior works, the explicit expression of the ergodic link capacity in Poisson wireless networks was not well studied and derived in a simple and general form. As we will show later, the type-$k$ ergodic link capacity can be derived in a very neat form by our new proposed mathematical derivation approach. Most importantly, our derived ergodic link capacity is able to explicitly indicate how it is affected by the random channel gain, transmit power and distance models, which provides very useful insight into devising the power control schemes in order to benefit the transmission performance by combating and/or exploiting the randomness in the SIR. 

\subsection{Preliminaries}
In this subsection, some preliminary results regarding the multiple independent homogeneous PPPs as well as the integral identity of the Shannon transform are introduced and discussed. These results are the underlying basis of paving a tractable way to analyze the success probability and ergodic link capacity in a very general manner. We first introduce the following theorem.
\begin{theorem}\label{Thm:SupBiasFunCDF}
Let $\Psi: \mathbb{R}_+\rightarrow \mathbb{R}_+$ be a Borel-measurable non-increasing function and it is positively scalable, i.e., for any $\beta, x\in\mathbb{R}_+$ we have $\Psi\left(\Psi^{-1}(\beta) x\right)=\beta\Psi(x)$ where $\Psi^{-1}(\cdot)$ denotes the inverse function of $\Psi(\cdot)$.  Suppose all $B_{k_i}$'s are independent nonnegative RVs for all $k\in\mathcal{K}$ and $i\in\mathbb{N}_+$ and they are i.i.d. for the same subscript $k$. If $\mathbb{E}\left\{\left[\Psi^{-1}(B_k)\right]^{-2}\right\}<\infty$ for all $k\in\mathcal{K}$, the following identity 
\begin{align}
\mathbb{P}\left[\sup_{X_{k_i}\in \Phi}B_{k_i}\Psi\left(\|X_i\|\right)\leq \Psi(x)\right]=\exp\left(-\pi x^2\lambda'\right),\label{Eqn:SupBiasFunCDF}
\end{align}
holds for the whole transmitter set $\Phi=\bigcup_{k\in\mathcal{K}} \Phi_k$ where $\lambda'\defn \sum_{k=1}^{K}\lambda'_k$  and $\lambda'_k\defn \lambda_k\mathbb{E}\left\{\left[\Psi^{-1}(B_k)\right]^{-2}\right\}$.
\end{theorem}
\begin{IEEEproof}
First we know the following identity
\begin{align*}
\mathbb{P}\left[\sup_{X_{k_i}\in\Phi}B_{k_i}\Psi\left(\|X_{k_i}\|\right)\leq \Psi(x)\right]&=\mathbb{E}_{\Phi}\left\{\prod_{X_{k_i}\in\Phi}\mathbb{P}\left[B_{k_i}\Psi\left(\|X_{k_i}\|\right)\leq \Psi(x)\right]\right\}\\\
&\stackrel{(a)}{=}\mathbb{E}_{\Phi}\left\{\prod_{X_{k_i}\in\Phi}\mathbb{P}\left[\Psi\left(\Psi^{-1}(B_{k_i})\|X_{k_i}\|\right)\leq \Psi(x)\right]\right\}\\
&\stackrel{(b)}{=}\exp\left(-2\pi\sum_{k=1}^{K}\lambda'_k\int_{0}^{\infty}\mathbb{P}\left[\Psi\left(\Psi^{-1}(B_k)r\right)\geq \Psi(x)\right] r\dif r\right)\\
&\stackrel{(c)}{=}\exp\left(-\pi\sum_{k=1}^{K}\lambda'_k\int_{0}^{\infty}\mathbb{P}\left[r^2\leq \left(\frac{x}{\Psi^{-1}(B_k)}\right)^2\right] \dif r^2\right),
\end{align*}
where $(a)$ following from the assumption that $\Psi(\cdot)$ is positively scalable, $(b)$ is obtained by using the probability generation functional (PGF) of $K$ independent homogeneous PPPs \cite{DSWKJM13}\cite{MH12}, (c) is due to the fact that $\Psi(\cdot)$ is non-increasing and invertible. Thus, carrying out the integral inside $\exp(\cdot)$ in $(c)$ yields the result in \eqref{Eqn:SupBiasFunCDF}.
\end{IEEEproof}

\noindent Theorem \ref{Thm:SupBiasFunCDF} implicitly reveals an important fact that the biased and transformed supreme distance between the origin and the TXs in $\Phi$ has the same distribution as the nearest distance between the origin and a single homogeneous PPP of intensity $\lambda'$. To elaborate on this point, letting $\Psi(x)=x^{-\alpha}$ and $B_{k_i}=P_{k_i}H_{k_i}$ gives $\sup_{X_{k_i}\in\Phi} P_{k_i}H_{k_i}\|X_{k_i}\|^{-\alpha}$ and we have
\begin{align*}
\mathbb{P}\left[\sup_{X_{k_i}\in\Phi} \frac{P_{k_i}H_{k_i}}{\|X_{k_i}\|^{\alpha}}\leq x^{-\alpha}\right]&=\mathbb{P}\left[\inf_{X_{k_i}\in\Phi} \frac{\|X_i\|}{(P_iH_i)^{\frac{1}{\alpha}}}\geq x\right]\\
&=\mathbb{P}\left[\inf_{X'_i\in\Phi'} \|X'_i\|\geq x\right]=\exp\left(-\pi x^2\lambda'\right),
\end{align*}
which indeed shows that $\inf_{X_{k_i}\in\Phi} (P_{k_i}H_{k_i})^{-\frac{1}{\alpha}}\|X_{k_i}\|$ is the biased shortest distance from the origin to $\Phi$ and it has the same distribution as the nearest distance from the origin to $\Phi'$. As we will see later, Theorem \ref{Thm:SupBiasFunCDF} facilitates the derivation processes in the analysis of interference cancellation. 

Another important result that needs to be introduced here is the identity of the Shannon transform. The Shannon transform of a nonnegative RV $Z$ for a nonnegative  $\varrho\in\mathbb{R}_+$ is defined as \cite{AMTSV04}
\begin{align}\label{Eqn:DefnShannonTransform}
\mathcal{S}_{Z}(\varrho) = \mathbb{E}\left[\ln(1+\varrho Z)\right],
\end{align}
which has an integral identity as shown in the following theorem. 
\begin{theorem}[The integral identity of the Shannon transform]\label{Thm:ShannonTransform}
Consider a nonnegative RV $Z$ and the Laplace transform of its reciprocal always exists, i.e., $\mathcal{L}_{Z^{-1}}(s)\defn\mathbb{E}[e^{-s Z^{-1}}]<\infty$. If $\mathcal{S}_{Z}(\varrho)$ defined in \eqref{Eqn:DefnShannonTransform} exists for any $\varrho\in\mathbb{R}_+$, the following identity
\begin{align}\label{Eqn:IdenShannonTrans}
\mathcal{S}_{Z}(\varrho) =\int_{0^+}^{\infty}\frac{(1-e^{-\varrho s})}{s}\mathcal{L}_{Z^{-1}}(s) \dif s 
\end{align}
always holds. Furthermore, we can have
\begin{align}\label{Eqn:IdenShannonTrans2}
\mathbb{E}\left[\mathcal{S}_{Z}(\varrho)\right] =\int_{0^+}^{\infty}\frac{[1-\mathcal{L}_{\varrho}(s)]}{s}\mathcal{L}_{Z^{-1}}(s) \dif s
\end{align}
if $\varrho$ is a nonnegative RV independent of $Z$ and its Laplace transform exists.
\end{theorem}
\begin{IEEEproof}
The Shannon transformation of nonnegative RV $Z$ defined in \eqref{Eqn:DefnShannonTransform} can be rewritten as
\begin{align*}
\mathcal{S}_Z(\varrho)=\int_{0}^{1} \mathbb{E}\left[\frac{\varrho Z}{1+y\varrho Z}\right]\dif y=\int_{0}^{1} \mathbb{E}\left[\frac{1}{1/\varrho Z+y}\right]\dif y.
\end{align*}
If $\mathcal{L}_{Z^{-1}}(s)$ always exists,  for any $y\in[0,1]$ we have
\begin{align*}
\mathbb{E}\left[\frac{1}{1/\varrho Z+y}\right]&=\int_{0}^{\infty} e^{-uy}\mathbb{E}\left[e^{-u/\varrho Z}\right]\dif u\\
&=\int_{0}^{\infty} e^{-uy}\mathcal{L}_{Z^{-1}}(u/\varrho) \dif u=\int_{0}^{\infty} e^{-\varrho sy}\mathcal{L}_{Z^{-1}}(s)\varrho \dif s
\end{align*}
and then substituting this result into $\mathcal{S}_Z(\varrho)$ yields
\begin{align*}
\mathcal{S}_Z(\varrho)&=\varrho\int_{0}^{\infty}\int_{0}^{1} e^{-\varrho sy}\mathcal{L}_{ Z^{-1}}(s)\dif y \dif s=\int_{0^+}^{\infty}\int_{0}^{\infty} \frac{(1-e^{-\varrho s})}{se^{\frac{s}{z}}}f_Z(z)\dif z \dif s \text{ (Letting $s=\varrho z$)}.\\
&=\int_{0^+}^{\infty}\frac{(1-e^{-\varrho s})}{s}\mathcal{L}_{Z^{-1}}(s) \dif s,
\end{align*}
which is exactly the result in \eqref{Eqn:IdenShannonTrans}. The result in \eqref{Eqn:IdenShannonTrans2} readily follows from \eqref{Eqn:IdenShannonTrans} and the definition of the Laplace transform of a nonnegative RV. 
\end{IEEEproof}
The ergodic link capacity in \eqref{Eqn:DefnErgodicRate} can be expressed in terms of the integral identity of the Shannon transform as
\begin{align}
c_k= \frac{\mathbb{E}[\mathcal{S}_{I^{-1}_k}(S_k)]}{\ln(2)}=\frac{1}{\ln(2)}\int_{0^+}^{\infty}\frac{1}{s}\left[1-\mathcal{L}_{S_k}(s)\right]\mathcal{L}_{I_k}(s)\dif s
\end{align}
since $S_k$  and $I_k$ are independent. This demonstrates that the integral identity of the Shannon transform is very useful in deriving the explicit expression of the ergodic link capacity if the Laplace transforms of the received signal power and interference are analytically tractable. More details about how to use the integral identity of the Shannon transform to find the ergodic link capacity of each type will be demonstrated in the following section. Next, we will first study the general distribution of the type-$k$ SIR that is regarding the statistical properties under general random channel gain, transmit power and distance models. The fundamental theory pertaining to the general distribution of the type-$k$ SIR will be established without specifying these random models involved in the SIR so that it is of the model-independent nature and valid for the distribution of the type-$k$ SIR with any specific random signal models. Most importantly, the theory not only straightforwardly indicates how different random models involved in the SIR influence the performance metrics regarding the SIR, but also gives us insight into exploiting the model randomness in order to enhance the SIR.  

\section{The Statistics of the Type-$k$ SIR with a General Distribution}
Prior works on the distribution of the SIR in Poisson wireless networks are channel-model-dependent and the majority of the prior works reached the closed-form distribution of the SIR with assuming communication channel gains are exponentially distributed (i.e., Rayleigh fading), whereas the distribution of the SIR for  the channel gains without an exponential distribution is generally intractable. As a result, the prior results cannot thoroughly reveal how the distribution of the SIR is impacted as the random models involved in the SIR are changed. In this section, our goal is to generally characterize the distribution expressions of the type-$k$ SIR with a general (unknown) distribution. The fundamental approach to fulling our goal is to first study the Laplace transform of the interference of the type-$k$ RXs since it plays a pivotal role in deriving the general distribution of the type-$k$ SIR. Surprisingly, we will see that some closed-form (or near closed-form) expressions of the distribution of the type-$k$ SIR indeed exist and they intuitively show how the statistical properties of the SIR are influenced by the randomness existing the SIR. 

\subsection{The Laplace Transforms of Interferences $I_k$ and $I_{k,L}$ }
Let $\Phi_L$ denote the set of the first $L$ strongest interferers in set $\{\Phi\setminus X_k\}$ for the type-$k$ typical RX. Hence, the type-$k$ interference $I_k$ can be rewritten as
\begin{align}\label{Eqn:InterferenceWithKsep}
I_k=\sum_{X_i\in\Phi_L}\frac{P_iH_i}{\|X_i\|^{\alpha}}+I_{k,L},
\end{align}
where $I_{k,L}\defn \sum_{X_{k_i}\in\Phi\setminus(\Phi_L\cup X_k)} \frac{H_{k_i}P_{k_i}}{\|X_{k_i}\|^{\alpha}}$ denotes the residual type-$k$ interference by removing the first $L$ strongest interferers in $I_k$. For arbitrary random power-law channel and transmit power models, the Laplace transforms of $I_k$ and $I_{k,L}$ are shown in the following theorem.
\begin{theorem}\label{Thm:LaplaceFunCanKinterfer}
According to the type-$k$ interference $I_k$ given in \eqref{Eqn:InterferenceWithKsep}, its Laplace transform can be shown in closed-form as\footnote{Note that the Laplace transform of the interference generated by a wireless ad hoc network consisting of single-type TXs has been shown in \cite{MHRKG09} for constant transmit power and distance, whereas $\mathcal{L}_{I_k}(s)$ in \eqref{Eqn:LapTranInter} generalizes it to the case of $K$ types of TXs with random transmit powers.}
\begin{align}\label{Eqn:LapTranInter}
\mathcal{L}_{I_k}(s) = \exp\left\{-\pi\Gamma\left(1-\frac{2}{\alpha}\right) s^{\frac{2}{\alpha}}\widetilde{\lambda}\right\},
\end{align}
where $\widetilde{\lambda}=\sum_{k=1}^{K}\widetilde{\lambda}_k$, $\widetilde{\lambda}_k\defn\lambda_k\mathbb{E}\left[H^{\frac{2}{\alpha}}_k\right]\mathbb{E}\left[P_k^{\frac{2}{\alpha}}\right]$ and $\Gamma(a)=\int_{0}^{\infty} t^{a-1}e^{-t}\dif t$ for $a>0$ is the gamma function. The Laplace transform of the residual interference $I_{k,L}$ defined in \eqref{Eqn:InterferenceWithKsep} can be found as
\begin{align}\label{Eqn:LaplaceCanKinterference}
\mathcal{L}_{I_{k,L}}(s)=\mathcal{L}_{I_k}(s)\cdot\mathcal{M}_{\ell_{D_L}(s)}\left(\pi\widetilde{\lambda}\right),
\end{align}
where $D_L\sim\text{Erlang}\left(L,\pi\widetilde{\lambda}\right)$ is an Erlang RV with parameters $L\in\mathbb{N}_+$ and $\pi\widetilde{\lambda}$, $\mathcal{M}_{\ell_{D_L}(s)}(\pi\widetilde{\lambda})\defn \mathbb{E}\left[e^{\pi\widetilde{\lambda}\ell_{D_L}(sD_L^{-\frac{\alpha}{2}},\frac{2}{\alpha})}\right]$,  and $\ell_z(y,x)$ with $z,y\in\mathbb{R}_+$  and $x\in(0,1)$
is defined as 
\begin{align}\label{Eqn:ellfun}
\ell_z\left(y,x\right)\defn z\left[1-x y^{x}\Gamma\left(-x,y\right)\right]
\end{align}
in which $\Gamma(a,y)=\int_{y}^{\infty}t^{a-1}e^{-t}\dif t$ is the upper incomplete gamma function. Also, there exists an $\omega\in(0,1)$ for each sample of $D_L$ such that $\mathcal{L}_{I_{k,L}}(s)$ is upper-bounded as
\begin{align}\label{Eqn:LaplaceInterUpperBound}
\mathcal{L}_{I_{k,L}}(s)\leq \mathcal{L}_{I_k}(s)\cdot\mathcal{M}_{\omega^{-\frac{\alpha}{2}}D_L^{1-\frac{\alpha}{2}}}\left(\pi\widetilde{\lambda}s\right),
\end{align}
where $\mathcal{M}_{\omega^{-\frac{\alpha}{2}}D^{1-\frac{\alpha}{2}}}(\pi\widetilde{\lambda}s)\defn\mathbb{E}\left[\exp(\pi\widetilde{\lambda}s\omega^{-\frac{\alpha}{2}}D^{1-\frac{\alpha}{2}}_L)\right]$. 
\end{theorem}
\begin{IEEEproof}
See Appendix \ref{App:ProofLaplaceFunCanKinterfer}.
\end{IEEEproof}
There are a couple of crucial implications about $\mathcal{L}_{I_{k,L}}(s)$ in \eqref{Eqn:LaplaceCanKinterference} that can be explained in more detail as follows. First, $\mathcal{L}_{I_{k,L}}(s)$ characterizes the statistical property of the interference which is partially cancelable at the RX side, and $\ell_{D_L}(sD_L^{-\frac{\alpha}{2}},\frac{2}{\alpha})$ in $\mathcal{M}_{\ell_{D_L}(s)}(\pi\widetilde{\lambda})$ for the case of $L=0$ reduces to 0 so that $\mathcal{L}_{I_{k,L}}(s)$ is exactly equal to $\mathcal{L}_{I_k}(s)$ in \eqref{Eqn:LapTranInter}. In other words, the second term of the right hand side in \eqref{Eqn:LaplaceCanKinterference} compensates the Laplace transform of the canceled interference for $\mathcal{L}_{I_{k,L}}(s)$. Second, although $\mathcal{L}_{I_{k,L}}(s)$ can be theoretically written as a neat result as shown in \eqref{Eqn:LaplaceCanKinterference}, $\mathcal{M}_{\ell_{D_L}(s)}(\pi\widetilde{\lambda})$ is actually somewhat complicate in practical applications and calculations. As a result, a low-complexity upper bound on $\mathcal{L}_{I_{k,L}}(s)$ is shown in \eqref{Eqn:LaplaceInterUpperBound}, and it is  very tight if $\widetilde{\lambda}s\omega^{-\frac{\alpha}{2}}D^{1-\frac{\alpha}{2}}_L$ is small with high probability  and it becomes tighter as $L$ gets larger since removing more interferers makes $D_L$ increase so that the upper bound gets closer to the Laplace transform of $I_{k,L}$.  In the following subsection, we will show how to apply Theorem \ref{Thm:LaplaceFunCanKinterfer} to characterize the distribution of the type-$k$ SIR without and with interference cancellation, which is the foundation of developing the generalized analytical approach to the success probability and ergodic link capacity.

\subsection{Analysis of the Distributions of $\sinr_k$ and $\sinr_{k,L}$}

The definition of the SIR of the type-$k$ typical RX with interference $I_k$ is already given in \eqref{Eqn:DefnSINR}. Similarly, if the type-$k$ typical RX is able to cancel the aggregated interference contributed by the first $L$ strongest interferers, its SIR in this case is defined as
\begin{align}\label{Eqn:SIRwKcan}
\sinr_{k,L}\defn \frac{S_k}{I_{k,L}},
\end{align} 
where interference $I_{k,L}$ is already defined in \eqref{Eqn:InterferenceWithKsep}. In this subsection,  our first focus is on the Laplace transforms of the reciprocals of $\sinr_k$ and $\sinr_{k,L}$ that play a pivotal role in deriving some important performance metrics of Poisson wireless networks, such as success probability, ergodic link capacity, etc.  The explicit distribution expressions regarding to $\sinr_k$ and $\sinr_{k,L}$ are summarized in the following theorem.
\begin{theorem}\label{Thm:LapTransInvSIR}
Let $f_Z(\cdot)$ and $F_z(\cdot)$ denote the probability density function (pdf) and the cumulative density function (CDF) of RV $Z$, respectively. The Laplace transform of the reciprocal of the type-$k$ SIR defined in \eqref{Eqn:DefnSINR} can be explicitly expressed as
\begin{align}\label{Eqn:LapInvSIR}
\mathcal{L}_{\sinr^{-1}_k}(s) =\int_{0}^{\infty}s\mathcal{L}_{I_k}\left(\frac{1}{t\mathbb{E}[S_k]}\right)f_{\widehat{S}_k}(s t)\dif t,
\end{align}
where $\mathcal{L}_{I_k}(\cdot)$ is given in \eqref{Eqn:LapTranInter} and $\widehat{S}_k\defn S_k/\mathbb{E}[S_k]=P_kH_kR^{-\alpha}_k/\mathbb{E}[P_kH_kR^{-\alpha}_k]$ is called the type-$k$ received signal power with unit mean. The  CDF of $\sinr_k$ can be shown as
\begin{align}\label{Eqn:CDFSIRTypek}
F_{\sinr_k}\left(\theta\right)=1-\mathcal{L}^{-1}\left\{ \int_{0}^{\infty} \mathcal{L}_{I_k}\left(\frac{1}{t\mathbb{E}[S_k]}\right)f_{\widehat{S}_k}(st)\dif t\right\}\left(\theta^{-1}\right),
\end{align}
where $\theta\in\mathbb{R}_+$. If each type-$k$ RX can cancel its first $L$ strongest interferers, the Laplace transform of the reciprocal of $\sinr_{k,L}$ in \eqref{Eqn:SIRwKcan} can be expressed as
\begin{align}\label{Eqn:LapInvSIRcanK}
\mathcal{L}_{\sinr^{-1}_{k,L}}(s)= \int_{0}^{\infty}s\mathcal{L}_{I_k}\left(\frac{1}{t\mathbb{E}[S_k]}\right)\mathcal{M}_{\ell_{D_L}(s)}\left(\pi\widetilde{\lambda}\right)f_{\widehat{S}_k}(st)  \dif t.
\end{align}
Also, the CDF of $\sinr_{k,L}$ can be shown as
\begin{align}\label{Eqn:CDFSIRTypekKcan}
F_{\sinr_{k,L}}\left(\theta\right)=1-\mathcal{L}^{-1}\left\{\int_{0}^{\infty}\mathcal{L}_{I_k}\left(\frac{1}{t\mathbb{E}[S_k]}\right)\mathcal{M}_{\ell_{D_L}(\frac{1}{t\mathbb{E}[S_k]})}\left(\pi\widetilde{\lambda}\right) f_{\widehat{S}_k}(st)  \dif t \right\}\left(\theta^{-1}\right),
\end{align}
where $\mathcal{M}_{\ell_{D_L}(\frac{1}{t\mathbb{E}[S_k]})}\left(\pi\widetilde{\lambda}\right)=\mathbb{E}\left\{\exp\left[\pi\widetilde{\lambda}\ell_{D_L}\left(1/D^{\frac{\alpha}{2}}_L\mathbb{E}[S_k]t,\frac{2}{\alpha}\right)\right]\right\}$.
\end{theorem}
\begin{IEEEproof}
See Appendix \ref{App:ProofLapTransInvSIR}.
\end{IEEEproof}

Theorem \ref{Thm:LapTransInvSIR} demonstrates the general expressions of the Laplace transforms of $\sinr^{-1}_k$ and $\sinr^{-1}_{k,L}$ as well as the CDFs of $\sinr_k$ and $\sinr_{k,L}$ without assuming any specific random channel gain, transmit power and distance models. Although in general the expressions in Theorem \ref{Thm:LapTransInvSIR} cannot be completely found in closed-form, they can be calculated by using the numerical inverse Laplace transform. Nonetheless, as shown in the following corollary we still can characterize the low-complexity bounds on $F_{\sinr_k}(\theta)$ and $F_{\sinr_k}(\theta)$ and the near closed-form of $F_{\sinr_k}(\theta)$ and $F_{\sinr_k}(\theta)$ for $\alpha=4$ without specifying the distribution of $S_k$.
\begin{corollary}\label{Cor:BoundCDFSIRk}
For a general $\alpha>2$, the CDF of $\sinr_k$ in \eqref{Eqn:CDFSIRTypek} can be bounded as shown in the following:
\begin{align}\label{Eqn:BoundsCDFSIR}
\min\left\{1,\pi\widetilde{\lambda}\mathbb{E}\left[S^{-\frac{2}{\alpha}}_k\right]\theta^{\frac{2}{\alpha}}\right\}\geq F_{\sinr_k}(\theta) \geq \mathcal{L}^{-1}\left\{\frac{\pi\Gamma(1-\frac{2}{\alpha})\widetilde{\lambda}}{s^{1-\frac{2}{\alpha}}\left(\pi\Gamma(1-\frac{2}{\alpha})\widetilde{\lambda}s^{\frac{2}{\alpha}}+\mathbb{E}\left[S^{\frac{2}{\alpha}}_k\right]\right)}\right\}(\theta^{-1}).
\end{align}
In particular, if $\alpha=4$, then $F_{\sinr_k}(\theta)$ can be simply found as
\begin{align}\label{Eqn:CDFSIRkAlpha4}
F_{\sinr_k}(\theta) = \mathbb{E}\left[\mathrm{erf}\left(\frac{\pi^{\frac{3}{2}}\widetilde{\lambda}\sqrt{\theta}}{2\sqrt{S_k}}\right)\right]
\end{align}
in which $\mathrm{erf}(x)\defn \frac{2}{\sqrt{\pi}}\int_{0}^{x} e^{-t^2} \dif t$ is the error function and  $\widetilde{\lambda}=\sum_{j=1}^{K}\lambda_j\mathbb{E}[\sqrt{P_j}]\mathbb{E}[\sqrt{H_j}]$, and thus we have the following closed-form bounds on $F_{\sinr_k}(\theta)$ in \eqref{Eqn:CDFSIRkAlpha4}
\begin{align}\label{Eqn:BoundsCDFSIRAlpha4}
\mathrm{erf}\left(\frac{\pi^{\frac{3}{2}}\widetilde{\lambda}\sqrt{\theta}}{2}\mathbb{E}\left[\frac{1}{\sqrt{S_k}}\right]\right) \geq F_{\sinr_k}(\theta) \geq \exp\left\{\left(\frac{\mathbb{E}\left[\sqrt{S_k}\right]}{\pi^{\frac{3}{2}}\widetilde{\lambda}\sqrt{\theta}}\right)^2\right\}\mathrm{erfc}\left(\frac{\mathbb{E}\left[\sqrt{S_k}\right]}{\pi^{\frac{3}{2}}\widetilde{\lambda}\sqrt{\theta}}\right),
\end{align}
where $\mathrm{erfc}(x)=1-\mathrm{erf}(x)$. Whereas there exists a lower bound on the CDF of $\sinr_{k,L}$ given by
\begin{align}\label{Eqn:LowBoundCDFKCan}
F_{\sinr_{k,L}}(\theta)\geq
1- \mathcal{L}^{-1}\left\{\frac{1}{s}\mathbb{E}\left[\mathcal{L}_{I_k}\left(\frac{s}{S_k}\right)\mathcal{M}_{\omega^{-\frac{\alpha}{2}}D^{1-\frac{\alpha}{2}}_L}\left(\frac{\pi\widetilde{\lambda}s}{S_k}\right)\right]\right\}(\theta^{-1}).
\end{align}
\end{corollary}
\begin{IEEEproof}\label{App:ProofBoundCDFSIRk}
The CDF of $\sinr_k$ in \eqref{Eqn:CDFSIRTypek} can be rewritten as
\begin{align}
F_{\sinr_k}(\theta) &= \mathcal{L}^{-1}\left\{\mathbb{E}\left[\frac{1}{s}\left(1-e^{-\pi\Gamma(1-\frac{2}{\alpha})\widetilde{\lambda}(s/S_k)^{\frac{2}{\alpha}}}\right)\right]\right\}(\theta^{-1})\nonumber\\
&=\mathbb{E}\left[\mathcal{L}^{-1}\left\{\frac{1}{s}\left(1-e^{-\pi\Gamma(1-\frac{2}{\alpha})\widetilde{\lambda}(s/S_k)^{\frac{2}{\alpha}}}\right)\right\}(\theta^{-1})\right].\label{Eqn:DefCDFSIRkInvLap}
\end{align}
Using the inequality $\frac{x}{1+x}\leq 1-e^{-x}\leq x$ for $x>0$, the upper bound on the result in \eqref{Eqn:DefCDFSIRkInvLap} is
\begin{align*}
F_{\sinr_k}(\theta) \leq  \mathbb{E}\left[\mathcal{L}^{-1}\left\{\frac{\pi\Gamma(1-\frac{2}{\alpha})\widetilde{\lambda}}{s^{1-\frac{2}{\alpha}}S^{\frac{2}{\alpha}}_k}\right\}(\theta^{-1})\right]=\pi\widetilde{\lambda}\mathbb{E}\left[S^{-\frac{2}{\alpha}}_k\right]\theta^{\frac{2}{\alpha}}.
\end{align*}
and
\begin{align*}
F_{\sinr_k}(\theta) &\geq \mathcal{L}^{-1}\left\{\mathbb{E}\left[\frac{\pi\Gamma(1-\frac{2}{\alpha})\widetilde{\lambda}}{s^{1-\frac{2}{\alpha}}\left(\pi\Gamma(1-\frac{2}{\alpha})\widetilde{\lambda}s^{\frac{2}{\alpha}}+S^{\frac{2}{\alpha}}_k\right)}\right]\right\}(\theta^{-1})\\
&\geq \mathcal{L}^{-1}\left\{\frac{\pi\Gamma(1-\frac{2}{\alpha})\widetilde{\lambda}}{s^{1-\frac{2}{\alpha}}\left(\pi\Gamma(1-\frac{2}{\alpha})\widetilde{\lambda}s^{\frac{2}{\alpha}}+\mathbb{E}\left[S^{\frac{2}{\alpha}}_k\right]\right)}\right\}(\theta^{-1})\nonumber\\
&\stackrel{\alpha=4}{=} \exp\left\{\left(\frac{\mathbb{E}\left[\sqrt{S_k}\right]}{\pi^{\frac{3}{2}}\widetilde{\lambda}\sqrt{\theta}}\right)^2\right\}\text{erfc}\left(\frac{\mathbb{E}\left[\sqrt{S_k}\right]}{\pi^{\frac{3}{2}}\widetilde{\lambda}\sqrt{\theta}}\right),
\end{align*}
where the second inequality holds due to the convexity of $1/(a+x)$ for $a,x>0$ and the final equality follows from solving the inverse Laplace transform for $\alpha=4$. Therefore, the upper and lower bounds in \eqref{Eqn:BoundsCDFSIR} and \eqref{Eqn:BoundsCDFSIRAlpha4} are acquired. For $\alpha=4$, the inverse Laplace transform in \eqref{Eqn:DefCDFSIRkInvLap} can be found in closed-form so that we have
\begin{align}
F_{\sinr_k}(\theta) = \mathbb{E}\left[\text{erf}\left(\frac{\pi^{\frac{3}{2}}\widetilde{\lambda}\sqrt{\theta}}{2\sqrt{S_k}}\right)\right]\leq \mathrm{erf}\left(\frac{\pi^{\frac{3}{2}}\widetilde{\lambda}\sqrt{\theta}}{2}\mathbb{E}\left[\frac{1}{\sqrt{S_k}}\right]\right),
\end{align}
where the upper bound is obtained by applying Jensen's inequality to $\mathrm{erf}(x)$ that is concave for $x>0$. For the CDF of $\sinr_{k,L}$, it can be written as
\begin{align*}
F_{\sinr_{k,L}}(\theta) =1- \mathcal{L}^{-1}\left\{\frac{1}{s}\mathbb{E}\left[\mathcal{L}_{I_{k,L}}\left(\frac{s}{S_k}\right)\right]\right\}(\theta^{-1})
\end{align*}
and its lower bound in \eqref{Eqn:LowBoundCDFKCan} is readily obtained by replacing $s$ in \eqref{Eqn:LaplaceInterUpperBound} with $s/S_k$. 
\end{IEEEproof}

We can elaborate on a couple of implications of Corollary \ref{Cor:BoundCDFSIRk} as follows.
\begin{itemize}
\item When $\widetilde{\lambda}\mathbb{E}\left[S^{-\frac{2}{\alpha}}_k\right]\ll 1$ (e.g., the mean of the interference-to-signal power ratio is fairly small), $F_{\sinr_k}(\theta)$ is accurately approximated by the inverse Laplace transform of the Taylor's expansion of the $1-\exp(\cdot)$ term in \eqref{Eqn:DefCDFSIRkInvLap} as
\begin{align}\label{Eqn:CDFSIRApprox}
F_{\sinr_k}(\theta)\approx \sum_{n=1}^{\lfloor\alpha/2 \rfloor} \frac{(-1)^{n+1}}{\Gamma(1-\frac{2n}{\alpha})}\left[\Gamma\left(1-\frac{2}{\alpha}\right)\pi\theta^{\frac{2}{\alpha}}\widetilde{\lambda}\right]^n \mathbb{E}\left[S_k^{-\frac{2n}{\alpha}}\right],
\end{align}
where $\lfloor x \rfloor\defn\max\{y\in\mathbb{Z}: y\leq x\}$. Namely, we have  $F_{\sinr_k}(\theta)\in\Theta\left(\widetilde{\lambda}\mathbb{E}\left[S^{-\frac{2}{\alpha}}_k\right]\right)$ for a given $\theta>0$ as $\widetilde{\lambda}\mathbb{E}\left[S^{-\frac{2}{\alpha}}_k\right]$ approaches zero\footnote{Throughout this paper, we use the standard asymptotic notations to denote the scaling results in this paper:  $O(\cdot)$, $\Omega(\cdot)$ and $\Theta(\cdot)$ correspond to (asymptotic) upper, lower, and tight bounds, respectively. For instance, given two real-valued functions $f(x)$ and $g(x)$, we use $f(x)\in\Theta(g(x))$ to mean that there exist two positive constants $c_1$ and $c_2$ such that $c_1g(x)\leq f(x)\leq c_2 g(x)$ for $x\rightarrow 0$ or $x\rightarrow\infty$.}. In other words, $F_{\sinr_k}(\theta)$ in \eqref{Eqn:CDFSIRApprox} is very accurate in this case and the bounds in \eqref{Eqn:BoundsCDFSIR} are very tight since they coverage to each other eventually. 
\item For  $\alpha=4$, the neat expression of $F_{\sinr_k}(\theta)$ and its closed-form bounds exist, and they can precisely reveal how much the disparate distributions of $S_k$ affect $F_{\sinr_k}$. Third, in general, the lower bound in \eqref{Eqn:LowBoundCDFKCan} is very tight if $\widetilde{\lambda}D^{1-\frac{\alpha}{2}}_L/\omega^{\frac{\alpha}{2}}S_k\ll 1$ with high probability so that using   $e^x\approx 1+x$ for $x\ll 1$ and the Laplace transform table in \cite{MAIAS72} makes $F_{\sinr_{k,L}}$ tightly lower-bounded for $\alpha=4$ as
\begin{align}\label{Eqn:LowBoundCDFKCanAlpha4}
F_{\sinr_{k,L}}(\theta)\gtrapprox\mathbb{E}\left[\mathrm{erf}\left(\frac{\pi^{\frac{3}{2}}\widetilde{\lambda}\sqrt{\theta}}{2\sqrt{S_k}}\right)\right]-\mathbb{E}\left[\frac{(\pi\widetilde{\lambda})^2\theta^{\frac{3}{2}}}{2S^{\frac{3}{2}}_k\exp\left(\frac{\pi^3\widetilde{\lambda}\theta}{4S_k}\right)}\right]\mathbb{E}\left[(\omega^2D_L)^{-1}\right],
\end{align}
where $a\gtrapprox b$ ($a \lessapprox b$) denotes that $a$ is the \textit{tight} lower (upper) bound on $b$. This tight lower bound implies $F_{\sinr_{k,L}}(\theta)\in\Omega\left(\widetilde{\lambda}\mathbb{E}\left[1/\sqrt{S_k}\right]\right)$ as well as the effect of interference cancellation is offset by strong received signal power $S_k$. \textit{Interference cancellation benefits more the RXs with a weaker received signal power}.
\end{itemize}
 
For the case of received signal power $S_k$ having an Erlang distribution, the closed-form results of $F_{\sinr_k}(\theta)$ and $F_{\sinr_{k,L}}(\theta)$ in Theorem \ref{Thm:LapTransInvSIR} indeed exist, as shown in following corollary.
\begin{corollary}\label{Cor:CDFSIRErlang}
If the type-$k$ received signal power $\widehat{S}_k$ with unit mean is an Erlang RV (i.e., $\widehat{S}_k\sim\text{Erlang}(\mu,\mu)$ where $\mu\in\mathbb{N}_+$), then we have
\begin{align}\label{Eqn:CDFSIRErlang}
 F_{\sinr_k}\left(\theta\right)=1-\frac{1}{(\mu-1)!}\frac{\dif^{\mu-1} }{\dif v^{\mu-1}}\left[v^{\mu-1}\mathcal{L}_{I_k}\left(\frac{\mu}{v\mathbb{E}[S_k]}\right)\right]\bigg|_{v=\theta^{-1}}
\end{align}
and
\begin{align}\label{Eqn:CDFSIRErlangKcan}
F_{\sinr_{k,L}}\left(\theta\right)=1-\frac{1}{(\mu-1)!}\frac{\dif^{\mu-1}}{\dif v^{\mu-1}}\left[v^{\mu-1}\mathcal{L}_{I_k}\left(\frac{\mu}{v\mathbb{E}[S_k]}\right)\mathcal{M}_{\ell_{D_L}(\frac{\mu}{v\mathbb{E}[S_k]})}\left(\pi\widetilde{\lambda}\right)\right]\bigg|_{v=\theta^{-1}}.
\end{align} 
\end{corollary}
\begin{IEEEproof}
Since we assume $\widehat{S}_k\sim\text{Erlang}(\mu,\mu)$, $F_{\sinr_k}(\theta)$ in \eqref{Eqn:CDFSIRTypek} can be written as
\begin{align*}
F_{\sinr_k}(\theta)&= 1- \mathcal{L}^{-1}\left\{\frac{\mu^{\mu}s^{\mu-1}}{(\mu-1)!} \int_{0}^{\infty} \mathcal{L}_{I_k}\left(\frac{1}{t\mathbb{E}[S_k]}\right)t^{\mu-1}e^{-\mu st}\dif t\right\}\left(\theta^{-1}\right)\\
&=1- \frac{1}{(\mu-1)!}\mathcal{L}^{-1}\left\{s^{\mu-1} \int_{0}^{\infty} \left[\mathcal{L}_{I_k}\left(\frac{\mu}{v\mathbb{E}[S_k]}\right)v^{\mu-1}\right]e^{-sv}\dif v\right\}\left(\theta^{-1}\right),
\end{align*}
and using the identity $\mathcal{L}\left\{\frac{\dif^{\mu}}{\dif t^{\mu}}g(t)\right\}(s)=s^{\mu}\int_{0}^{\infty}g(t) e^{-st}\dif t$ to simplify $F_{\sinr}(\theta)$ in above yields the result in \eqref{Eqn:CDFSIRErlang}.
Similarly, $F_{\sinr_{k,L}}(\theta)$ in \eqref{Eqn:CDFSIRTypekKcan} can be further expressed as
\begin{align*}
F_{\sinr_{k,L}}(\theta)=1-\frac{1}{(\mu-1)!}\mathcal{L}^{-1}\left\{s^{\mu-1}\int_{0}^{\infty} v^{\mu-1}\mathcal{L}_{I_k}\left(\frac{\mu}{v\mathbb{E}[S_k]}\right)\mathcal{M}_{\ell_{D_L}(\frac{\mu}{v\mathbb{E}[S_k]})}\left(\pi\widetilde{\lambda}\right)e^{-sv} \dif v \right\}\left(\theta^{-1}\right),
\end{align*}
which is exactly equal to the result in \eqref{Eqn:CDFSIRErlangKcan} due to the identity $\mathcal{L}\left\{\frac{\dif^{\mu}}{\dif t^{\mu}}g(t)\right\}(s)=s^{\mu}\int_{0}^{\infty}g(t) e^{-st}\dif t$.
\end{IEEEproof}

\noindent  For any particular value of $\mu$, the explicit closed-form expressions of $F_{\sinr_k}(\theta)$ and $F_{\sinr_{k,L}}(\theta)$ can be easily found by carrying out the $\mu$th-order derivatives in \eqref{Eqn:CDFSIRErlang} and \eqref{Eqn:CDFSIRErlangKcan}, respectively. For instance, in the special case of $\mu=1$, i.e., $\widehat{S}_k\sim \exp(1,1)$ is an exponential RV with unit mean and variance\footnote{This could happen in the case that the transmit power and distance are constant and the communication channel undergoes Rayleigh fading so that its gain distribution is $\exp(1,1)$.}, $F_{\sinr_k}(\theta)$ in \eqref{Eqn:CDFSIRErlang} and $F_{\sinr_{k,L}}(\theta)$ in \eqref{Eqn:CDFSIRErlangKcan} respectively reduce to
\begin{align}\label{Eqn:CDFSIRExpDis}
F_{\sinr_k}(\theta)=1-e^{-\pi\Gamma\left(1-\frac{2}{\alpha}\right)\widetilde{\lambda}(\theta/\mathbb{E}[S_k])^{\frac{2}{\alpha}}}=1-\mathcal{L}_{I_k}\left(\frac{\theta}{\mathbb{E}[S_k]}\right).
\end{align}
and
\begin{align}\label{Eqn:CDFSIRExpDisKcan}
F_{\sinr_{k,L}}(\theta) =1-F^c_{\sinr_k}(\theta)\mathcal{M}_{\ell_{D_L}(\frac{\theta}{\mathbb{E}[S_k]})}\left(\pi\widetilde{\lambda}\right)=1-\mathcal{L}_{I_k}\left(\frac{\theta}{\mathbb{E}[S_k]}\right)\mathcal{M}_{\ell_{D_L}(\frac{\theta}{\mathbb{E}[S_k]})}\left(\pi\widetilde{\lambda}\right),
\end{align}
where $F^c_{\sinr_k}(\theta)=1-F_{\sinr_k}(\theta)$ denotes the CCDF of $\sinr_k$. Note that $F_{\sinr_k}(\theta)$ in \eqref{Eqn:CDFSIRExpDis} obviously shows that $\sinr_k$ has a Weibull distribution with parameters $\frac{2}{\alpha}$ and $\mathbb{E}[S_k]/(\pi\Gamma(1-\frac{2}{\alpha})\widetilde{\lambda})^{\frac{\alpha}{2}}$, which was shown in \cite{MHRKG09} for the network with a single PPP. Accordingly, we can say that $\sinr_{k,L}$ has a ``modified'' Weibull distribution with parameters $\frac{2}{\alpha}$ and $\mathbb{E}[S_k]/(\pi\Gamma(1-\frac{2}{\alpha})\widetilde{\lambda})^{\frac{\alpha}{2}}$ since $F_{\sinr_{k,L}}(\theta)$ can be expressed in terms of $F_{\sinr_k}(\theta)$.

Another case that $F_{\sinr_k}(\theta)$ and $F_{\sinr_{k,L}}(\theta)$ in Theorem \ref{Thm:LapTransInvSIR} can be found in a simpler form is when the received signal power $S_k$ does not possess any randomness, as shown in the following corollary.
\begin{corollary}\label{Cor:CDFSIRNoFading}
If the received signal power of a type-$k$ RX is not a random variable, i.e., $S_k$ in \eqref{Eqn:DefnSINR} and \eqref{Eqn:SIRwKcan} is deterministic, the CDFs of $\sinr_k$ in \eqref{Eqn:CDFSIRTypek} and $\sinr_{k,L}$ in \eqref{Eqn:CDFSIRTypekKcan} reduce to
\begin{align}\label{Eqn:CDFSIRNoRamSig}
F_{\sinr_k}(\theta) = 1-\mathcal{L}^{-1}\left\{\frac{1}{s}\mathcal{L}_{I_k}\left(\frac{s}{S_k}\right)\right\}\left(\theta^{-1}\right)
\end{align}
and
\begin{align}\label{Eqn:CDFSIRNoRamSigKcan}
F_{\sinr_{k,L}}(\theta) =1-\mathcal{L}^{-1}\left\{\frac{1}{s}\mathcal{L}_{I_k}\left(\frac{s}{S_k}\right)\mathcal{M}_{\ell_{D_L}(\frac{s}{S_k})}\left(\pi\widetilde{\lambda}\right)  \right\}\left(\theta^{-1}\right),
\end{align}
respectively.
\end{corollary}
\begin{IEEEproof}
Notice that $F_{\sinr_k}(\theta)$ in \eqref{Eqn:CDFSIRTypek} can be rewritten as follows
\begin{align*}
F_{\sinr_k}\left(\theta\right)&=1-\mathcal{L}^{-1}\left\{ \int_{0}^{\infty}\frac{1}{s} \mathcal{L}_{I_k}\left(\frac{s}{u\mathbb{E}[S_k]}\right)f_{\widehat{S}_k}(u)\dif u\right\}\left(\theta^{-1}\right)\\
&=1-\mathcal{L}^{-1}\left\{ \mathbb{E}_{S_k}\left[\frac{1}{s}\mathcal{L}_{I_k}\left(\frac{s}{S_k}\right)\right]\right\}\left(\theta^{-1}\right).
\end{align*}
Thus, if $S_k$ is a constant, we readily obtain \eqref{Eqn:CDFSIRNoRamSig}. Similarly, $F_{\sinr_{k,L}}(\theta)$ in \eqref{Eqn:CDFSIRTypekKcan} also can be rewritten as
\begin{align*}
F_{\sinr_{k,L}}\left(\theta\right)=1-\mathcal{L}^{-1}\left\{\frac{1}{s}\mathbb{E}_{S_k}\left[\mathcal{L}_{I_k}\left(\frac{s}{S_k}\right)\mathcal{M}_{\ell_{D_L}(\frac{s}{S_k})}\left(\pi\widetilde{\lambda}\right)\right] \right\}\left(\theta^{-1}\right),
\end{align*}
which readily reduces to \eqref{Eqn:CDFSIRNoRamSigKcan} if $S_k$ is a constant. 
\end{IEEEproof}
Although the inverse Laplace transforms in \eqref{Eqn:CDFSIRNoRamSig}  and \eqref{Eqn:CDFSIRNoRamSigKcan} in general still cannot be explicitly calculated, they can be evaluated by the numerical inverse Laplace transform for any particular value of $\theta$. For $\pi\widetilde{\lambda}/S^{\frac{2}{\alpha}}_k\ll 1$, the closed-form approximation of $F_{\sinr_k}(\theta)$ also can be inferred from \eqref{Eqn:CDFSIRApprox} as
\begin{align}
F_{\sinr_k}(\theta) \approx \sum_{n=1}^{\lfloor\alpha/2 \rfloor} \frac{(-1)^{n+1}}{\Gamma(1-\frac{2n}{\alpha})}\left[\Gamma\left(1-\frac{2}{\alpha}\right)\pi\widetilde{\lambda}\left(\frac{\theta}{S_k}\right)^{\frac{2}{\alpha}}\right]^n.
\end{align}
Furthermore, for the special case of $\alpha=4$, \eqref{Eqn:CDFSIRNoRamSig} has a closed-form expression directly obtained from \eqref{Eqn:CDFSIRkAlpha4} as
\begin{align}\label{Eqn:CDFSIDNoFadingAlpha4}
F_{\sinr_k}(\theta) =\text{erf}\left(\frac{\pi^{\frac{3}{2}}\widetilde{\lambda}}{2}\sqrt{\frac{\theta}{S_k}}\right),
\end{align}
where  $\widetilde{\lambda}=\sum_{j=1}^{K}\lambda_j\mathbb{E}\left[\sqrt{H_j}\right]\mathbb{E}\left[\sqrt{P_j}\right]$and $S_k=P_kH_kR^{-4}_k$ is a constant, and the closed-form tight lower bound on $F_{\sinr_{k,L}}(\theta)$ for $\alpha=4$ also can be found as
\begin{align}\label{Eqn:LowBoundCDFKCanAlpha4NoFading}
F_{\sinr_{k,L}}(\theta)\gtrapprox \mathrm{erf}\left(\frac{\pi^{\frac{3}{2}}\widetilde{\lambda}\sqrt{\theta}}{2\sqrt{S_k}}\right)-\frac{(\pi\widetilde{\lambda})^2\theta^{\frac{3}{2}}}{2S^{\frac{3}{2}}_k\exp\left(\frac{\pi^3\widetilde{\lambda}\theta}{4S_k}\right)}\mathbb{E}\left[\frac{1}{\omega^2D_L}\right],
\end{align}
which is directly inferred from \eqref{Eqn:LowBoundCDFKCanAlpha4}. The result in \eqref{Eqn:CDFSIDNoFadingAlpha4} is fascinating (even though it is only valid for $\alpha=4$) since it shows that the outage probability with constant transmit power, distance and no channel fading has a closed-form result and it is only characterized by bounds in \cite{SWXYJGAGDV05}.

Although the CDFs of $\sinr_k$ and $\sinr_{k,L}$ in Theorem \ref{Thm:LapTransInvSIR} are characterized for an arbitrary distribution of received signal power $S_k$, directly using them to find the moments of the SIRs cannot acquire a neat or tractable expression. In the following theorem, we show that the fractional moment of the $\sinr_k$ and $\sinr_{k,L}$ can be explicitly found in a closed/neat form for a general distribution of $S_k$.
\begin{theorem}\label{Thm:MomentSIR}
For any $\delta\in\mathbb{R}_{++}$, the fractional moment of $\sinr_k$ in \eqref{Eqn:DefnSINR} can be shown as
\begin{align}\label{Eqn:NthMomSIR}
\mathbb{E}\left[\sinr_k^{\delta}\right] =\frac{\Gamma\left(1+\frac{\delta\alpha}{2}\right)(\mathbb{E}[S_k])^{\delta}}{\left[\pi\Gamma\left(1-\frac{2}{\alpha}\right)\widetilde{\lambda}\right]^{\frac{\delta\alpha}{2}}}.
\end{align}
and the fractional moment of $\sinr_{k,L}$ is given by
\begin{align}\label{Eqn:NthMomSIRKcan}
\mathbb{E}\left[\sinr^{\delta}_{k,L}\right] = \bigintsss_{0}^{\infty}\mathcal{L}_{I_k}\left(\frac{t^{\frac{1}{\delta}}}{\mathbb{E}[S_k]}\right)\mathcal{L}_{\widetilde{\ell}^L_{k,t^{-1/\delta}}}\left(\pi\widetilde{\lambda}\right) \dif t.
\end{align}
\end{theorem}
\begin{IEEEproof}
See Appendix \ref{App:ProofMeanSIR}.
\end{IEEEproof}
\noindent The results in Theorem \ref{Thm:MomentSIR} can be applied to estimate the success probability and the ergodic link capacity in some special contexts, as we will expound these applications in the following sections. Moreover, they also indicate that the fractional moment of the type-$k$ SIR increases as long as $\widetilde{\lambda}/(\mathbb{E}[S_k])^{\frac{2}{\alpha}}$ decreases by maintaining some randomness in the SIR. To clarify this point easily, for example, assuming transmit power $P_k$ is a constant and transmit distance $R_k$ is unity, the term $\mathbb{E}[(P_kH_k)^{2/\alpha}]/\mathbb{E}[S_k]$ in $\widetilde{\lambda}/(\mathbb{E}[S_k])^{\frac{2}{\alpha}}$ reduces to $\mathbb{E}[H^{2/\alpha}_k]/(\mathbb{E}[H_k])^{2/\alpha}$ which is less than or equal to one based on Jensen's inequality. This manifests that channel randomness benefits the fractional moment of the SIR. Exploiting the signal randomness in the SIR can increase its fractional moment. 

\section{Application of the SIR Statistics(I): Success Probability}
The CDF of the SIR found in the previous section has a paramount application in evaluating the success probability defined in \eqref{Eqn:DefnSuccessProb}. For the successful decoding threshold $\theta>0$, the success probability of a type-$k$ RX (called type-$k$ success probability) without or with interference cancellation can be simply written as
\begin{align}\label{Eqn:SuccessProb}
p_k(\theta)=F^c_{\sinr_k}(\theta)\quad\text{and}\quad p_{k,L}(\theta)=F^c_{\sinr_{k,L}}(\theta).
\end{align}
Now we first specify how to apply the CDF of the SIR to find the success probabilities in some practical application contexts of the random received signal power models. Afterwards, some numerical results are provided to validate our analytical findings.  

\subsection{Success Probability with General Random Models of the Received Signal Power}\label{SubSec:SuccProbWithGenRXSigPower}
 In general, the success probability $p_k(\theta)$ cannot be derived in an explicit closed form based on \eqref{Eqn:CDFSIRTypek} if  $S_k$ does not have an Erlang distribution. Nonetheless, the bounds on $p_k(\theta)$ and $p_{k,L}(\theta)$ can be characterized as shown in the following corollary. 
\begin{corollary}
The type-$k$ success probability  without interference cancellation can be bounded as follows
\begin{align}\label{Eqn:BoundSuccProbNoIntCan}
\left(1-\pi\widetilde{\lambda}\mathbb{E}\left[S^{-\frac{2}{\alpha}}_k\right]\theta^{\frac{2}{\alpha}}\right)^+ \leq p_k(\theta) \leq \mathcal{L}^{-1}\left\{\frac{\mathbb{E}\left[S^{\frac{2}{\alpha}}_k\right]}{s\left(\pi\Gamma(1-\frac{2}{\alpha})\widetilde{\lambda}s^{\frac{2}{\alpha}}+\mathbb{E}\left[S^{\frac{2}{\alpha}}_k\right]\right)}\right\}(\theta^{-1}).
\end{align}
If $\alpha=4$, $p_k(\theta)$ has a nearly closed-form expression given by
\begin{align}
 p_k(\theta)&=\mathbb{E}\left[ \mathrm{erfc}\left(\frac{\pi^{\frac{3}{2}} \widetilde{\lambda}}{2}\sqrt{\frac{\theta}{S_k}}\right)\right],\label{Eqn:ExactSuccProbAlpha4}
 \end{align}  
 where $\widetilde{\lambda}=\sum_{j=1}^{K}\lambda_j\mathbb{E}[\sqrt{P_j}]\mathbb{E}[\sqrt{H_j}]$ and $S_k=P_kH_kR^{-4}_k$. 
\end{corollary}
\begin{IEEEproof}
The proof is omitted since it is similar to the proof of Corollary \ref{Cor:BoundCDFSIRk}.
\end{IEEEproof}
\noindent In addition, using the error function's Maclaurin series \eqref{Eqn:ExactSuccProbAlpha4} can be further written as
 \begin{align}
 p_k(\theta)&= 1-\frac{2}{\sqrt{\pi}}\sum_{n=0}^{\infty} \frac{(-1)^n}{n!(2n+1)}\left(\frac{\pi^{\frac{3}{2}}\widetilde{\lambda}\sqrt{\theta}}{2}\right)^{2n+1}\mathbb{E}\left[\left(\frac{1}{\sqrt{S_k}}\right)^{2n+1}\right]\label{Eqn:SuccProbAlpha4} \\
 &\geq \text{erfc}\left(\frac{\pi^{\frac{3}{2}}\widetilde{\lambda}\sqrt{\theta}}{2}\mathbb{E}\left[\frac{1}{\sqrt{S_k}}\right]\right),\label{Eqn:LowBoundSuccProbAlpha4} 
 \end{align}
 where the lower bound in \eqref{Eqn:LowBoundSuccProbAlpha4} is obtained by applying Jensen's inequality to the erfc function with a positive argument that is convex. Although the result in \eqref{Eqn:ExactSuccProbAlpha4} is derived by considering the special case of $\alpha=4$, it is still very important since it is applicable to any random channel gain, transmit power and distance models and able to directly provide some insight into how the randomness of the received signal power affects the success probability. Note that $p_k(\theta)$ in \eqref{Eqn:SuccProbAlpha4} reduces to \eqref{Eqn:LowBoundSuccProbAlpha4} if $S_k$ is constant. 

In order to have a more tractable result of $p_k(\theta)$ in practically applicable contexts with a general pathloss exponent, we consider normalized received signal power $\widehat{S}_k$ as a Gamma random variable with mean 1 and variance $1/m_k$, i.e., $\widehat{S}_k\sim\text{Gamma}(m_k,1/m_k)$, for $m_k\in\mathbb{N}_+$. Such a received signal power model is somewhat general because it characterizes the different randomness levels\footnote{For constant transmit power $P_k$ and distance $R_k$, $\widehat{S}_k\sim\mathrm{Gamma}(m_k, 1/m_k)$ means the communication channel of the type-$k$ RX suffers Nakagami-$m_k$ fading and $H_k\sim\mathrm{Gamma}(m_k, 1/m_k)$} of $S_k$. According to the results in Corollary \ref{Cor:CDFSIRErlang} and $f_{\widehat{S}_k}(x)=\frac{m_k^{m_k}x^{m_k-1}}{e^{m_k x}\Gamma(m_k)}$, $p_k(\theta)$ without interference cancellation based on \eqref{Eqn:CDFSIRErlang} for a positive integer $m_k$ can be readily obtained by
 \begin{align}\label{Eqn:SuccProbGammaDis}
 p_k(\theta)=\frac{1}{(m_k-1)!}\frac{\dif^{m_k-1} }{\dif v^{m_k-1}}\left[v^{m_k-1}\mathcal{L}_{I_k}\left(\frac{m_k}{v\mathbb{E}[S_k]}\right)\right]\bigg|_{v=\theta^{-1}},
 \end{align}
 whose closed-form expression can be explicitly calculated once $m_k$ is designated. For the special case of $\widehat{S}_k\sim\exp(1,1)$ and Rayleigh fading interference channels, $p_k(\theta)$ in \eqref{Eqn:SuccProbGammaDis} reduces to  
 \begin{align}\label{Eqn:SuccProbFixDisRayFadNoCan}
 p_k(\theta)=\exp\left(-\frac{2\pi^2 \theta^{\frac{2}{\alpha}}\sum_{j=1}^{K}\lambda_j\mathbb{E}\left[P^{\frac{2}{\alpha}}_j\right]}{\alpha\sin(2\pi/\alpha)(\mathbb{E}[S_k])^{\frac{2}{\alpha}}}\right),
 \end{align}
 which reduces to the seminal result firstly shown in \cite{FBBBPM06} for $K=1$, constant transmit power and distance. The success probability in \eqref{Eqn:SuccProbGammaDis} reveals a very important fact that \textit{the closed-form success probability exists as long as the received signal power has an Erlang distribution}. This overthrows the traditional impression that the success probability only has a closed-form result for constant transmit power, distance and Rayleigh fading channels. 
 
 When the type-$k$ RXs can cancel their first $L$ strongest interferers, the type-$k$ success probability in \eqref{Eqn:SuccessProb}, i.e., $p_{k,L}(\theta)$, is readily obtained by $F_{\sinr_{k,L}}(\theta)$ in \eqref{Eqn:CDFSIRTypekKcan} for a general distribution of $S_k$ and has an upper bound as shown in the following corollary.
 \begin{corollary}\label{Lem:UppBoundSuccProfwithCan}
The type-$k$ success probability with canceling the first $L$ strongest interferers is given by
\begin{align}
p_{k,L}(\theta) = \mathcal{L}^{-1}\left\{\bigintsss_{0}^{\infty}\mathcal{L}_{I_k}\left(\frac{1}{t\mathbb{E}[S_k]}\right)\mathcal{M}_{\ell_{D_L}(\frac{1}{t\mathbb{E}[S_k]})}\left(\pi\widetilde{\lambda}\right)f_{\widehat{S}_k}(st)  \dif t \right\}\left(\theta^{-1}\right),\label{Eqn:SuccProbKCan}
\end{align}\
and its upper bound is given by
\begin{align}\label{Eqn:UppBoundSuccProbKCan}
p_{k,L}(\theta)\leq \mathcal{L}^{-1}\left\{\frac{1}{s}\mathbb{E}\left[\mathcal{L}_{I_k}\left(\frac{s}{S_k}\right)\mathcal{M}_{\omega^{-\frac{\alpha}{2}}D^{1-\frac{\alpha}{2}}_L}\left(\frac{\pi\widetilde{\lambda}s}{S_k}\right)\right]\right\}(\theta^{-1}).
\end{align}
 \end{corollary}
 \begin{IEEEproof}
Since we know $p_{k,L}(\theta)=1-F_{\sinr_{k,L}}(\theta)$, \eqref{Eqn:SuccProbKCan} and \eqref{Eqn:UppBoundSuccProbKCan} are directly inferred from \eqref{Eqn:CDFSIRTypekKcan} and \eqref{Eqn:LowBoundCDFKCan}, respectively.
 \end{IEEEproof}
\noindent Furthermore, if $\widehat{S}_k\sim\Gamma(m_k,1/m_k)$ with $m_k\in\mathbb{N}_+$, then $p_{k,L}(\theta)$ can be readily found by \eqref{Eqn:CDFSIRErlangKcan} as
 \begin{align}
 p_{k,L}(\theta)= \frac{1}{(m_k-1)!}\frac{\dif^{m_k-1}}{\dif v^{m_k-1}}\left[v^{m_k-1}\mathcal{L}_{I_k}\left(\frac{m_k}{v\mathbb{E}[S_k]}\right)\mathcal{M}_{\ell_{D_L}(\frac{m_k}{v\mathbb{E}[S_k]})}\left(\pi\widetilde{\lambda}\right)\right]\bigg|_{v=\theta^{-1}},
 \end{align}
and then its low-complexity upper bound can be found as
 \begin{align}\label{Eqn:SuccProbNakagamiKCanBound}
 p_{k,L}(\theta)\leq \frac{1}{(m_k-1)!}\frac{\dif^{m_k-1}}{\dif v^{m_k-1}}\left[v^{m_k-1}\mathcal{L}_{I_k}\left(\frac{m_k}{v\mathbb{E}[S_k]}\right)\mathcal{M}_{\omega^{-\frac{\alpha}{2}}D^{1-\frac{\alpha}{2}}_L}\left(\frac{m_k\pi\widetilde{\lambda}}{v\mathbb{E}[S_k]}\right)\right]\bigg|_{v=\frac{1}{\theta}}
 \end{align}
by using Corollary \ref{Lem:UppBoundSuccProfwithCan} and the closed-form upper bound can be found once the value of $m_k$ is designated. For example, if $m_k=1$ (i.e., $\widehat{S}_k\sim\exp(1,1)$) for all $k\in\mathcal{K}$, the results in \eqref{Eqn:SuccProbNakagamiKCanBound} and  \eqref{Eqn:SuccProbNakagamiKCanBound} reduce to a simple closed form given by
 \begin{align}
 p_{k,L}(\theta)&=\mathcal{L}_{I_k}\left(\frac{\theta}{\mathbb{E}[S_k]}\right)\times \mathcal{M}_{\ell_{D_L}(\frac{\theta}{\mathbb{E}[S_k]})}\left(\pi\widetilde{\lambda}\right)\label{Eqn:SuccProbRayleighKCan}\\
 &\lessapprox\mathcal{L}_{I_k}\left(\frac{\theta}{\mathbb{E}[S_k]}\right) \mathcal{M}_{\omega^{-\frac{\alpha}{2}}D^{1-\frac{\alpha}{2}}_L}\left(\frac{\pi\widetilde{\lambda}\theta }{\mathbb{E}[S_k]}\right).\label{Eqn:SuccProbRayleighKCanBound}
 \end{align}
Note that \eqref{Eqn:SuccProbRayleighKCan} is somewhat complicate in practical computation even though it is the exact expression of $p_{k,L}(\theta)$ without specifying the distribution of $S_k$, whereas the tight upper bound in \eqref{Eqn:SuccProbRayleighKCanBound} presents a low-complex formula of evaluating how different received signal random models impact the success probability with interference cancellation. 

\subsection{Does the Randomness of the Received Signal Power Jeopardize the Success Probability?}\label{SubSec:ChannelRanHelpSuccProb}
The randomness of the received signal power could be induced by random channel gain, transmit distance and power. When it can definitely benefit or jeopardize the SIR has not yet been analytically shown and explained. In the following theorem, we show that the randomness of the received signal power does not necessarily jeopardize the success probability and it even could benefit the success probability if certain condition is satisfied. 
\begin{theorem}\label{Thm:FadingImproveSuccProb}
Suppose the type-$k$ received signal power with unit mean is a Gamma RV with mean 1 and variance $1/m_k$, i.e., $\widehat{S}_k\sim\text{Gamma}(m_k,1/m_k)$ for all $k\in\mathcal{K}$. For $m_k\in\mathbb{N}_+$ and a given $\theta$, define set $\widetilde{\mathbf{\Lambda}}_k(\theta)$ as follows
\begin{align}
\widetilde{\mathbf{\Lambda}}_k(\theta)\defn&\left\{\widetilde{\Lambda}_k\in\mathbb{R}_{++}:\exp\left(-\Gamma\left(1-\frac{2}{\alpha}\right)\left(m_k\theta\right)^{\frac{2}{\alpha}}\widetilde{\Lambda}_k\right)\leq 1-\varsigma_k\theta^{\frac{2}{\alpha}}\widetilde{\Lambda}_k\right\},\label{Eqn:IntensitySetFadingGoodSuccProb}
\end{align}
where $\widetilde{\Lambda}_k\defn \pi\widetilde{\lambda}/(\mathbb{E}[S_k])^{\frac{2}{\alpha}}$, $\varsigma_k\defn \frac{(m_k-1)!}{\prod_{i=1}^{m_k-1} (i-\frac{2}{\alpha})}\mathds{1}(m_k>1)+\mathds{1}(m_k=1)$ and $\mathds{1}(\mathcal{A})$ is an indicator function which is one if event $\mathcal{A}$ is true and zero otherwise. If the network has $\pi\widetilde{\lambda}/(\mathbb{E}[S_k])^{\frac{2}{\alpha}}\in\widetilde{\mathbf{\Lambda}}_k(\theta)$ which is nonempty, the randomness of received signal power $S_k$ must reduce the success probability of the type-$k$ TXs.
\end{theorem}
\begin{IEEEproof}
See Appendix \ref{App:ProofFadingImproveSuccProb}.
\end{IEEEproof}
Theorem \ref{Thm:FadingImproveSuccProb} also indicates that the  randomness of $S_k$ could benefit $p_k$ if $\pi\widetilde{\lambda}(\mathbb{E}[S_k])^{-\frac{2}{\alpha}}\notin\widetilde{\mathbf{\Lambda}}_k(\theta)$.
To make this point more understandable, consider a simple example of the success probability with pathloss exponent $\alpha=4$, constant transmission distance $R_k$, constant transmit power $P_k$ and all channels with Rayleigh fading, i.e., $H_{k_i}\sim\exp(1,1)$. In this example, the success probability with Rayleigh fading is $p_k(\theta)=\exp\left(-\sqrt{\pi\theta}\widetilde{\Lambda}_k\right)$ inferred from \eqref{Eqn:SuccProbFixDisRayFadNoCan} and the success probability without fading is $p_k(\theta)=\mathrm{erfc}\left(\sqrt{\theta}\widetilde{\Lambda}_k\right)$ based on \eqref{Eqn:ExactSuccProbAlpha4} so that Rayleigh fading benefits the success probability as  $\exp\left(-\sqrt{\pi\theta}\widetilde{\Lambda}_k\right)<\mathrm{erfc}\left(\sqrt{\theta}\widetilde{\Lambda}_k\right)$, i.e., we have $\widetilde{\Lambda}_k=\frac{\pi^{3/2}R_k^2}{2\sqrt{P_k}}\sum_{j=1}^{K}\lambda_j\sqrt{P_j}<\frac{0.8951}{\sqrt{\theta}}$ and thus $\mathbf{\widetilde{\Lambda}}_k(\theta)=\left(0,\frac{0.8951}{\sqrt{\theta}}\right)$. Another crucial finding shown in Theorem \ref{Thm:FadingImproveSuccProb} is that \textit{we are able to improve the success probability by changing the setups of the transmit powers and transmitter intensities based on the channel fading status in the network}. That is, if channels do not suffer fading, the setups of the transmit powers and transmitter intensities that make $\widetilde{\mathbf{\Lambda}}_k(\theta)$ \textit{empty} help improve the success probability for the given threshold $\theta$. On the contrary, we should change the setups of the transmit powers and TX intensities that make $\widetilde{\mathbf{\Lambda}}_k(\theta)$ \textit{nonempty} to increase the type-$k$ success probability if channels suffer (severe) fading. 

\subsection{Success Probability with Stochastic Power Control}\label{SubSec:SuccProbPowerControl}
In this subsection, we would like to investigate how to improve the success probability by designing distributed stochastic power control schemes that change the distribution of the received signal power. The \textit{centralized} power control schemes are difficult to be implemented in this heterogeneous ad hoc network since all TXs only  know its own local information and cannot optimize their transmit powers jointly in order to maximize their success probabilities. According to the explicit results of the success probability in Section \ref{SubSec:SuccProbWithGenRXSigPower}, the key to maximizing the success probability of the type-$k$ TXs is how to minimize the term $\widetilde{\lambda}/(\mathbb{E}[S_k])^{\frac{2}{\alpha}}$ by optimally devising distributed power control schemes. Since each TX only has its local information available, we specifically propose the stochastic power control scheme for a type-$k$ TX as follows
\begin{align}\label{Eqn:StochasticPowerControl}
P_k=\frac{\overline{P}_k\left(H_kR^{-\alpha}_k\right)^{\gamma_k}}{\mathbb{E}\left[H^{\gamma_k}_k\right]\mathbb{E}\left[R^{-\alpha\gamma_k}_k\right]},
\end{align}
where $\overline{P}_k$ is the mean of transmit power $P_k$, $\gamma_k$ is the power control exponent needed to be designed. When there is no power control (i.e., constant transmit power is used), $P_k=\overline{P}_k$ (i.e., $\gamma_k=0$). This power control scheme is motivated by the fractional power control in \cite{NJSPWJGA08} and the fact that the randomness of the received signal power could improve the success probability, as already pointed out in Section \ref{SubSec:ChannelRanHelpSuccProb}. We can change $\gamma_k$ to adjust the randomness of $S_k$ to improve the success probability  in different network contexts. Therefore, the fundamental problem needed to be firstly studied is how the stochastic power control in \eqref{Eqn:StochasticPowerControl} changes/benefits the type-$k$ success probability. The success probability with stochastic power control was essentially intractable in prior works, whereas it becomes much more tractable if using the success probability results found in Section \ref{SubSec:ChannelRanHelpSuccProb}. The following theorem presents the type-$k$ success probability with the proposed stochastic power control, denoted by $p^{pc}_k(\theta)$. 

\begin{theorem}\label{Thm:SuccProbPowerCon}
Suppose all the type-$k$ TXs adopt the stochastic power control given in \eqref{Eqn:StochasticPowerControl}. Let $S_k=\overline{P}_kH_kR^{-\alpha}_k$ here be the received signal power without stochastic power control and the CCDF of $S_k$ has the property $\mathbb{E}[F^c_{S_k}(Z)]\leq F^c_{S_k}(\mathbb{E}[Z])$ for a nonnegative RV $Z$. For $\gamma_k>-1$, the bounds on the type-$k$ success probability with stochastic power control are shown as 
\begin{align}\label{Eqn:SuccProbPowerCon}
F^c_{S_k}\left(\frac{\Gamma(1+\frac{\alpha}{2(1+\gamma_k)})(\theta \mathbb{E}[S^{\gamma_k}_k])^{\frac{1}{1+\gamma_k}}}{\left[\pi\Gamma(1-\frac{2}{\alpha})\widetilde{\lambda}^{pc}\right]^{\frac{\alpha}{2(1+\gamma_k)}}}\right)\geq p^{pc}_k(\theta) \geq\left(1-\pi\widetilde{\lambda}^{pc}\left(\mathbb{E}\left[S^{\gamma_k}_k\right]\right)^{\frac{2}{\alpha}}\mathbb{E}\left[S^{-\frac{2(1+\gamma_k)}{\alpha}}_k\right]\theta^{\frac{2}{\alpha}}\right)^+,
\end{align}
where  superscript $pc$ means ``power control'' and $\widetilde{\lambda}^{pc}$ is given by
\begin{align}\label{Eqn:IntensityPowerControl}
\widetilde{\lambda}^{pc}=\sum_{j=1}^{K}\lambda_j\overline{P}^{\frac{2}{\alpha}}_j\mathbb{E}\left[H^{\frac{2}{\alpha}}_j\right]\frac{ \mathbb{E}\left[H^{\frac{2\gamma_j}{\alpha}}_j\right]}{(\mathbb{E}[H^{\gamma_j}_j])^{\frac{2}{\alpha}}}\frac{\mathbb{E}\left[R^{-2\gamma_j}_j\right]}{(\mathbb{E}[R^{-\alpha\gamma_j}_j])^{\frac{2}{\alpha}}},
\end{align}
which is smaller than $\widetilde{\lambda}=\sum_{j=1}^{K}\lambda_j\overline{P}^{\frac{2}{\alpha}}_j\mathbb{E}\left[H^{\frac{2}{\alpha}}_j\right]$. Furthermore, if $\alpha=4$, then $p^{pc}_k(\theta)$ has the following simple identity
\begin{align}\label{Eqn:SuccProbPowerContAlpha4}
p^{pc}_k(\theta) = \mathbb{E}\left[\mathrm{erfc}\left(\frac{\pi^{\frac{3}{2}}\widetilde{\lambda}^{pc}}{2}\sqrt{\frac{\theta\mathbb{E}[S^{\gamma_k}_k]}{S^{\gamma_k+1}_k}}\right)\right],
\end{align} 
where $\widetilde{\lambda}^{pc}$ is given in \eqref{Eqn:IntensityPowerControl} with $\alpha=4$ and $S_k=\overline{P}_kH_kR^{-4}_k$, and its lower bound is 
\begin{align}\label{Eqn:LowBoundSuccProbPowerContAlpha4}
p^{pc}_k(\theta) \geq \mathrm{erfc}\left(\frac{\pi^{\frac{3}{2}}\widetilde{\lambda}^{pc}\sqrt{\theta}}{2}\mathbb{E}\left[S^{-\frac{(\gamma_k+1)}{2}}_k\right]\sqrt{\mathbb{E}[S^{\gamma_k}_k]}\right).
\end{align}

\end{theorem}
\begin{IEEEproof}
See Appendix \ref{App:ProofSuccProbPowerCon}. 
\end{IEEEproof}
According to Theorem \ref{Thm:SuccProbPowerCon}, the stochastic power control scheme with nonzero $\gamma_k$ can reduce the interference since $\widetilde{\lambda}^{pc}<\widetilde{\lambda}$. This also implies that \textit{the ``randomness" of transmit power always results in less interference no matter if the power depends on the channel gain and/or pathloss}. Nonetheless, this does not mean the stochastic power control always benefits the success probability since it may not enhance the received signal power without using a proper value of $\gamma_k$. To make stochastic power control benefit the type-$k$ success probability, this condition $p^{pc}_k(\theta)>p_k(\theta)$ must hold, which poses the constraint on the values of $\gamma_k$ that are able to improve the type-$k$ success probability. Unfortunately, the explicitly constraints on $\gamma_k$'s for all $k\in\mathcal{K}$ are only tractably to be found for some special cases. 


To understand more about the context in which the stochastic power control in \eqref{Eqn:StochasticPowerControl} definitely benefits the success probability, consider the special case of $\alpha=4$ and the success probability in \eqref{Eqn:SuccProbPowerContAlpha4}. To make the stochastic power control benefit $p_k(\theta)$ in this case, the following inequalities must hold:
\begin{align}\label{Eqn:CondPowControlOutperformAlpha4}
p^{pc}_k(\theta) =\mathbb{E}\left[\mathrm{erfc}\left(\frac{\pi^{\frac{3}{2}}\widetilde{\lambda}^{pc}}{2}\sqrt{\frac{\theta\mathbb{E}[S^{\gamma_k}_k]}{S^{\gamma_k+1}_k}}\right)\right]&\geq\mathrm{erfc}\left(\frac{\pi^{\frac{3}{2}}\widetilde{\lambda}^{pc}\sqrt{\theta}}{2}\mathbb{E}\left[S^{-\frac{(\gamma_k+1)}{2}}_k\right]\sqrt{\mathbb{E}[S^{\gamma_k}_k]}\right)\nonumber\\ &>\mathbb{E}\left[\mathrm{erfc}\left(\frac{\pi^{\frac{3}{2}}\widetilde{\lambda}}{2}\sqrt{\frac{\theta}{S_k}}\right)\right].
\end{align}
By assuming $H_k\sim\exp(1,1)$, constant $R_k$ and $\gamma_k=\gamma$ for all $k\in\mathcal{K}$, we have $P_k=\overline{P}_kH^{\gamma+1}_k/\Gamma(1+\gamma)$, $\widetilde{\lambda}=\frac{\sqrt{\pi}}{2}\sum_{j=1}^{K}\lambda_j\sqrt{\overline{P}_j}$, $\widetilde{\lambda}^{pc}=\widetilde{\lambda}\Gamma(1+\frac{\gamma}{2})/\sqrt{\Gamma(1+\gamma)}$,$\sqrt{\mathbb{E}[S^{\gamma}_k]}=\sqrt{\Gamma(1+\gamma)}\overline{P}^{\gamma/2}_k/R^{2\gamma}_k$, $\sqrt{\mathbb{E}[S_k]}=\sqrt{\overline{P}_k}/R^{2}_k $, $\mathbb{E}[S^{-(1+\gamma)/2}_k]=\Gamma(\frac{1-\gamma}{2})R^{2(1+\gamma)}_k/\overline{P}^{\frac{(1+\gamma)}{2}}_k$ and thus \eqref{Eqn:CondPowControlOutperformAlpha4} reduces to
\begin{align}\label{Eqn:SimpCondPowControlOutperformAlpha4}
p^{pc}_k(\theta) &\geq\mathrm{erfc}\left(\frac{1}{2}\Gamma\left(1+\frac{\gamma}{2}\right)\Gamma\left(\frac{1-\gamma}{2}\right)\frac{\pi^{\frac{3}{2}}\sqrt{\theta}\widetilde{\lambda}}{\sqrt{\mathbb{E}[S_k]}}\right)\nonumber\\
&>\mathbb{E}\left[\mathrm{erfc}\left(\frac{\pi^{\frac{3}{2}}\widetilde{\lambda}}{2}\sqrt{\frac{\theta}{S_k}}\right)\right]= \exp\left(-\frac{\pi^{\frac{3}{2}}\sqrt{\theta}\widetilde{\lambda}}{\sqrt{\mathbb{E}[S_k]}}\right).
\end{align}
The stochastic power control with any $\gamma\in(-1,1)$ that satisfies this inequality outperforms no power control in terms of the success probability. This demonstrates that we need to appropriately use different $\gamma_k$ based on different values of $\widetilde{\lambda}/\sqrt{\mathbb{E}[S_k]}$ so that stochastic power control can always outperform no power control, which will be validated by the numerical results in Section \ref{Subsec:NumericalResultSuccProb}.

\subsection{Numerical Results}\label{Subsec:NumericalResultSuccProb}
\begin{table}[!t]
	\centering
	\caption{Network Parameters for Simulation}\label{Tab:SimPara}
	\begin{tabular}{|c|c|c|c|}
	\hline Parameter $\setminus$ TX Type $k$ & Type 1 & Type 2 & Type 3\\ \hline
	Transmit Power $P_k$ (W) & 1 & 0.5  & 0.05 \\ \hline
	Intensity $\lambda_k$ (TXs/m$^2$) & $\lambda_1$ & $5\lambda_1$ & $10\lambda_1$  \\ \hline
	Pathloss Exponent $\alpha$ &\multicolumn{3}{c|}{4}\\ \hline 
	Transmit Distance $R_k$ (m) &\multicolumn{3}{c|}{10}\\ \hline 
	Channel Gain $H_{k_i}$ & \multicolumn{3}{c|}{$\sim \exp(1,1)$} \\ \hline
	SIR Threshold $\theta$ &  \multicolumn{3}{c|}{1} \\ \hline
	Power control exponent $\gamma_k$ &  \multicolumn{3}{c|}{$\gamma$} \\ \hline
	\end{tabular} 
\end{table}

\begin{figure}[t!]
	\centering
	\includegraphics[scale=0.35]{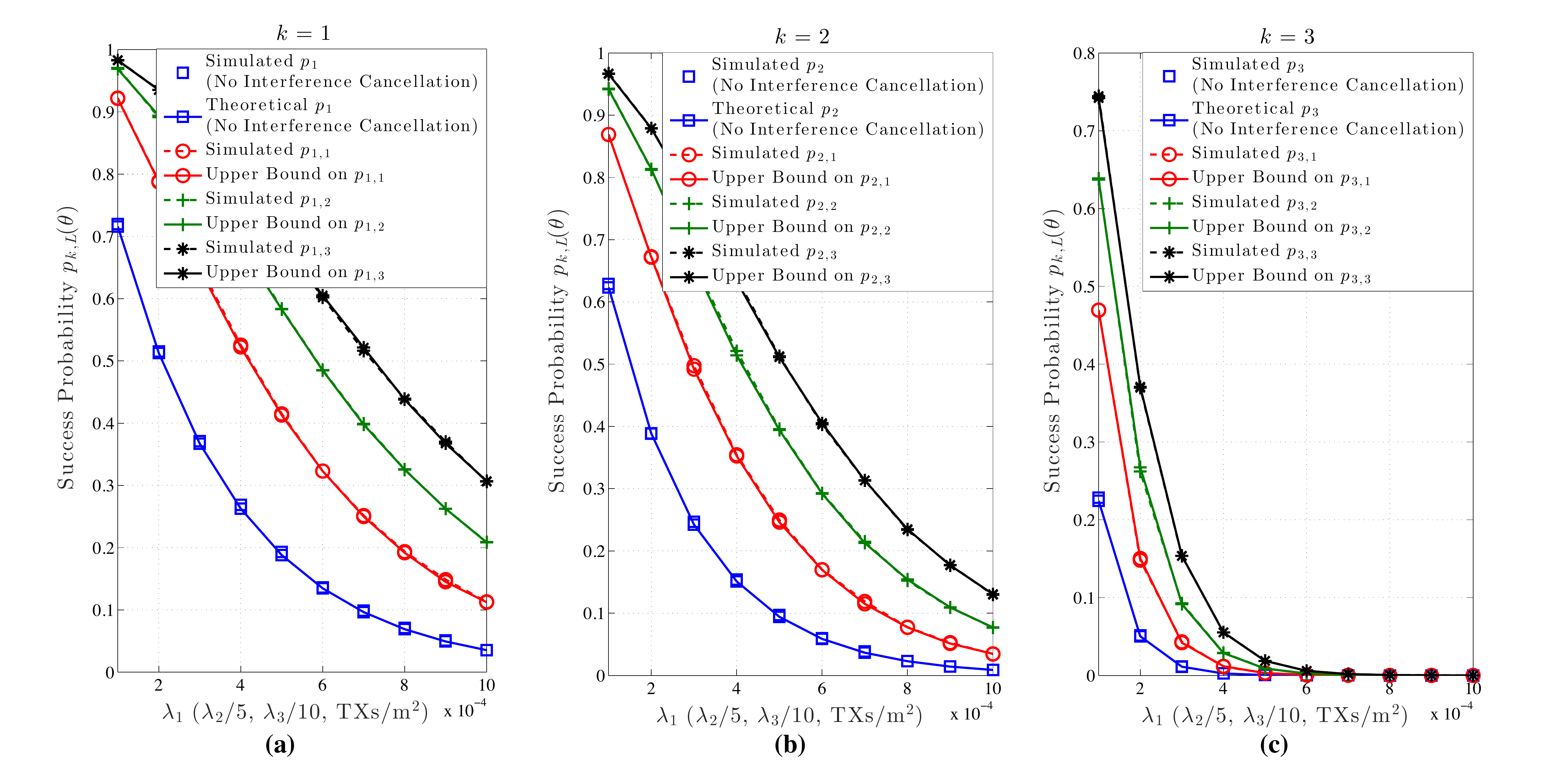}
	\caption{Simulation results of the success probabilities without and with interference cancellation: (a) $p_1(\theta)$ and $p_{1,L}(\theta)$ for $L= 1, 2, 3$. (b) $p_2(\theta)$ and $p_{2,L}(\theta)$ for $L=1, 2, 3$ (c) $p_3(\theta)$ and $p_{3,L}(\theta)$ for $L=1, 2, 3$. The upper bound on each type of the success probability is calculated by using \eqref{Eqn:SuccProbRayleighKCanBound}.}
	\label{Fig:SuccessProb}
\end{figure}

In this subsection, a few numerical results are provided to validate the success probabilities derived in the previous subsections. We consider the heterogeneous wireless ad hoc network consisting of three types of TXs and the simulation parameters for this heterogeneous network are listed in Table \ref{Tab:SimPara}. First, the simulated and theoretical results of the success probabilities without and with interference cancellation are shown in Fig. \ref{Fig:SuccessProb} and \eqref{Eqn:SuccProbRayleighKCanBound} is used to calculate the upper bound on the success probability with interference cancellation shown in the figure. As can be seen in Fig. \ref{Fig:SuccessProb}, the simulation results of the success probabilities without interference cancellation perfectly coincide with their corresponding theoretical results given in \eqref{Eqn:SuccProbFixDisRayFadNoCan}. With properly chosen $\omega$, the upper bounds are also very much close to their corresponding simulated success probabilities so that the tightness of \eqref{Eqn:SuccProbRayleighKCanBound} is validated. Thus, our analytical results in Section \ref{SubSec:SuccProbWithGenRXSigPower} regrading the success probabilities without and with interference cancellation are correct and accurate. Since all transmission distances are the same and the TXs with a higher type number $k$ have a much smaller transmit power, the mean of $S_k$ is always smaller than that of $S_{k-1}$ so that canceling interference should help to increase $p_k$ much more than $p_{k-1}$, as indicated in \eqref{Eqn:SuccProbRayleighKCanBound}. We can easily see this phenomenon by comparing the three subfigures in Fig. \ref{Fig:SuccessProb}.

\begin{figure}[t!]
	\centering
	\includegraphics[scale=0.42]{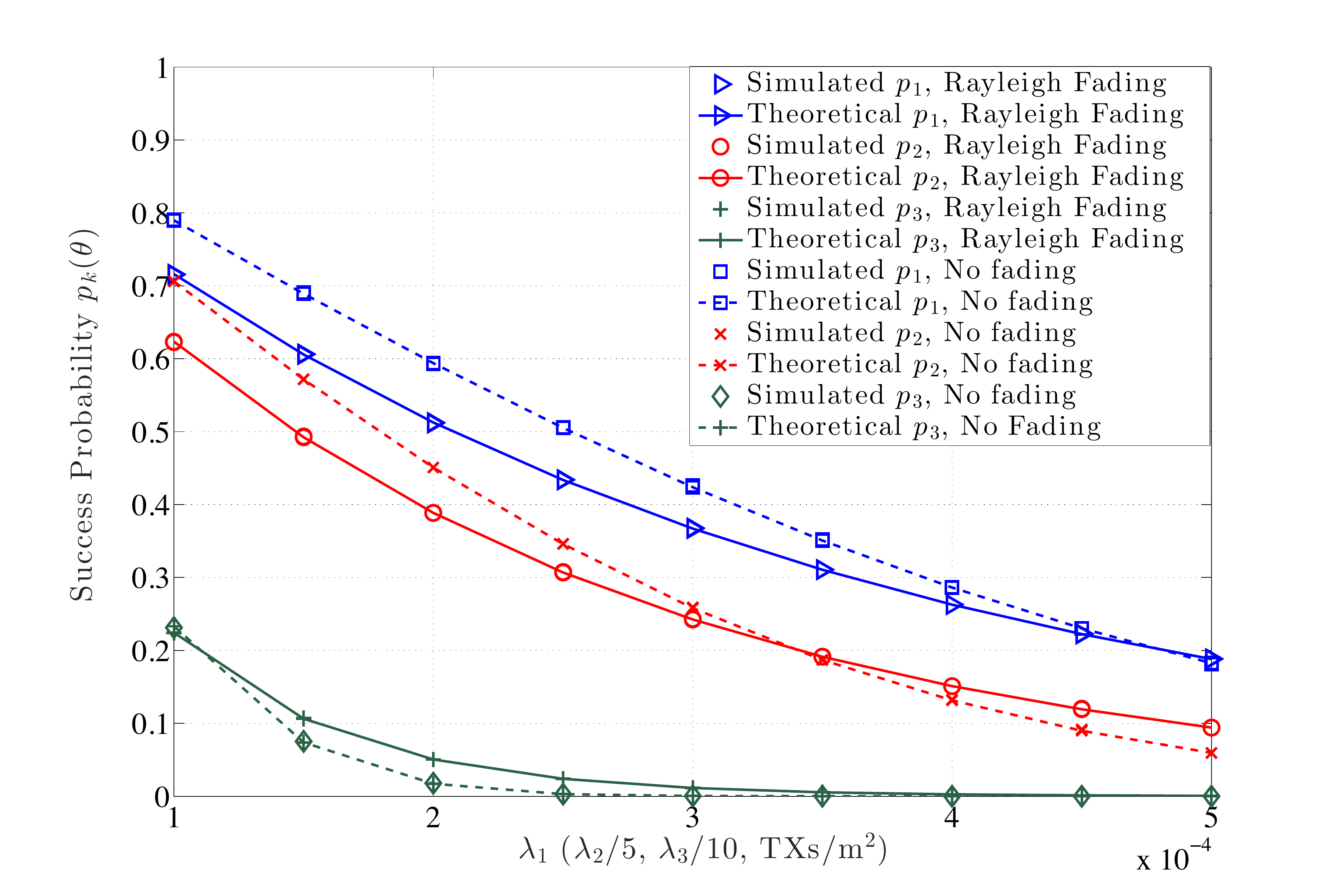}
	\caption{Simulation results of the success probabilities with and without Rayleigh fading.}
	\label{Fig:RanRXSigSucProb}
\end{figure}

In Section \ref{SubSec:ChannelRanHelpSuccProb}, we have pointed out that the randomness of the received signal power does not necessarily jeopardize the success probability. This has been demonstrated in Fig. \ref{Fig:RanRXSigSucProb} for the success probabilities for channels with or without Rayleigh fading and we certainly observe that Rayleigh fading does not always weaken the success probability under different TX intensities. In a dense network, usually channel randomness helps to improve the success probability since it weakens the interference channels much more than the communication channel. Also, we can exactly find the intensity region in which Rayleigh fading benefits the success probability. For example, in the simulation setting here we have $\widetilde{\Lambda}_2=2.662\times 10^3\lambda_1$, $p_2(\theta)=\exp(-\sqrt{\pi}\widetilde{\Lambda}_2)$ for Rayleigh fading and $p_2(\theta)=\mathrm{erfc}\left(\widetilde{\Lambda}_2\right)$ for no fading. According to the discussion in the paragraph right after Theorem \ref{Thm:FadingImproveSuccProb}, Rayleigh fading increases $p_2(\theta)$ when $\widetilde{\Lambda}_2>0.8951$, i.e., $\lambda_1>3.36\times 10^{-4}$, which is accurately illustrated by the curves of $p_2(\theta)$ in Fig. \ref{Fig:RanRXSigSucProb}. 

\begin{figure}[!t]
	\centering
	\includegraphics[scale=0.42]{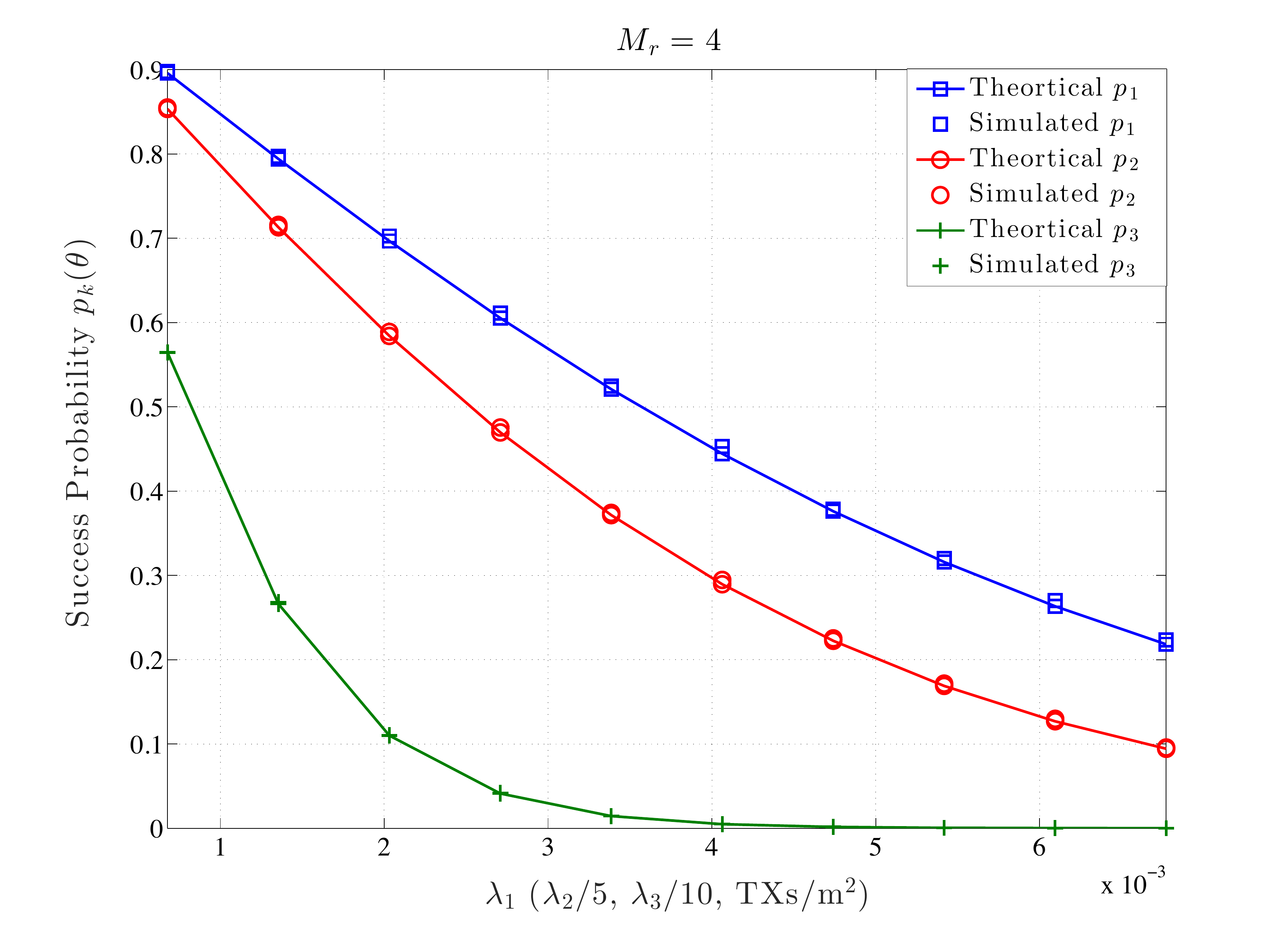}
	\caption{Simulation results of the success probability with $1\times 4$ SIMO Rayleigh fading channels. The theoretical results of $p_1$, $p_2$ and $p_3$ are based on the result in \eqref{Eqn:SIMOSuccProb} for $M_r=4$.}
	\label{Fig:SuccProbSIMO}
\end{figure}

\begin{figure}[!t]
	\centering
	\includegraphics[scale=0.4]{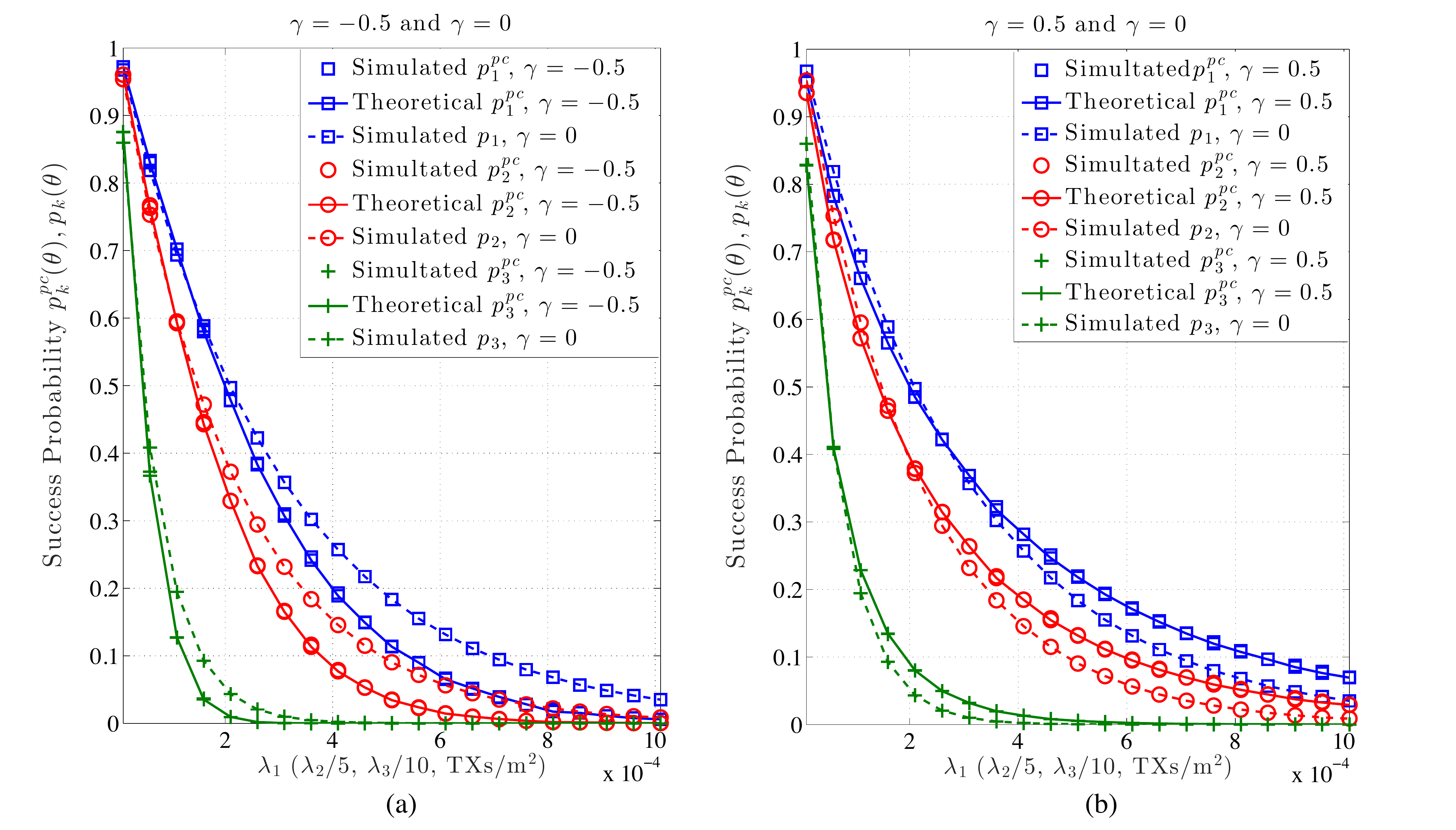}
	\caption{Simulation results of the success probabilities with and without the stochastic power control: (a) $p^{pc}_k(\theta)$ with $\gamma=-0.5$ and $p_k(\theta)=p^{pc}_k(\theta)$ with $\gamma=0$, (b) $p^{pc}_k(\theta)$ with $\gamma=-0.5$ and $p_k(\theta)=p^{pc}_k(\theta)$ with $\gamma=0$.  The theoretical results of $p^{pc}_k$ are found by using \eqref{Eqn:SuccProbPowerContAlpha4}.}
	\label{Fig:SuccProbPowerContorl}
\end{figure}

The success probability found in Section \ref{SubSec:SuccProbWithGenRXSigPower} is also applicable to the scenario of a TX or RX with multiple antennas. To demonstrate this point, consider the single-input-multiple-out (SIMO) case in which a TX has a single antenna and its intended RX has $M_r$ receive antennas. According to \eqref{Eqn:ExactSuccProbAlpha4}, the success probability for this $1\times M_r$ SIMO channel with independent Rayleigh fading can be written as
\begin{align}
p_k(\theta) &= \frac{M_r^{M_r}}{(M_r-1)!}\int_{0}^{\infty} \mathrm{erfc}\left(\frac{\pi^{\frac{3}{2}}\widetilde{\lambda}}{2\sqrt{\mathbb{E}[S_k]}}\sqrt{\frac{\theta}{x}}\right)x^{M_r-1}e^{-x}\dif x \nonumber\\
&= \frac{(-1)^{M_r-1}M_r^{M_r}}{(M_r-1)!}\left(\frac{\dif^{M_r-1}}{\dif y^{M_r-1}}\left[\int_{0}^{\infty} \mathrm{erfc}\left(\frac{\pi^2R_k^2\sum_{j=1}^{K}\lambda_j\sqrt{P_j}}{4\sqrt{M_rP_k}}\sqrt{\frac{\theta}{x}}\right)e^{-yx}\dif x\right]\right)\bigg|_{y=M_r}\nonumber\\
&=\frac{(-1)^{M_r-1}M_r^{M_r}}{(M_r-1)!}\left(\frac{\dif^{M_r-1}}{\dif y^{M_r-1}} \left[\frac{1}{y}\exp\left(-\frac{\pi^2R_k^2\sqrt{\theta}\sum_{j=1}^{K}\lambda_j\sqrt{P_j}}{2\sqrt{M_rP_k}}\sqrt{y}\right)\right]\right)\bigg|_{y=M_r}, \label{Eqn:SIMOSuccProb}
\end{align}
where $\mathbb{E}[S_k]=M_rP_kR^{-4}_k$, $\widetilde{\lambda}=\frac{\sqrt{\pi}}{2}\sum_{j=1}^{K}\lambda_j\sqrt{P_j}$ and  $\widehat{S}_k \sim \mathrm{Erlang}(M_r,M_r)$ (i.e., $H_k$ has a chi-square distribution with $2M_r$ degrees of freedom \cite{NJSPWJGA11,PXCHLJGA13}). For $M_r=4$, the theoretical result of $p_k(\theta)$ in \eqref{Eqn:SIMOSuccProb} and its corresponding simulated result are shown in Fig. \ref{Fig:SuccProbSIMO}. As can be seen, they perfectly coincide each other, and the SIMO channel significantly improves the success probability due to spatial diversity by comparing the corresponding single-antenna simulated results in Figs. \ref{Fig:RanRXSigSucProb} and \ref{Fig:SuccProbSIMO}. In addition, $p_k(\theta)$ with an MISO channel also can be derived by following the same technique used in \eqref{Eqn:SIMOSuccProb}.

In Fig. \ref{Fig:SuccProbPowerContorl}, we show the success probabilities when the stochastic power control schemes with $\gamma=-0.5$ and $\gamma=0.5$ in \eqref{Eqn:StochasticPowerControl} are adopted. In Fig \ref{Fig:SuccProbPowerContorl}(a) for $\gamma=-0.5$, we observe that stochastic power control (slightly) outperforms no power control in the low intensity region (roughly when $\lambda_1<0.0001$), whereas in Fig \ref{Fig:SuccProbPowerContorl}(b) for $\gamma=0.5$ stochastic power control outperforms no power control in the high intensity region (roughly when $\lambda_1>0.0001$). This validates our previous discussion in Section \ref{SubSec:SuccProbPowerControl} that the power control exponent $\gamma$ should change based on different TX intensities in order to make stochastic power control work better than no power control, and exploiting more randomness of the received signal power in a dense network 
(i.e., using a larger power control exponent) achieves a larger success probability. In addition, the correctness of $p^{pc}_k(\theta)$ in \eqref{Eqn:SuccProbPowerContAlpha4} is validated in Fig. \ref{Fig:SuccProbPowerContorl} since it is used to provide the theoretical results of $p^{pc}_k$ in the figure  that perfectly coincide with their corresponding simulated results.

\section{Application of the SIR Statistics (II): Ergodic Link Capacity and Spatial Throughput Capacity}
In this section, we study another main application of the distribution of the SIR in finding the explicit low-complexity expressions of the ergodic link capacity that is already defined in Definition \ref{Defn:LinkNetworkCapacity}. As we will show in the following, the expression is not derived by assuming any specific random channel gain, transmit power and distance models in advance so that they are generally applicable for many random signal models involved in the SIR. The type-$k$ ergodic link capacity with and without interference cancellation is studied first and then we derive and discuss the type-$k$ ergodic link capacity with stochastic power control. Afterwards, the spatial throughput capacity of the network is analyzed based on the derived success probability and ergodic link capacity of each type. Finally, some numerical results are provided to validate our analytical findings. 

\subsection{Ergodic Link Capacity with General Random Models of the Received Signal Power}\label{SubSec:ErgodicLinkCapacitywithGenRanModel}
The type-$k$ ergodic link capacity (bits/sec/Hz) in \eqref{Eqn:DefnErgodicRate} can be expressed in terms of the type-$k$ success probability as shown in the following:
\begin{align}
c_k &=\frac{1}{\ln(2)}\int_{0}^{\infty} \mathbb{P}\left[\sinr_k\geq 2^x-1\right] \dif x=\frac{1}{\ln(2)}\int_{0}^{\infty}\frac{ p_k(\vartheta)}{1+\vartheta}\dif \vartheta\nonumber\\
&=\frac{1}{\ln(2)}\bigintsss_{0}^{\infty}\frac{\mathcal{L}^{-1}\left\{\mathcal{L}_{\text{SIR}^{-1}_k}(s)/s\right\}\left(\vartheta^{-1}\right) }{(1+\vartheta)}\dif\vartheta.\label{Eqn:ErgodicLinkCap}
\end{align}
This result cannot be further simplified to another neat expression except in some special cases that either the inverse Laplace transform of  $\mathcal{L}_{\text{SIR}^{-1}_{\mathsf{c}}}(s)/s$ or $p_k(\vartheta)$ is able to be found in closed-form. Thus, using the result of the success probability to find the ergodic link capacity, which is the common approach used in prior works, in general, cannot yield a tractable and low-complexity result for the received signal power with a general distribution. A more tractable approach to deriving $c_k$ is shown in the following theorem.
 
\begin{theorem}\label{Thm:ErgodicLinkCapNoCan}
If the type-$k$ RXs do not cancel any interference and the Laplace transform of their received signal power exists, the type-$k$ ergodic link capacity in \eqref{Eqn:DefnErgodicRate} can be simply expressed as
\begin{align}\label{Eqn:ErgodicLinkCapNoCan}
c_k=\frac{1}{\ln(2)}\int_{0^+}^{\infty} \frac{1}{\vartheta}\left[1-\mathcal{L}_{\widehat{S}_k}(\vartheta)\right]\mathcal{L}_{I_k}\left(\frac{\vartheta}{\mathbb{E}[S_k]}\right) \dif \vartheta,\,k\in\mathcal{K}. 
\end{align}
On the other hand, if the type-$k$ RXs can cancel their first $L$ strongest interferers, then their ergodic link capacity can be found as
\begin{align}\label{Eqn:ErgodicLinkCapKCan}
c_{k,L}
&=\frac{1}{\ln(2)} \int_{0^+}^{\infty}\left[1-\mathcal{L}_{\widehat{S}_k}(\vartheta)\right]\mathcal{L}_{I_k}\left(\frac{\vartheta}{\mathbb{E}[S_k]}\right)\mathcal{M}_{\ell_{D_L}(\frac{\theta}{\mathbb{E}[S_k]})}\left(\pi\widetilde{\lambda}\right)\frac{\dif\vartheta}{\vartheta},\,\,k\in\mathcal{K}.
\end{align}
\end{theorem}
\begin{IEEEproof}
See Appendix \ref{App:ProofErgodicLinkCanNoCan}.
\end{IEEEproof}

The results in Theorem \ref{Thm:ErgodicLinkCapNoCan} are important since they show the fairly simple expressions of calculating the type-$k$ ergodic link capacity with and without interference cancellation for a general random model of $S_k$ as long as the Laplace transform of $S_k$ exists, which are firstly shown in this work to the best of our knowledge. Like the case of the type-$k$ success probability, the type-$k$ ergodic link capacity in \eqref{Eqn:ErgodicLinkCapNoCan} can reduce to a simpler expression for some special cases. For example, consider constant $S_k$ and $c_k$ in \eqref{Eqn:ErgodicLinkCapNoCan} becomes
\begin{align}\label{Eqn:ErgodicLinkCapNoFading}
c_k=\frac{1}{\ln(2)} \int_{0^+}^{\infty}\frac{\left(1-e^{-\vartheta}\right)}{\vartheta e^{\pi\widetilde{\lambda}\Gamma(1-\frac{2}{\alpha})(\vartheta/S_k)^{\frac{2}{\alpha}}}} \dif \vartheta
\end{align}
since $\widehat{S}_k=1$ and $\mathcal{L}_{\widehat{S}_k}(\theta)=e^{-\theta}$, which further can be expressed as
\begin{align}
c_k=\frac{\alpha}{2\ln(2)} \sum_{n=1}^{\infty} \frac{(-1)^{n+1}\Gamma\left(\frac{\alpha n}{2}\right)S^n_k}{n!\left[\pi\widetilde{\lambda}\Gamma\left(1-\frac{2}{\alpha}\right)\right]^{\frac{\alpha n}{2}}}
\end{align}
by using the Taylor's expansion of $e^{-x}$ for $x>0$ and the Gamma function. This result can be applied to any scenarios that make the received signal power constant, such as constant transmit power, distance as well as no channel (fading) randomness.  Another example is that
we have $c_k$ in \eqref{Eqn:ErgodicLinkCapNoCan} with $\widehat{S}_k\sim\mathrm{Gamma}(m_k,1/m_k)$ given by
\begin{align}\label{Eqn:ErgodicLinkCapNakagami}
c_k=\frac{1}{\ln(2)}\int_{0^+}^{\infty} \frac{1}{\vartheta}\left[1-\left(\frac{m_k}{m_k+\vartheta}\right)^{m_k}\right] \mathcal{L}_{I_k}\left(\frac{\vartheta}{\mathbb{E}[S_k]}\right)   \dif \vartheta,\,\,\text{for }m_k\in\mathbb{N}_+.
\end{align}
By comparing \eqref{Eqn:ErgodicLinkCapNakagami} with \eqref{Eqn:ErgodicLinkCapNoFading}, we immediately acquire the following important observations. 
\begin{itemize}
\item We realize that $1-(m_k/(m_k+\vartheta))^{m_k}$ monotonically increases up to $1-e^{-\vartheta}$ as $m_k\rightarrow\infty$. If $c_k$ in \eqref{Eqn:ErgodicLinkCapNoFading} and $c_k$ in \eqref{Eqn:ErgodicLinkCapNakagami} have the same $\widetilde{\lambda}=\sum_{j=1}^{K}\lambda_j\mathbb{E}\left[P^{\frac{2}{\alpha}}_j\right]$ (i.e., all interference channels do not suffer fading), $c_k$ in  \eqref{Eqn:ErgodicLinkCapNakagami} must be smaller than $c_k$ in \eqref{Eqn:ErgodicLinkCapNoFading} in this case. Hence,  \textit{$c_k$ in \eqref{Eqn:ErgodicLinkCapNoFading} is the maximum type-$k$ ergodic link capacity provided that only the communication channel of the type-$k$ RXs suffers fading, and the randomness of $S_k$ reduces the type-$k$ ergodic link capacity}. Thus, for the special case of $m_k=1$ for all $k\in\mathcal{K}$), $c_k$ in \eqref{Eqn:ErgodicLinkCapNakagami} reduces to
\begin{align}\label{Eqn:ErgodicLinkCapNoCanRayleigh}
c_k&=\frac{1}{\ln(2)}\int_{0^+}^{\infty}\mathcal{L}_{I_k}\left(\frac{\vartheta}{\mathbb{E}[S_k]}\right) \frac{\dif \vartheta}{(1+\vartheta)},
\end{align} 
which is the minimum type-$k$ ergodic link capacity for $\widehat{S}_k\sim\mathrm{Gamma}(m_k,1/m_k)$ and no fading interference channels.
\item If all interference channel gains also suffer Nakagami-$m_k$ fading such that we have $H_{k_i}\sim\mathrm{Gamma}(m_k,1/m_k)$ for all $k$ and $i$, $\widetilde{\lambda}$ in  \eqref{Eqn:ErgodicLinkCapNakagami} is equal to $\sum_{j=1}^{K} \lambda_j\mathbb{E}\left[P^{\frac{2}{\alpha}}_j\right]\frac{\Gamma(m_j+\frac{2}{\alpha})}{\Gamma(m_j)}$ that is always smaller than $\widetilde{\lambda}=\sum_{j=1}^{K}\lambda_j\mathbb{E}\left[P^{\frac{2}{\alpha}}_j\right]$ in \eqref{Eqn:ErgodicLinkCapNoFading} due to $\mathbb{E}\left[H^{\frac{2}{\alpha}}_j\right]=\frac{\Gamma(m_j+\frac{2}{\alpha})}{\Gamma(m_j)}<1$ for all $j\in\mathcal{K}$. Accordingly, $c_k$ in  \eqref{Eqn:ErgodicLinkCapNakagami} is not necessarily smaller/greater than $c_k$ in \eqref{Eqn:ErgodicLinkCapNoFading}. In other words, \textit{channel fading does not necessarily jeopardize or benefit the ergodic link capacity}. 
\item According to \eqref{Eqn:ErgodicLinkCapNoFading} and \eqref{Eqn:ErgodicLinkCapNakagami}, if $\widetilde{\Lambda}_k=\pi\widetilde{\lambda}/(\mathbb{E}[S_k])^{\frac{2}{\alpha}}$ and the following set $\widetilde{\mathbf{\Lambda}}^{\dagger}_k$
\begin{align}
\widetilde{\mathbf{\Lambda}}^{\dagger}_k \defn\left\{\widetilde{\Lambda}_k\in\mathbb{R}_{++}: c_k(\widetilde{\Lambda}_k) \text{ in \eqref{Eqn:ErgodicLinkCapNakagami}} > c_k(\widetilde{\Lambda}_k)\text{ in \eqref{Eqn:ErgodicLinkCapNoFading}}\right\}
\end{align}
is not empty, then the randomness of $S_k$ increases $c_k$ in the network with any $\pi\widetilde{\lambda}/(\mathbb{E}[S_k])^{\frac{2}{\alpha}}\in \widetilde{\mathbf{\Lambda}}^{\dagger}_k$. For given $m_k$ and $\alpha$, set $\widetilde{\mathbf{\Lambda}}^{\dagger}_k$ is generally not possible to be explicitly found. However, it can be determined by numerical methods. 
\end{itemize}

Although the explicit expression of the ergodic link capacity with interference cancellation is found in \eqref{Eqn:ErgodicLinkCapKCan}, it is somewhat complicate for practical computation. According to \eqref{Eqn:LaplaceInterUpperBound} and \eqref{Eqn:ErgodicLinkCapKCan}, the upper bound on $c_{k,L}$ can be easily inferred from \eqref{Eqn:LaplaceInterUpperBound} as follows
\begin{align}\label{Eqn:UppBoundLinkCapCanK}
c_{k,L}\leq\frac{1}{\ln(2)}\int_{0^+}^{\infty}\left[1-\mathcal{L}_{\widehat{S}_k}(\vartheta)\right]\mathcal{L}_{I_k}\left(\frac{\vartheta}{\mathbb{E}[S_k]}\right)\mathcal{M}_{\omega^{-\frac{\alpha}{2}}D_L^{1-\frac{\alpha}{2}}}\left(\frac{\pi\widetilde{\lambda}\vartheta}{\mathbb{E}[S_k]}\right) \frac{\dif \vartheta}{\vartheta },
\end{align}
which is much neater and easier computed. Most importantly, this upper bound shows that \textit{the larger mean of the received signal power offsets more the effect of interference cancellation} and it is usually very tight. Similarly, if we consider $\widehat{S}_k\sim\text{Gamma}(m_k,1/m_k)$, $c_{k,L}$ in \eqref{Eqn:ErgodicLinkCapKCan} becomes
\begin{align}\label{Eqn:ErgodicLinkCapKCanNakagami}
c_{k,L}=\frac{1}{\ln(2)}\int_{0^+}^{\infty} \left[1-\left(\frac{m_k}{m_k+\vartheta}\right)^{m_k}\right]\mathcal{L}_{I_k}\left(\frac{\vartheta}{\mathbb{E}[S_k]}\right)\mathcal{M}_{\ell_{D_L}(\frac{\vartheta}{\mathbb{E}[S_k]})}\left(\pi\widetilde{\lambda}\right)\frac{\dif\vartheta}{\vartheta }
\end{align}
and its upper bound in \eqref{Eqn:UppBoundLinkCapCanK}  is explicitly given by
\begin{align}\label{Eqn:UppBoundLinkCapCanKNakagami}
c_{k,L}\leq\frac{1}{\ln(2)}\int_{0^+}^{\infty} \left[1-\left(\frac{m_k}{m_k+\vartheta}\right)^{m_k}\right]\mathcal{L}_{I_k}\left(\frac{\vartheta}{\mathbb{E}[S_k]}\right)\mathcal{M}_{\omega^{-\frac{\alpha}{2}}D_L^{1-\frac{\alpha}{2}}}\left(\frac{\pi\widetilde{\lambda}\vartheta}{\mathbb{E}[S_k]}\right)\frac{\dif\vartheta}{\vartheta}.
\end{align}
Hence, we can find $c_{k,L}$ and its upper bound for the different randomness levels of $S_k$.  

\subsection{Ergodic Link Capacity with Stochastic Power Control}\label{SubSec:ErgodicLinkCapacityPowerControl}
According to the analytical results and discussions in Section \ref{SubSec:SuccProbPowerControl}, the stochastic power control proposed in \eqref{Eqn:StochasticPowerControl} does not always improve the success probability if the power control exponent is not properly chosen. This conclusion may also happen at the ergodic link capacity with the stochastic power control in that it can be found by using the integral formula in \eqref{Eqn:ErgodicLinkCap} that essentially contains the success probability. To gain a better understanding about when the stochastic power control benefits $c_k$, we need to first study the explicit expression of $c_k$ with the stochastic power control and it is shown in the following theorem.
\begin{theorem}\label{Thm:ErgodicRatePowerControl}
If the stochastic power control in \eqref{Eqn:StochasticPowerControl} is adopted by the type-$k$ TXs and let $S_k=\overline{P}_kH_kR^{-\alpha}_k$  be the type-$k$ received signal power without stochastic power control, the type-$k$ ergodic link capacity without interference cancellation is given by
\begin{align}\label{Eqn:ErgodicRatePowerControl}
c^{pc}_k = \frac{1}{\ln(2)}\bigintssss_{0^+}^{\infty} \mathcal{L}_{I^{pc}_k}\left(\frac{\vartheta\mathbb{E}[S^{\gamma_k}_k]}{\mathbb{E}[S^{1+\gamma_k}_k]}\right)\left[1-\mathcal{L}_{\widehat{S}^{\gamma_k+1}_k}\left(\frac{\vartheta(\mathbb{E}[S_k])^{\gamma_k+1}}{\mathbb{E}\left[S^{\gamma_k+1}_k\right]}\right)\right] \frac{\dif \vartheta}{\vartheta},
\end{align}
where   $\mathcal{L}_{I^{pc}_k}\left(\frac{\vartheta\mathbb{E}[S^{\gamma_k}_k]}{\mathbb{E}[S^{1+\gamma_k}_k]}\right)$ is given by 
\begin{align}
\mathcal{L}_{I^{pc}_k}\left(\frac{\vartheta\mathbb{E}[S^{\gamma_k}_k]}{\mathbb{E}[S^{1+\gamma_k}_k]}\right)=\exp\left(-\pi \Gamma\left(1-\frac{2}{\alpha}\right)\left(\frac{\vartheta\mathbb{E}[S^{\gamma_k}_k]}{\mathbb{E}[S^{1+\gamma_k}_k]}\right)^{\frac{2}{\alpha}}\widetilde{\lambda}^{pc}\right),
\end{align}
and $\widetilde{\lambda}^{pc}$ is already defined in \eqref{Eqn:IntensityPowerControl}. Also, the bounds on $c^{pc}_k$ can be shown as
\begin{align}\label{Eqn:BoundsErgodicLinkCapPowerCon}
 \int_{0^+}^{\infty} \mathcal{L}_{I^{pc}_k}\left(\frac{\vartheta\mathbb{E}[S^{\gamma_k}_k]}{\mathbb{E}[S^{1+\gamma_k}_k]}\right) \frac{(1-e^{-\vartheta})}{(\ln 2)\vartheta}\dif\vartheta \geq c_k\geq  \int_{0^+}^{\infty} \mathcal{L}_{I^{pc}_k}\left(\frac{\vartheta\mathbb{E}[S^{\gamma_k}_k]}{\mathbb{E}[S^{1+\gamma_k}_k]}\right) \frac{\dif\vartheta}{(\ln 2)(1+\vartheta)}.
\end{align}
\end{theorem}  
\begin{IEEEproof}
See Appendix \ref{App:ProofErgodicRatePowerControl}.
\end{IEEEproof}

The expression of $c^{pc}_k$ in \eqref{Eqn:ErgodicRatePowerControl} is a fairly general and neat result that is never found in the prior works. It straightforwardly indicates under which circumstances the stochastic power control is superior/inferior to no power control by comparing $c^{pc}_k$ with $c_k$. To let $c^{pc}_k\geq c_k$ hold, we can let  
\begin{align}\label{Eqn:CondPowerControl1}
\left(\frac{\widetilde{\lambda}^{pc}}{\widetilde{\lambda}}\right)^{\frac{\alpha}{2}}\leq \frac{\mathbb{E}\left[S^{1+\gamma_k}_k\right]}{\mathbb{E}[S_k]\mathbb{E}[S^{\gamma_k}_k]}
\end{align}
and
\begin{align}\label{Eqn:CondPowerControl2}
\bigintssss_{0^+}^{\infty} \mathcal{L}_{I_k}\left(\frac{\vartheta}{\mathbb{E}[S_k]}\right)\left[\mathcal{L}_{\widehat{S}_k}\left(\vartheta\right)-\mathcal{L}_{\widehat{S}^{\gamma_k+1}_k}\left(\frac{\vartheta(\mathbb{E}[S_k])^{\gamma_k+1}}{\mathbb{E}\left[S^{\gamma_k+1}_k\right]}\right)\right]\frac{\dif\vartheta}{\vartheta}\geq 0 
\end{align}
hold because \eqref{Eqn:CondPowerControl2} can be rewritten as
 \begin{align*}
 \int_{0}^{\infty} \mathcal{L}_{I_k}\left(\frac{\vartheta}{\mathbb{E}[S_k]}\right)\left[-1+\mathcal{L}_{\widehat{S}_k}\left(\vartheta\right)+1-\mathcal{L}_{\widehat{S}^{\gamma_k+1}_k}\left(\frac{\vartheta(\mathbb{E}[S_k])^{\gamma_k+1}}{\mathbb{E}\left[S^{\gamma_k+1}_k\right]}\right)\right]\frac{\dif\vartheta}{\vartheta}>0,
 \end{align*}
which leads to $c^{pc}_k \geq c_k$ if $\mathcal{L}_{I^{pc}_k}(\vartheta\mathbb{E}[S_k]/\mathbb{E}[S^{1+\gamma_k}_k])\geq \mathcal{L}_{I_k}(\vartheta/\mathbb{E}[S_k])$ and \eqref{Eqn:CondPowerControl2} both hold. Whereas $\mathcal{L}_{I^{pc}_k}(\vartheta\mathbb{E}[S_k]/\mathbb{E}[S^{1+\gamma_k}_k])\geq \mathcal{L}_{I_k}(\vartheta/\mathbb{E}[S_k])$ holds once \eqref{Eqn:CondPowerControl1} is valid. Hence, \textit{stochastic power control outperforms no power control in terms of the ergodic link capacity if \eqref{Eqn:CondPowerControl2} and \eqref{Eqn:CondPowerControl1} both satisfy}. Similarly, we can show that no power control is superior to stochastic power control in terms of the ergodic link capacity if the following
\begin{align}\label{Eqn:CondNoPowerControl1}
\left(\frac{\widetilde{\lambda}^{pc}}{\widetilde{\lambda}}\right)^{\frac{\alpha}{2}}\geq \frac{\mathbb{E}\left[S^{1+\gamma_k}_k\right]}{\mathbb{E}[S_k]\mathbb{E}[S^{\gamma_k}_k]}
\end{align}
and 
\begin{align}\label{Eqn:CondNoPowerControl2}
\bigintssss_{0^+}^{\infty} \mathcal{L}_{I^{pc}_k}\left(\frac{\vartheta\mathbb{E}[S^{\gamma_k}_k]}{\mathbb{E}[S^{1+\gamma_k}_k]}\right)\left[\mathcal{L}_{\widehat{S}^{\gamma_k+1}_k}\left(\frac{\vartheta(\mathbb{E}[S_k])^{\gamma_k+1}}{\mathbb{E}\left[S^{\gamma_k+1}_k\right]}\right)-\mathcal{L}_{\widehat{S}_k}\left(\vartheta\right)\right]\frac{\dif\vartheta}{\vartheta}\geq 0 
\end{align}
both satisfy.

The bounds in \eqref{Eqn:BoundsErgodicLinkCapPowerCon} indicate the fundamental limits on $c_k$ that could be achieved by the stochastic power control. They reveal three important implications:
\begin{itemize}
\item The upper bound can be interpreted as $s^{pc}_k$ achieved by the stochastic power control that makes $\widehat{S}_k$ be one (i.e., $S_k$ be a constant).
\item The lower bound can be interpreted as $s^{pc}_k$ achieved by the stochastic power control that results in $\widehat{S}_k\sim\exp(1,1)$.
\item The performance of the stochastic power control is dominated by the  $\mathcal{L}_{I^{pc}_k}\left(\frac{\vartheta\mathbb{E}[S^{\gamma_k}_k]}{\mathbb{E}[S^{1+\gamma_k}_k]}\right)$ term in the bounds.
\end{itemize}  
As a result,  we should properly choose the power control exponent $\gamma_k$ that is able to increase the ergodic link capacity by increasing $\mathcal{L}_{I^{pc}_k}\left(\frac{\vartheta\mathbb{E}[S^{\gamma_k}_k]}{\mathbb{E}[S^{1+\gamma_k}_k]}\right)$. In a Rayleigh fading environment, for example, the stochastic power control with $\gamma_k=-1$ (i.e., channel inversion power control) is not welcome since it leads to $c^{pc}_k=0$ due to $\mathbb{E}[S_k^{-1}]=\infty$ and  $\mathcal{L}_{I^{pc}_k}\left(\frac{\vartheta\mathbb{E}[S^{\gamma_k}_k]}{\mathbb{E}[S^{1+\gamma_k}_k]}\right)=\mathcal{L}_{I^{pc}_k}\left(\vartheta\mathbb{E}[S^{-1}_k]\right)=0$. Another example of considering the special case of $\gamma_k=\gamma$, $H_k\sim\exp(1,1)$ and constant $R_k$ can easily show how to choose $\gamma_k$ so that $c^{pc}_k$ is increased. In this case, since $\widetilde{\lambda}^{pc}=\widetilde{\lambda}\Gamma(1+\frac{2\gamma}{\alpha})(\Gamma(1+\gamma))^{-\frac{2}{\alpha}}$ and $\mathbb{E}[S^{\gamma+1}_k]/\mathbb{E}[S_k]\mathbb{E}[S^{\gamma}_k]=1/(1+\gamma)$, \eqref{Eqn:CondNoPowerControl1} reduces to $[\Gamma(1+\frac{2\gamma}{\alpha})]^{\frac{\alpha}{2}}\leq \Gamma(2+\gamma)$ and \eqref{Eqn:CondNoPowerControl2} holds for any $\gamma$ since $\mathcal{L}_{\widehat{S}^{\gamma_k+1}_k}\left(\frac{\vartheta(\mathbb{E}[S_k])^{\gamma_k+1}}{\mathbb{E}\left[S^{\gamma_k+1}_k\right]}\right)\geq \mathcal{L}_{\widehat{S}_k}\left(\vartheta\right)=\frac{\vartheta}{1+\vartheta}$ is always true based on the proof of Theorem \ref{Thm:ErgodicRatePowerControl}. Note that constraint $[\Gamma(1+\frac{2\gamma}{\alpha})]^{\frac{\alpha}{2}}\leq \Gamma(2+\gamma)$ always holds as long as $\gamma\geq 0$.  Hence,  \textit{in this special case the stochastic power control with  $\gamma>0$ always benefits the ergodic link capacity}. We will numerically verify this interesting and important finding in Section \ref{SubSec:ErgodicCapacityNumResults}.

\subsection{Analysis of Spatial Throughput Capacity}\label{SubSec:ThroughputCapacity}
According to the type-$k$ success probability and type-$k$ ergodic link capacity, the (spatial) throughput capacity of the heterogeneous wireless ad hoc network without interference cancellation can be defined as follows.
\begin{definition}[Spacial Throughput Capacity]
The spatial throughput capacity of the heterogeneous wireless ad hoc network with $K$ different types of TXs and SIR threshold $\theta$ is defined as
\begin{align}\label{Eqn:ThroughputCapacity}
C(\theta)\defn \sum_{k=1}^{K} c_k\lambda_k p_k(\theta),\,\,(\text{bps/Hz/m}^2)
\end{align}
which measures the successful transmitted data amount of $K$ different types per unit bandwidth and area (i.e., the successful area spectrum efficiency of the network.). 
\end{definition}
\noindent Spacial throughput capacity $C(\theta)$ can be explicitly expressed in terms of the explicit expressions of $p_k(\theta)$ and $c_k$ that are already derived in the previous sections. The salient feature of the throughput capacity defined in  \eqref{Eqn:ThroughputCapacity} is to realistically characterize how much per-unit-area data (area spectrum efficiency) can be successfully transported in the network with heterogeneity. Previous similar works on defining the network capacity, such as network throughput and transmission capacity \cite{FBBBPM06,SWJGAXYGDV07,CHLJGA12}, would be somewhat conservative and inaccurate in that they simply use the minimum link capacity $\log_2(1+\theta)$ to define their network capacity metrics due to having the difficulty in finding the explicit result of the ergodic link capacity. 

Note that  there must exist a set of optimal $\lambda_k$'s that maximizes $C(\theta)$ based on the Weierstrass theorem since $\lim_{\lambda_k\rightarrow 0, \forall  k\in\mathcal{K}}C(\theta)=0$, $\lim_{\lambda_k\rightarrow \infty, \forall  k\in\mathcal{K}}C(\theta)=0$ and $C(\theta)\geq 0$ is a continuous function of all $\lambda_k$'s \cite{DPB99}. We can find optimal $\lambda^*_k$ and optimal $C^*(\theta)\defn\sup_{\lambda_k} C(\theta)$ for $\widehat{S}_k$ (or $S_k$) with some special distribution. For example, consider the special case of $\widehat{S}_k\sim\exp(1,1)$ for all $k\in\mathcal{K}$ and $C(\theta)$ is explicitly given by
\begin{align}
C(\theta) = \frac{1}{\ln 2}\sum_{k=1}^{K} \lambda_k\left(\int_{0}^{\infty} \mathcal{L}_{I_k}\left(\frac{\vartheta}{\mathbb{E}[S_k]}\right)\frac{\dif \vartheta}{1+\vartheta}\right)\mathcal{L}_{I_k}\left(\frac{\theta}{\mathbb{E}[S_k]}\right).
\end{align} 
By solving $\frac{\partial C(\theta)}{\partial\lambda_k}=0$ for $\lambda_k$, the unique optimal $\lambda^*_k$ that maximizes $C(\theta)$ can be found as
\begin{align}\label{Eqn:OptIntenMaxThrputCap}
\lambda_k^*=\frac{(\mathbb{E}[S_k])^{\frac{2}{\alpha}}}{\pi\Gamma\left(1-\frac{2}{\alpha}\right)\left(\vartheta^{\frac{2}{\alpha}}+\theta^{\frac{2}{\alpha}}\right)\mathbb{E}\left[H^{\frac{2}{\alpha}}_k\right]\mathbb{E}\left[P^{\frac{2}{\alpha}}_k\right]}.
\end{align}
Then substituting $\lambda^*_k$ in \eqref{Eqn:OptIntenMaxThrputCap} yields $C^*(\theta)$ given by
\begin{align}\label{Eqn:OptimalThrputCap}
C^*(\theta) = \frac{\xi(\alpha,\theta)}{\pi(\ln 2)\Gamma(1-\frac{2}{\alpha})} \sum_{k=1}^{K}\frac{(\mathbb{E}[S_k])^{\frac{2}{\alpha}} e^{-\sum_{j=1}^{K}(\frac{\mathbb{E}[S_j]}{\mathbb{E}[S_k]})^{\frac{2}{\alpha}}}}{\mathbb{E}[H^{\frac{2}{\alpha}}_k]\mathbb{E}[P^{\frac{2}{\alpha}}_k]},
\end{align}
where $\xi(\alpha,\theta)\defn \int_{0}^{\infty}\left(\vartheta^{\frac{2}{\alpha}}+\theta^{\frac{2}{\alpha}}\right)^{-1}(1+\vartheta)^{-1}\dif \vartheta$
which further reduces to 
\begin{align}\label{Eqn:ThrputCapConstPwoDis}
C^*(\theta) = \left[\frac{\alpha\sin(2\pi/\alpha)}{(2\ln 2)\pi^2}\xi(\alpha,\theta)\right] \sum_{k=1}^{K} R^{-2}_k e^{-\sum_{j=1}^{K}\left(\frac{P_j}{P_k}\right)^{\frac{2}{\alpha}}\left(\frac{R_k}{R_j}\right)^2}
\end{align}
for $H_k\sim\exp(1,1)$, constant $P_k$ and constant $R_k$. In addition, the spatial throughput capacity with canceling the first $L$ strongest interferers, denoted by $C_L(\theta)$, can be obtained by substituting the results of $p_{k,L}(\theta)$ and $c_{k,L}$ into \eqref{Eqn:OptimalThrputCap}. Likewise, the spatial throughput capacity with stochastic power control, denoted by $C^{pc}(\theta)$, also can be acquired by substituting the results of $p^{pc}_k(\theta)$ and $c^{pc}_k$ into \eqref{Eqn:OptimalThrputCap}. The maximum of $C_L(\theta)$ and $C^{pc}(\theta)$ should also exist, but in general they are not easily found in closed-form. We will present some numerical results of $C(\theta)$, $C_L(\theta)$ and $C^{pc}(\theta)$ in Section \ref{SubSec:ErgodicCapacityNumResults} and verify their maximum values indeed exist.  

\subsection{Numerical Results}\label{SubSec:ErgodicCapacityNumResults}
\begin{figure}[t!]
\centering
\includegraphics[scale=0.36]{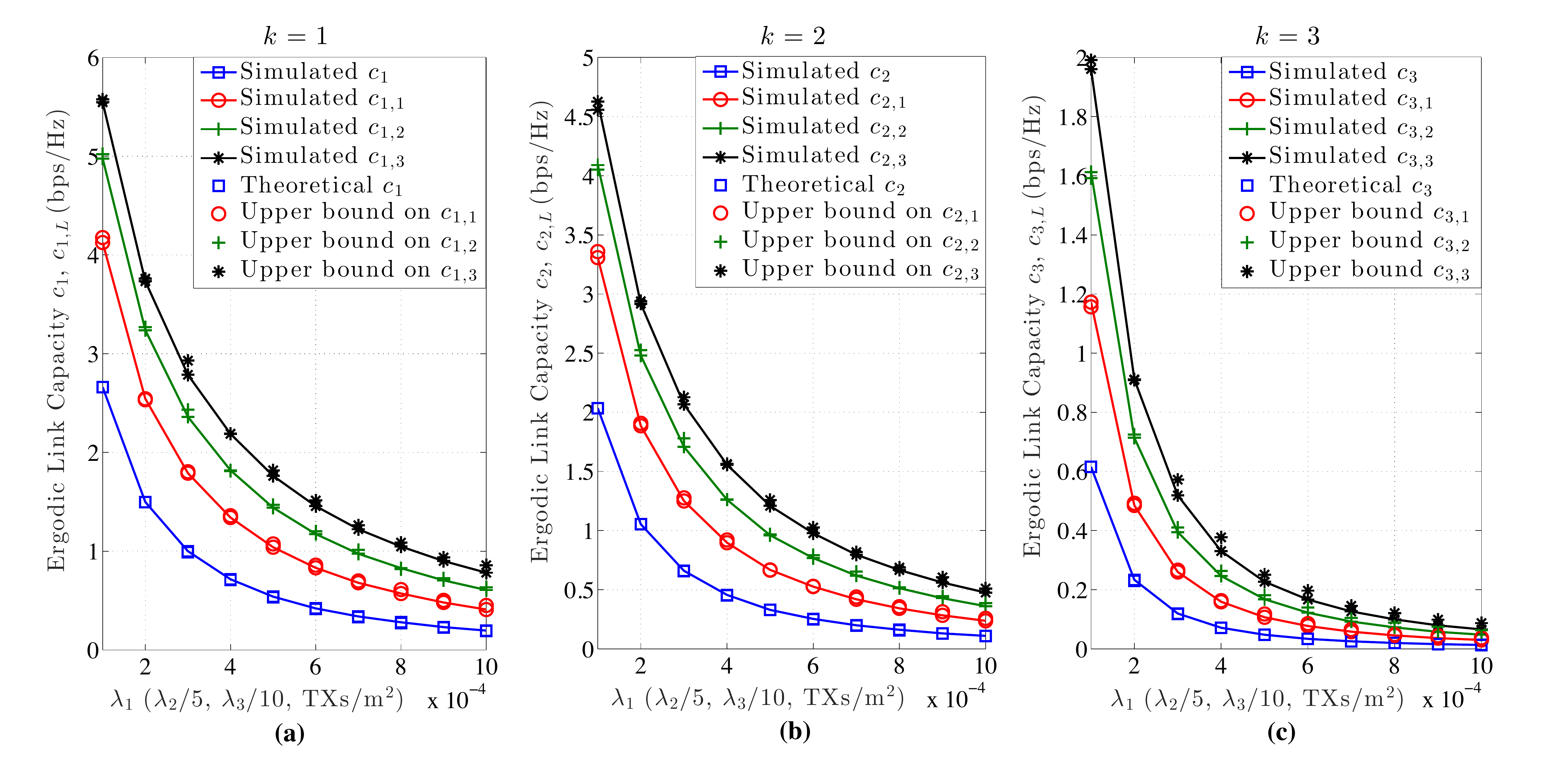}
\caption{Simulation results of the ergodic link capacities without and with  interference cancellation for Rayleigh fading: (a) $c_1$ and $c_{1,L}$ for $L=1, 2, 3$ (b) $c_2$ and $c_{2,L}$ for $L=1, 2, 3$ (c) $c_3$ and $c_{3,L}$ for $L=1, 2, 3$.}
\label{Fig:ErgodicLinkCap}
\end{figure}

\begin{figure}[t!]
\centering
\includegraphics[scale=0.35]{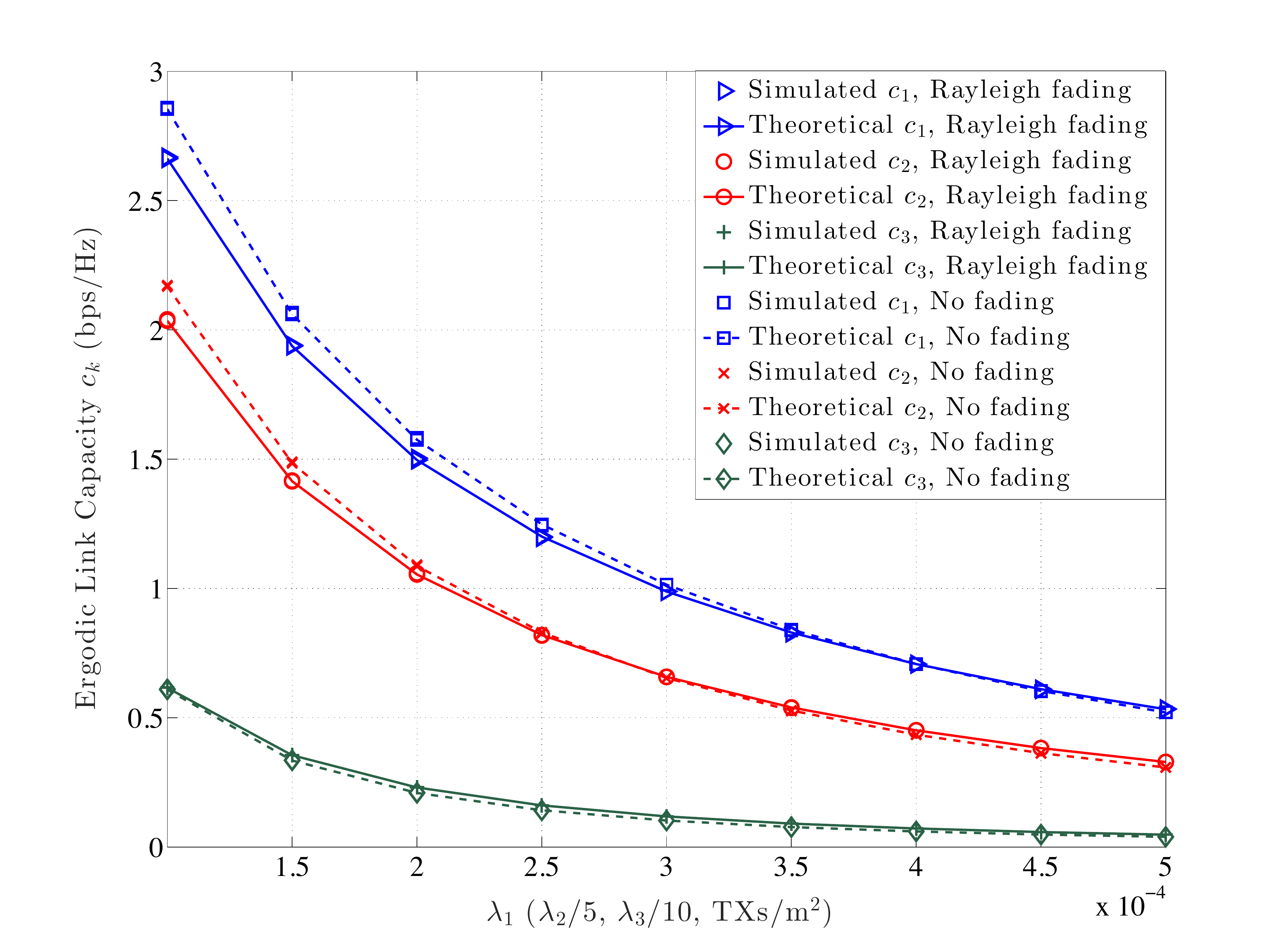}
\caption{The simulation results of the ergodic link capacities with and without Rayleigh fading. They show that channel fading does not always jeopardize the ergodic link capacity of each type.}
\label{Fig:RanRXErgodicLinkCap}
\end{figure}

In this subsection, the simulation results of the ergodic link capacity and spatial throughput capacity are presented to validate the theoretical analyses in the previous subsections. The main network parameters used for simulation here are the same as those in Table \ref{Tab:SimPara}. In Fig. \ref{Fig:ErgodicLinkCap}, we show the simulation results of the type-$k$ ergodic link capacity with and without interference in order to compare the simulated $c_k$ with its theoretical result in \eqref{Eqn:ErgodicLinkCapNakagami} and compare the simulated $c_{k,L}$ with its upper bound found in \eqref{Eqn:UppBoundLinkCapCanKNakagami} for $m_k=1$ (Rayleigh fading), and we can see the simulated $c_k$ and theoretical $c_k$ perfectly coincide with each other, and the derived upper bound on $c_{k,L}$ for each $k$ is also very tight and almost coincides with its  simulated result, and interference cancellation suffer the diminishing returns problem even through it is able to significantly improve the ergodic link capacity by removing a few strong interfere.  Fig. \ref{Fig:RanRXErgodicLinkCap} presents the comparison results for the type-$k$ ergodic link capacities with and without Rayleigh fading. As we can see, the ergodic link capacity with Rayleigh fading is not always smaller than that without Rayleigh fading and fading actually improves the ergodic link capacity when the network is getting dense, which validates our discussion in Section \ref{SubSec:ErgodicLinkCapacitywithGenRanModel}. 

\begin{figure}
\centering
\includegraphics[scale=0.38]{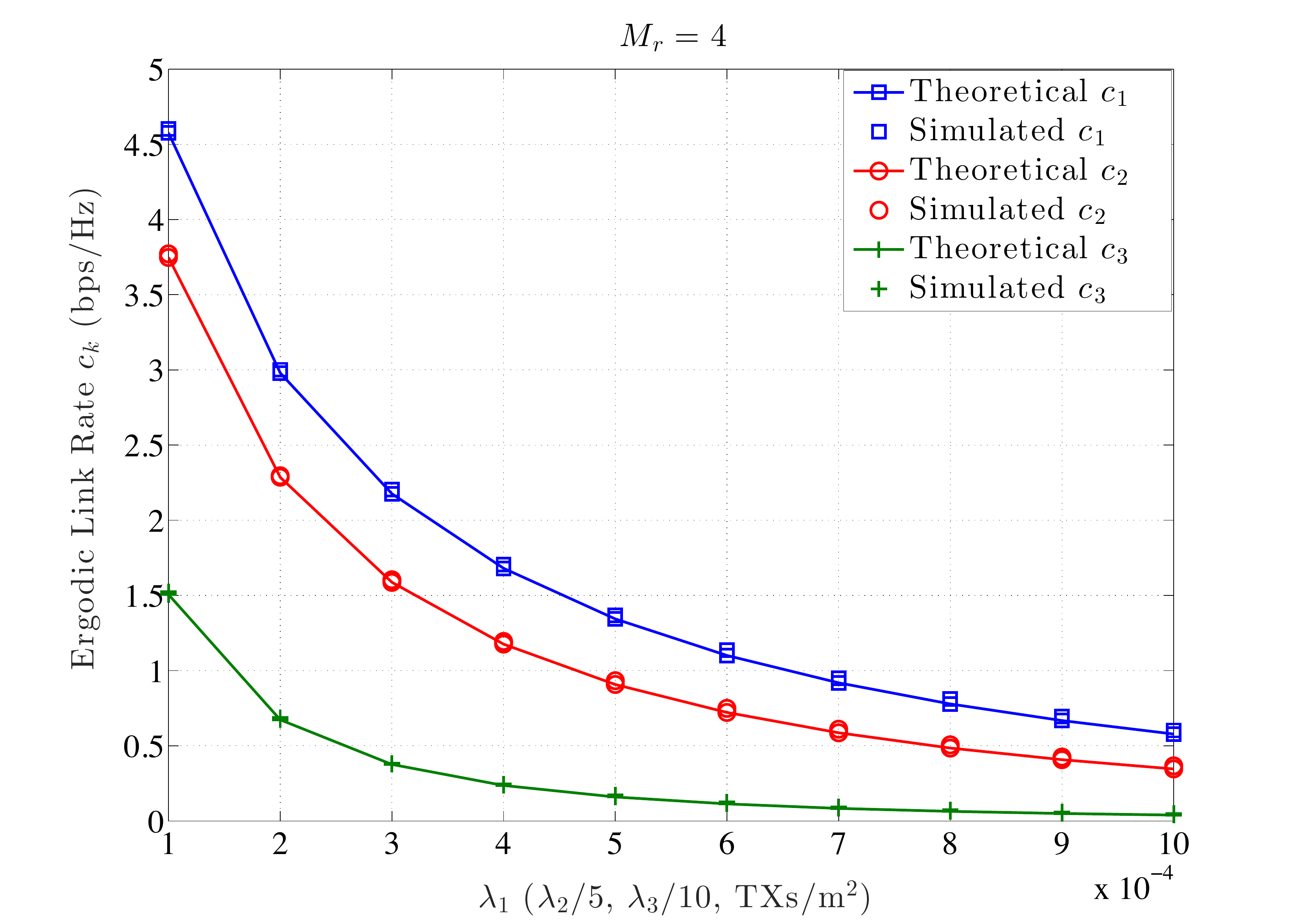}
\caption{Simulation results of the ergodic link capacity with SIMO Rayleigh fading channels. Each RX has 4 receive antennas and the theoretical results of $c_k$ are calculated based on \eqref{Eqn:ErgodicLinkRateSIMO}.}
\label{Fig:ErgodicLinkCapSIMO}
\end{figure}

The trait of $c_k$ derived in \eqref{Eqn:ErgodicLinkCapNoCan} is that $c_k$ works for any distribution of $\widehat{S}_k$ as long as the Laplace transform of $\widehat{S}_k$ exists. To demonstrate the generality of $c_k$ in \eqref{Eqn:ErgodicLinkCapNoCan}, consider an SIMO communication link in a type-$k$ TX-RX pair in which the TX has a single antenna and the RX has $M_r$ antennas. For this SIMO channel with receive beamforming, we have $\widehat{S}_k\sim\text{Erlang}(M_r,M_r)$, $\mathbb{E}[S_k]=M_rP_kR^{-\alpha}_k$ for constants $P_k$ and $R_k$ and the distribution of $I_k$ does not change. Using \eqref{Eqn:ErgodicLinkCapNoCan}, the type-$k$ ergodic link capacity with an $1\times M_r$ SIMO channel can be shown as
\begin{align}\label{Eqn:ErgodicLinkRateSIMO}
c_k=\frac{1}{\ln (2)}\int_{0^+}^{\infty}\frac{1}{\theta}\left[1-\left(1+\frac{\theta}{M_r}\right)^{-M_r}\right] \mathcal{L}_{I_k}\left(\frac{\theta R^{\alpha}_k}{M_rP_k}\right)\dif \theta.
\end{align} 

\begin{figure}
\centering
\includegraphics[scale=0.4]{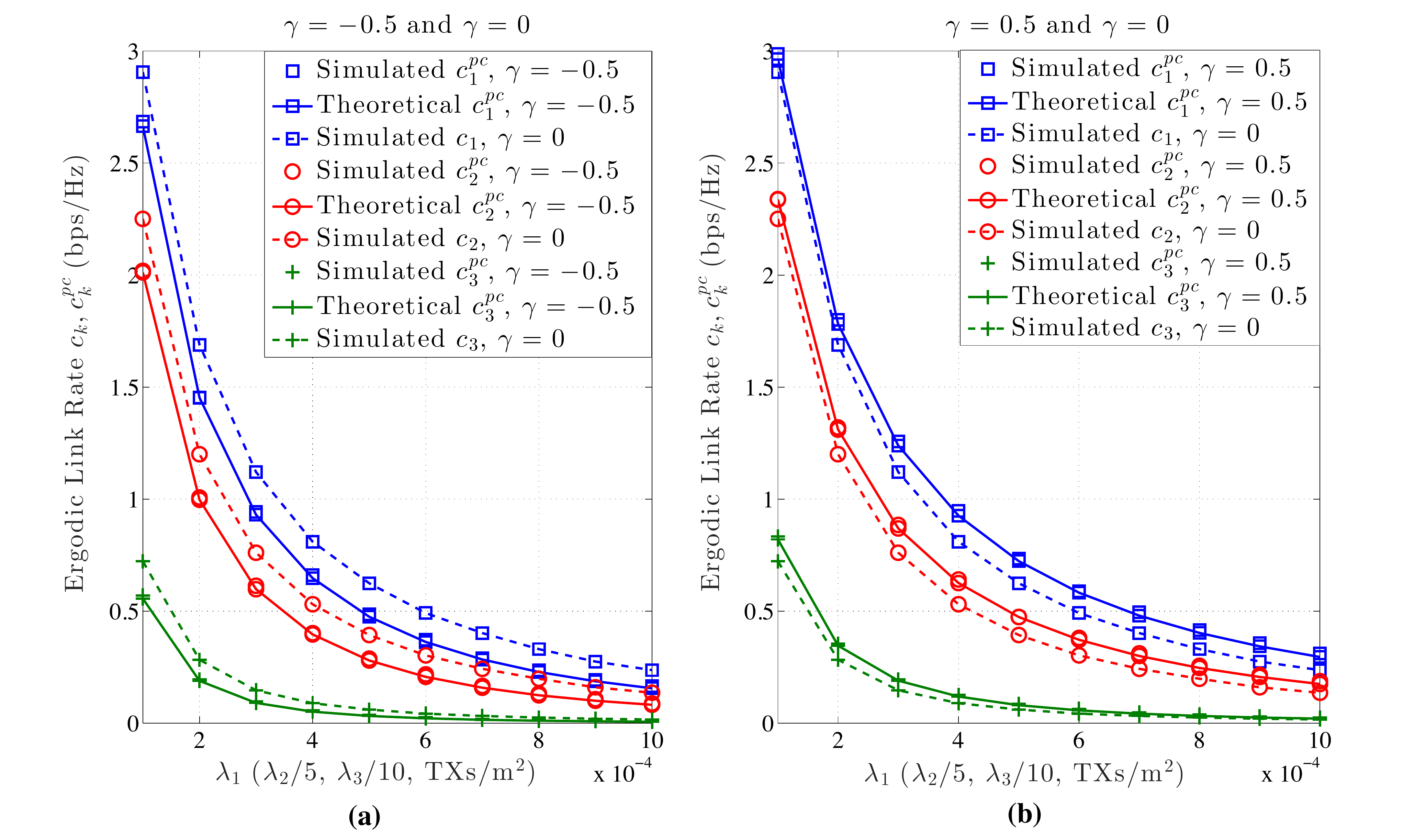}
\caption{Simulation results of the ergodic link capacity with stochastic power control: (a) $c_k$ and $c^{pc}_k$ with $\gamma=-0.5$, (b)}
\label{Fig:ErgodicLinkCapPowerControl}
\end{figure}

\begin{figure}[t!]
\centering
\includegraphics[scale=0.42]{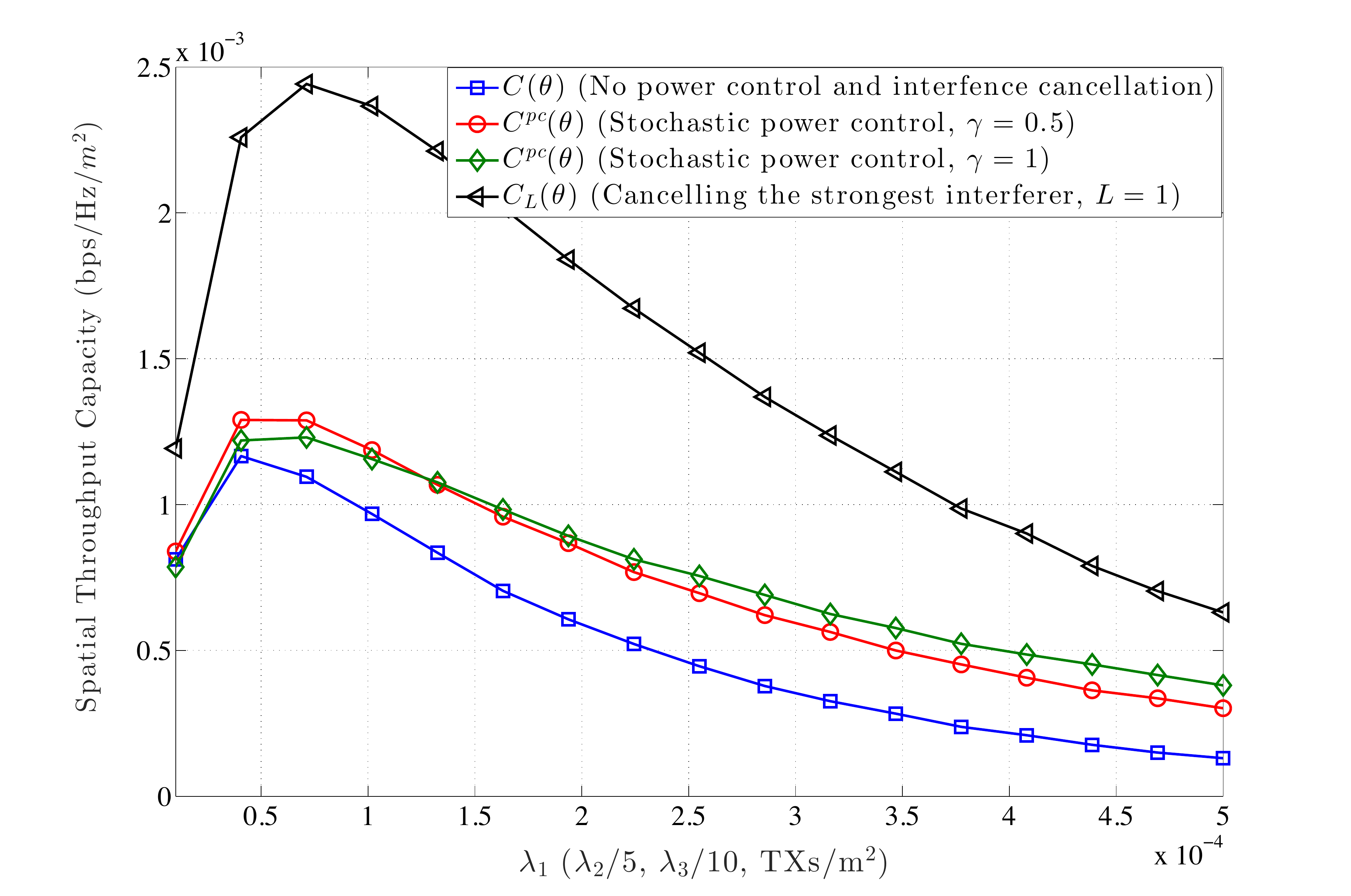}
\caption{Simulation results of the spatial throughput capacities for $\theta=1$:  $C(\theta)$ (spatial throughput capacity without interference cancellation and power control), $C^{pc}(\theta)$ (spatial throughput capacity with stochastic power control for $\gamma=0.5, 1$),  $C_2(\theta)$ (spatial throughput capacity with canceling the first two strongest interferers).}
\label{Fig:SpaThrCap}
\end{figure}

The simulation results of $c_k$ with an $1\times 4$ SIMO  Rayleigh fading channel are shown in Fig. \ref{Fig:ErgodicLinkCapSIMO} and the theoretical results obtained from \eqref{Eqn:ErgodicLinkRateSIMO} perfectly matches their corresponding simulated results. In Fig. \ref{Fig:ErgodicLinkCapPowerControl}, we can see the simulation results of the ergodic link capacity with and without stochastic power control. The theoretical results in the figure are found based on \eqref{Eqn:ErgodicRatePowerControl} and they completely coincide with their corresponding simulated results. According to the discussions in Section \ref{SubSec:ErgodicLinkCapacityPowerControl}, we know the stochastic power control with $\gamma>0$ always benefits the ergodic link capacity with Rayleigh fading channels and such a conclusion can be easily perceived by comparing the results in Fig. \ref{Fig:ErgodicLinkCapPowerControl}(a) of $\gamma=-0.5$ with those in Fig. \ref{Fig:ErgodicLinkCapPowerControl}(b) of $\gamma=0.5$. 

The simulation results of the spatial throughput capacities $C(\theta)$, $C^{pc}(\theta)$, and $C_L(\theta)$ are shown in Fig. \ref{Fig:SpaThrCap}. We see that the spatial throughput capacities with stochastic power control (i.e., $C^{pc}(\theta)$ with $\gamma=0.5,1$) and interference cancellation (i.e., $C_L(\theta)$ with $L=1$) all are higher than that without stochastic power control and interference cancellation. All spatial throughput capacities have a maximum vale, as expected. In addition, we observe that $C^{pc}(\theta)$ with $\gamma=1$ is not always better than that with $\gamma=0.5$ so that $\gamma$ should be carefully chosen based on the TX intensities, like the similar phenomenon observed in the case of the success probability. Finally, from Fig. \ref{Fig:SpaThrCap} we can infer that interference cancellation does not always outperform stochastic power control and it works much better in the low intensity region. This is because stochastic power control can reduce a lot of interference when the network is dense, while canceling one or more interferers may not reduce interference as much as stochastic power control.

\section{Concluding Remarks}
In prior works, the distribution of the SIR in a Poisson wireless ad hoc network is analyzed by presuming some specific random models used in the SIR. Such a model-dependent distribution is unable to provide some insight into how much the statistical properties of the SIR are impacted once the random models involved in the SIR change. Accordingly, in this paper we introduce a Laplace-transform-based framework of analyzing the distribution of the SIR with a general distribution. This framework successfully helps us find the model-free general expression of  the CDF of the SIR without and with interference cancellation, and this expression can be significantly simplified to a nearly closed-form result for the pathloss exponent of four. We apply the CDF results of the SIR in finding the success probabilities, ergodic link capacities and spatial throughput capacities in the scenarios of interference cancellation, stochastic power control and SIMO/MISO transmission. These results are traditionally intractable due to the model-dependent limit on the distribution of the SIR previously derived. Nonetheless, they become much more tractable with the aid of the analytical framework proposed in this work. The proposed analytical framework of the SIR can be also applied to many other analyses pertaining to the distribution of the SIR, such as SIR with or without channel-aware opportunistic scheduling, SIR with multi-hop transmissions, link and network energy efficiency, and SIR analysis in heterogeneous cellular networks, etc. 

\appendix[Proofs of Theorems]

\subsection{Proof of Theorem \ref{Thm:LaplaceFunCanKinterfer}}\label{App:ProofLaplaceFunCanKinterfer}
Let $\widetilde{I}_k$ be the interference at the type-$k$ typical RX generated by the homogeneous PPP $\widetilde{\Phi}$ of intensity $\widetilde{\lambda}=\sum_{k=1}^{K}\widetilde{\lambda}_k$. According to Theorem \ref{Thm:SupBiasFunCDF} and the discussion right after it, $I_k$ has the same distribution as $\widetilde{I}_k$. Namely, the interference $I_k$ can be rewritten as
\begin{align} I_k\stackrel{d}{=}\widetilde{I}_k\defn\sum_{\widetilde{X}_i\in\widetilde{\Phi}}\|\widetilde{X}_i\|^{-\alpha}=\sum_{\widetilde{X}_i\in\tilde{\Phi}}(\|\widetilde{X}_i\|^2)^{-\frac{\alpha}{2}},\label{Eqn:ProofInterference}
\end{align}
where $\stackrel{d}{=}$ means equivalence in distribution. Thus, the Laplace transform of $I_k$ is equal to the Laplace transform of $\widetilde{I}_k$  that can be readily obtained as shown in \eqref{Eqn:LapTranInter} according to the result in \cite{MHRKG09}.  Without loss of generality, we can assume $\widetilde{X}_i$ in \eqref{Eqn:ProofInterference} is the $i$th nearest node in $\tilde{\Phi}$ to the type-$k$ typical RX. Therefore, $\widetilde{X}_1$ is the nearest node to the type-$k$ typical RX and the probability density function (pdf) of $\|X_1\|^2$ is $f_{\|\tilde{X}_1\|^2}(x)=\pi\widetilde{\lambda}\exp(-\pi\widetilde{\lambda} x)$ and thus $\|\tilde{X}_1\|^2$ is an exponential RV with parameter $\pi\widetilde{\lambda}$. Also, we know  $\|\widetilde{X}_{i+1}\|^2=\|\widetilde{X}_i\|^2+\|\widetilde{X}_1\|^2$ for $i>1$ where $\|\widetilde{X}_i\|^2$ and $\|\widetilde{X}_1\|^2$ are independent \cite{MH05,CHLLCW16}.  Accordingly, $I_{k,L}$ has an identity in distribution given by
\begin{align*}
I_{k,L}&\stackrel{d}{=} \sum_{\widetilde{X}_{L+j}\in\widetilde{\Phi}\setminus\tilde{\Phi}_L}\|\widetilde{X}_{L+j}\|^{-\alpha}=\sum_{\widetilde{X}_{L+j}\in\widetilde{\Phi}\setminus\widetilde{\Phi}_L}\left(\|\widetilde{X}_L\|^2+\|\widetilde{X}_j\|^2\right)^{-\frac{\alpha}{2}},\,j=1,2,3,\cdots,
\end{align*} 
where $\tilde{\Phi}_L$ consists of the first $L$ nearest nodes in $\tilde{\Phi}$ to the type-$k$ typical RX. 

For a given $\widetilde{X}_L$, the Laplace transform of $I_{k,L}$ can be found as shown in the following:
\begin{align*}
\mathcal{L}_{I_{k,L}|\widetilde{X}_L}(s) &= \mathbb{E}\left[e^{-s\sum_{\widetilde{X}_{L+j}\in\tilde{\Phi}\setminus\widetilde{\Phi}_L}(\|\widetilde{X}_L\|^2+\|\widetilde{X}_j\|^2)^{-\frac{\alpha}{2}}}\right]=\mathbb{E}\left\{\prod_{\widetilde{X}_j\in\widetilde{\Phi}\setminus\widetilde{\Phi}_L}e^{-s (\|\widetilde{X}_L\|^2+\|\widetilde{X}_j\|^2)^{-\frac{\alpha}{2}}}\bigg| \widetilde{X}_L \right\}\\
&\stackrel{(a)}{=}\exp\left\{-\pi\widetilde{\lambda}\underbrace{\int_{0}^{\infty} \left(1-e^{-s\|\widetilde{X}_L\|^{-\alpha}(1+\frac{r}{\|\widetilde{X}_L\|^2})^{-\frac{\alpha}{2}}}\right) \dif r}_{(b)} \right\},
\end{align*}
where $(a)$ follows from the PGF of $K$ independent homogeneous PPPs \cite{DSWKJM13,MH12}. Also, by letting $Y\sim\exp(1,1)$, the integral in $(b)$ can be simplified as follows
\begin{align*}
(b)&=\int_{0}^{\infty} \mathbb{P}\left[\left(\|\widetilde{X}_L\|^2+r\right)\leq\left(\frac{s}{Y}\right)^{\frac{2}{\alpha}} \bigg|\|\widetilde{X}_L\|\right]\dif r\stackrel{u=r+\|\widetilde{X}_L\|^2}{=}\int_{\|\widetilde{X}_L\|^2}^{\infty} \mathbb{P}\left[u\leq\left(\frac{s}{Y}\right)^{\frac{2}{\alpha}} \right]\dif u\\
&=\int_{0}^{\infty} \mathbb{P}\left[u\leq\left(\frac{s}{Y}\right)^{\frac{2}{\alpha}} \right]\dif u-\int_{0}^{\|\widetilde{X}_L\|^2}\mathbb{P}\left[u\leq\left(\frac{s}{Y}\right)^{\frac{2}{\alpha}} \right]\dif u\\
&=s^{\frac{2}{\alpha}}\Gamma\left(1-\frac{2}{\alpha}\right)+\|\widetilde{X}_L\|^2\left[\int_{0}^{1}e^{-s(\omega\|\widetilde{X}_L\|^2)^{-\frac{\alpha}{2}}}\dif \omega-1\right]\\
&=s^{\frac{2}{\alpha}}\Gamma\left(1-\frac{2}{\alpha}\right)-\|\widetilde{X}_L\|^2\left[1-\left(\frac{2}{\alpha}\right)\left(\frac{s^{\frac{2}{\alpha}}}{\|\widetilde{X}_L\|^2}\right)\Gamma\left(-\frac{2}{\alpha},\frac{s^{\frac{2}{\alpha}}}{\|\widetilde{X}_L\|^2}\right)\right]\\
&=s^{\frac{2}{\alpha}}\Gamma\left(1-\frac{2}{\alpha}\right)-\ell_{\|\widetilde{X}_L\|^{2}}\left(s\|\widetilde{X}_L\|^{-\alpha},\frac{2}{\alpha}\right).
\end{align*}
As a result, letting $D_L=\|\tilde{X}_L\|^2$ yields $\mathcal{L}_{I_{k,L}|D_L}(s)$ given by 
\begin{align*}
\mathcal{L}_{I_{k,L}|D_L}(s)=\mathcal{L}_{I_k}(s)\cdot \exp\left[\pi\widetilde{\lambda}\ell_{D_L}\left(sD_L^{-\frac{\alpha}{2}},\frac{2}{\alpha}\right)\right].
\end{align*}
Then note that the pdf of $D_L$ is equal to the pdf of the sum of $L$ i.i.d. $(\|\widetilde{X}_1\|^2)$'s so that it is an Erlang distribution with parameters $L$ and $\pi\widetilde{\lambda}$, i.e., $f_{D_L}(x) = \frac{(\pi\widetilde{\lambda})^Lx^{L-1}e^{-\pi\widetilde{\lambda} x}}{(L-1)!}.$
Averaging $\mathcal{L}_{I_{k,L}|D_L}(s)$ over $D_L$ exactly yields the result in \eqref{Eqn:LaplaceCanKinterference}. 

Also, we know the following
\begin{align*}
-\ell_{D_L}\left(sD_L^{-\frac{\alpha}{2}},\frac{2}{\alpha}\right) &= \int_{0}^{D_L}\left(1-e^{-su^{-\frac{\alpha}{2}}}\right) \dif u\stackrel{(c)}{=}D_L\left(1-e^{-\frac{s}{(\omega D_L)^{\frac{\alpha}{2}}}}\right)\stackrel{(d)}{\leq} s \omega^{-\frac{\alpha}{2}} D_L^{1-\frac{\alpha}{2}},
\end{align*}
where $(c)$ follows from the mean value theorem in calculus for $\omega\in(0,1)$ and $(d)$ is due to $1-e^{-x}\leq x$ for $x\geq 0$. Thus, it follows that
\begin{align*}
\mathcal{L}_{I_{k,L}}(s)\leq \mathcal{L}_{I_k}(s)\cdot\mathbb{E}\left[\exp\left(\pi\widetilde{\lambda}s \omega^{-\frac{\alpha}{2}} D_L^{1-\frac{\alpha}{2}}\right)\right],
\end{align*}
which yields \eqref{Eqn:LaplaceInterUpperBound} since $\mathcal{M}_{\omega^{-\frac{\alpha}{2}}D_L^{1-\frac{\alpha}{2}}}(\pi\widetilde{\lambda}s)=\mathbb{E}\left[\exp\left(\pi\widetilde{\lambda}s \omega^{-\frac{\alpha}{2}}D_L^{1-\frac{\alpha}{2}}\right)\right]$.

\subsection{Proof of Theorem \ref{Thm:LapTransInvSIR}}\label{App:ProofLapTransInvSIR}
According to \eqref{Eqn:DefnSINR} and \eqref{Eqn:LapTranInter}, the Laplace transform of the reciprocal of $\sinr_k$ in \eqref{Eqn:DefnSINR} can be expressed as
\begin{align}
\mathcal{L}_{\sinr^{-1}_k}(s)&=\mathbb{E}_{S_k}\left[\mathcal{L}_{I_k}\left(\frac{s/\mathbb{E}[S_k]}{S_k/\mathbb{E}[S_k]}\right)\right]=\mathbb{E}_{\widehat{S}_k}\left[\exp\left(-\pi\Gamma\left(1-\frac{2}{\alpha}\right)\widetilde{\lambda} \left(\frac{\widehat{s}}{\widehat{S}_k}\right)^{\frac{2}{\alpha}}\right)\right]\nonumber\\
&=\mathcal{L}_{\widehat{S}^{-\frac{2}{\alpha}}_k}\left(\pi\Gamma\left(1-\frac{2}{\alpha}\right)\widehat{s}^{\frac{2}{\alpha}}\widetilde{\lambda}\right)=\int_{0}^{\infty} e^{-\pi\Gamma\left(1-\frac{2}{\alpha}\right)\widetilde{\lambda}(\widehat{s}/x)^{\frac{2}{\alpha}}}f_{\widehat{S}_k}(x) \dif x\nonumber\\
&=s\int_{0}^{\infty} e^{-\pi\Gamma\left(1-\frac{2}{\alpha}\right)\widetilde{\lambda}(t\mathbb{E}[S_k])^{-\frac{2}{\alpha}}}f_{\widehat{S}_k}(ts)\dif t,\,\,\text{(letting $t=x/s$)},
\end{align}
where $\widehat{s}\defn s/\mathbb{E}[S_k]$. By the definition of $\mathcal{L}_{\sinr^{-1}_k}(s)$, we also know
\begin{align*}
\mathcal{L}_{\sinr^{-1}_k}(s)&=\int_{0}^{\infty} f_{\sinr^{-1}_k}(t) e^{-st} \dif t=\int_{0}^{\infty} \frac{\dif F_{\sinr^{-1}_k}(t)}{\dif t} e^{-st} \dif t=\int_{0}^{\infty} s F_{\sinr^{-1}_k}(t) e^{-st} \dif t\\
&=s\int_{0}^{\infty} e^{-\pi\Gamma\left(1-\frac{2}{\alpha}\right)\widetilde{\lambda}(t\mathbb{E}[S_k])^{-\frac{2}{\alpha}}}f_{\widehat{S}_k}(ts)\dif t,
\end{align*}
which indicates 
\begin{align}\label{Eqn:LTof InverseType-kSIR}
\int_{0}^{\infty} F_{\sinr^{-1}_k}(t) e^{-st} \dif t = \int_{0}^{\infty} e^{-\pi\Gamma\left(1-\frac{2}{\alpha}\right)\widetilde{\lambda}(t\mathbb{E}[S_k])^{-\frac{2}{\alpha}}}f_{\widehat{S}_k}(ts)\dif t
\end{align}
and then taking the inverse Laplace transform of the both sides of  \eqref{Eqn:LTof InverseType-kSIR} yields
\begin{align*}
F_{\sinr^{-1}_k}(t)=1-F_{\sinr_k}\left(t^{-1}\right)=\mathcal{L}^{-1}\left\{ \int_{0}^{\infty} \mathcal{L}_{I_k}\left(\frac{1}{t\mathbb{E}[S_k]}\right)f_{\widehat{S}_k}(ts)\dif t\right\}(t)
\end{align*}
and then setting the argument of $F_{\sinr_k}\left(t^{-1}\right)$ as $t^{-1}=\theta$ results in \eqref{Eqn:CDFSIRTypek}. 

Now consider the Laplace transform of $\sinr^{-1}_{k,L}$. Using the result in \eqref{Eqn:LaplaceCanKinterference}, we can have the following
\begin{align*}
\mathcal{L}_{\sinr^{-1}_{k,L}}(s)&=\mathbb{E}\left[ \mathcal{L}_{I_k}\left(\frac{s}{S_k}\right)\mathcal{M}_{\ell_{D_L}(s/S_k)}\left(\pi\widetilde{\lambda}\right)\right]=\int_{0}^{\infty}e^{-\pi\Gamma\left(1-\frac{2}{\alpha}\right)\widetilde{\lambda}(\widehat{s}/x)^{\frac{2}{\alpha}}}\mathcal{M}_{\ell_{D_L}(\frac{\widehat{s}}{x})}\left(\pi\widetilde{\lambda}\right)f_{\widehat{S}_k}(x) \dif x\\
&\stackrel{(x=st)}{=}\int_{0}^{\infty} s\mathcal{L}_{I_k}\left(\frac{1}{t\mathbb{E}[S_k]}\right)\mathcal{M}_{\ell_{D_L}(\frac{1}{\mathbb{E}[S_k]t})}\left(\pi\widetilde{\lambda}\right) f_{\widehat{S}_k}(st) \dif t,
\end{align*}
which is exactly the result in \eqref{Eqn:LapInvSIRcanK}. According to the steps of deriving $F_{\sinr^{-1}_k}(t)$ in above, we also can show
\begin{align*}
F_{\sinr^{-1}_{k,L}}(t) = \mathcal{L}^{-1}\left\{\int_{0}^{\infty}s\mathcal{L}_{I_k}\left(\frac{1}{t\mathbb{E}[S_k]}\right)\mathcal{M}_{\ell_{D_L}(\frac{1}{\mathbb{E}[S_k]t})}\left(\pi\widetilde{\lambda}\right)f_{\widehat{S}_k}(st)\dif t\right\}(t)=1-F_{\sinr_{k,L}}\left(t^{-1}\right),
\end{align*}
which yields \eqref{Eqn:CDFSIRTypekKcan}.

\subsection{Proof of Theorem \ref{Thm:MomentSIR}}\label{App:ProofMeanSIR}
According to the proof of Theorem \ref{Thm:LaplaceFunCanKinterfer}, we know
\begin{align*}
\mathcal{L}_{\sinr^{-1}_k}(s)&=\int_{0}^{\infty}s F_{\sinr^{-1}_k}(t) e^{-st} \dif t=\int_{0}^{\infty}s F^c_{\sinr_k}\left(t^{-1}\right) e^{-st} \dif t\\
&=\bigintssss_{0}^{\infty} \frac{sf_{\widehat{S}_k}(s t)\dif t}{e^{\pi\Gamma\left(1-\frac{2}{\alpha}\right)\widetilde{\lambda}(t\mathbb{E}[S_k])^{-\frac{2}{\alpha}}}},
\end{align*}
where $F^c_{Z}(\cdot)$ is the CCDF of RV $Z$. This result implies
\begin{align*}
\int_{0}^{\infty} F^c_{\sinr_k}\left(t^{-1}\right)\int_{0}^{\infty} e^{-st}\dif s \dif t=\bigintssss_{0}^{\infty} \frac{\int_{0}^{\infty}f_{\widehat{S}_k}(s t)\dif s\dif t}{e^{\pi\Gamma\left(1-\frac{2}{\alpha}\right)\widetilde{\lambda}(t\mathbb{E}[S_k])^{-\frac{2}{\alpha}}}}
\end{align*}
and this leads to
\begin{align*}
\int_{0}^{\infty} \frac{1}{t} F^c_{\sinr_k}\left(t^{-1}\right) \dif t=\bigintssss_{0}^{\infty} \frac{\dif t}{te^{\pi\Gamma\left(1-\frac{2}{\alpha}\right)\widetilde{\lambda}(t\mathbb{E}[S_k])^{-\frac{2}{\alpha}}}},
\end{align*}
which yields
\begin{align*}
\int_{0}^{\infty} F^c_{\sinr_k}\left(t^{-1}\right)\dif t = \int_{0}^{\infty}e^{-\pi\Gamma\left(1-\frac{2}{\alpha}\right)\widetilde{\lambda}(t\mathbb{E}[S_k])^{-\frac{2}{\alpha}}} \dif t.
\end{align*}
Also, we know $\mathbb{E}[\sinr^{\delta}_k]=\int_{0}^{\infty} F^c_{\sinr_k}(t^{1/{\delta}})\dif t$ so that $\mathbb{E}[\sinr^{\delta}_k]$ can be found by
\begin{align*}
\mathbb{E}[\sinr^{\delta}_k] &= \int_{0}^{\infty}e^{-\pi\Gamma\left(1-\frac{2}{\alpha}\right)\widetilde{\lambda}(t^{\frac{1}{\delta}}/\mathbb{E}[S_k])^{\frac{2}{\alpha}}} \dif t=\left(\frac{(\mathbb{E}[S_k])^{\frac{2}{\alpha}}}{\pi\Gamma(1-\frac{2}{\alpha})\widetilde{\lambda}}\right)^{\frac{\delta\alpha}{2}}\left(\frac{\delta\alpha}{2}\right) \int_{0}^{\infty} e^{-u} u^{\frac{\delta\alpha}{2}-1}\dif u,
\end{align*} 
whose last term is $\Gamma(\frac{\delta\alpha}{2})$ and the result in \eqref{Eqn:NthMomSIR} is acquired accordingly. Similarly, we also can show
\begin{align*}
\int_{0}^{\infty} F^c_{\sinr_{k,L}}\left(t\right)\dif t = \bigintsss_{0}^{\infty}\dfrac{\mathcal{L}_{\widetilde{\ell}_{D_L,1/t}}\left(\pi\widetilde{\lambda}\right)}{e^{\pi\Gamma\left(1-\frac{2}{\alpha}\right)\widetilde{\lambda}(t/\mathbb{E}[S_k])^{\frac{2}{\alpha}}}} \dif t
\end{align*}
and using it to find $\mathbb{E}[\sinr^{\delta}_{k,L}]=\int_{0}^{\infty} F^c_{\sinr_{k,L}}(t^{1/{\delta}})\dif t$ yields \eqref{Eqn:NthMomSIRKcan}.  

\subsection{Proof of Theorem \ref{Thm:FadingImproveSuccProb}}\label{App:ProofFadingImproveSuccProb}
Since $\widehat{S}_k\sim\text{Gamma}(m_k,1/m_k)$, the success probability, $p_k(\theta)$, based on the result in Corollary \ref{Cor:CDFSIRErlang}, is given by
\begin{align*}
p_k\left(v^{-1}\right)&=\frac{1}{(m_k-1)!}\frac{\dif^{m_k-1} }{\dif v^{m_k-1}}\left(\frac{v^{m_k-1}}{e^{\Gamma\left(1-\frac{2}{\alpha}\right)(m_k/vP_k\mathbb{E}[R^{-\alpha}_k])^{\frac{2}{\alpha}}\widetilde{\Lambda}^{\dagger}}}\right).
\end{align*}
In the case of constant $S_k$, we know $\widehat{S}_k=1$ and $S_k=\mathbb{E}[S_k]$. Thus,  the success probability based on the result in \eqref{Eqn:CDFSIRNoRamSig} is
\begin{align*}
p_k(v^{-1}) &= \mathcal{L}^{-1}\left\{\frac{1}{s} \exp\left[-\pi\Gamma\left(1-\frac{2}{\alpha}\right)\left(\frac{s}{\mathbb{E}[S_k]}\right)^{\frac{2}{\alpha}}\widetilde{\lambda}\right]\right\}\left(v\right)\\
&\stackrel{(a)}{\geq} 1-\Gamma\left(1-\frac{2}{\alpha}\right)\frac{\pi\widetilde{\lambda}}{(\mathbb{E}[S_k])^{\frac{2}{\alpha}}}\mathcal{L}^{-1}\left\{s^{\frac{2}{\alpha}-1}\right\}(v)=\left[1-\frac{\pi\widetilde{\lambda}}{(v\mathbb{E}[S_k])^{\frac{2}{\alpha}}}\right]^+,
\end{align*} 
where $(a)$ follows from the assumption that $e^{-x}\geq 1-x$ for $x\geq 0$ and $(x)^+\defn \max\{x,0\}$.  Hence, when the following inequality
\begin{align*}
\frac{\dif^{m_k-1} }{\dif v^{m_k-1}}\left(\frac{v^{m_k-1}}{e^{\pi\Gamma\left(1-\frac{2}{\alpha}\right)(m_k/v\mathbb{E}[S_k])^{\frac{2}{\alpha}}\widetilde{\lambda}}}\right)\leq (m_k-1)!\left[1-\frac{\pi\widetilde{\lambda}}{(v\mathbb{E}[S_k])^{\frac{2}{\alpha}}}\right]^+
\end{align*}
holds, the randomness of the received signal power does not benefit the success probability. By integrating both sides of this inequality $m_k-1$ times and letting  $\theta=v^{-1}$, we have
\begin{align*}
\pi\widetilde{\lambda}\left(\frac{\theta}{\mathbb{E}[S_k]}\right)^{\frac{2}{\alpha}}\varsigma_k\leq 1-\exp\left\{-\pi\widetilde{\lambda}\Gamma\left(1-\frac{2}{\alpha}\right)\left(\frac{m_k\theta}{\mathbb{E}[S_k]}\right)^{\frac{2}{\alpha}}\right\},
\end{align*}
which can be simplified as
\begin{align*}
\exp\left\{-\Gamma\left(1-\frac{2}{\alpha}\right)\left(m_k\theta\right)^{\frac{2}{\alpha}}\widetilde{\Lambda}_k\right\}\leq 1-\theta^{\frac{2}{\alpha}}\varsigma_k\widetilde{\Lambda}_k
\end{align*}
due to $\widetilde{\Lambda}_k= \pi\widetilde{\lambda}/(\mathbb{E}[S_k])^{\frac{2}{\alpha}}$.
Thus, the randomness of the received signal power does not benefit the success probability in the network with any $\widetilde{\Lambda}\in\mathbf{\widetilde{\Lambda}}_k(\theta)$ if $\mathbf{\widetilde{\Lambda}}_k(\theta)$ is nonempty for any given $\theta>0$.

\subsection{Proof of Theorem \ref{Thm:SuccProbPowerCon}}\label{App:ProofSuccProbPowerCon}
Since the power control scheme in \eqref{Eqn:StochasticPowerControl} is adopted, the received signal power with power control is
\begin{align*}
S^{pc}_k=\frac{\overline{P}_k\left(H_kR^{-\alpha}_k\right)^{\gamma_k+1}}{\mathbb{E}\left[H^{\gamma_k}_k\right]\mathbb{E}\left[R^{-\alpha\gamma_k}_k\right]}=\frac{S_k^{\gamma_k+1}}{\mathbb{E}\left[S^{\gamma_k}_k\right]}.
\end{align*}
Therefore, $\mathbb{E}\left[(S^{pc}_k)^{-\frac{2}{\alpha}}\right]$ and $\mathbb{E}\left[(S^{pc}_k)^{\frac{2}{\alpha}}\right]$ are readily obtained as
\begin{align*}
\mathbb{E}\left[(S^{pc}_k)^{-\frac{2}{\alpha}}\right]&=\left(\mathbb{E}\left[S^{\gamma_k}_k\right]\right)^{\frac{2}{\alpha}}\mathbb{E}\left[S^{-\frac{2(1+\gamma_k)}{\alpha}}_k\right],\\
\mathbb{E}\left[(S^{pc}_k)^{\frac{2}{\alpha}}\right]&=\left(\mathbb{E}\left[S^{\gamma_k}_k\right]\right)^{-\frac{2}{\alpha}}\mathbb{E}\left[S^{\frac{2(1+\gamma_k)}{\alpha}}_k\right].
\end{align*}
Also, $\widetilde{\lambda}^{pc}$ can be found by
\begin{align*}
\widetilde{\lambda}^{pc}=\sum_{j=1}^{K} \lambda_j \mathbb{E}\left[H^{\frac{2}{\alpha}}_j\right]\mathbb{E}\left[P^{\frac{2}{\alpha}}_j\right]=\sum_{j=1}^{K}\lambda_j\overline{P}^{\frac{2}{\alpha}}_j\mathbb{E}\left[H^{\frac{2}{\alpha}}_j\right]\frac{ \mathbb{E}\left[H^{\frac{2\gamma_j}{\alpha}}_j\right]}{(\mathbb{E}[H^{\gamma_j}_j])^{\frac{2}{\alpha}}}\frac{\mathbb{E}\left[R^{-2\gamma_j}_j\right]}{(\mathbb{E}[R^{-\alpha\gamma_j}_j])^{\frac{2}{\alpha}}}
\end{align*}
in which $\mathbb{E}\left[H^{\frac{2\gamma_j}{\alpha}}_j\right]/(\mathbb{E}[H^{\gamma_j}_j])^{\frac{2}{\alpha}}\leq 1$ and $\mathbb{E}\left[R^{-2\gamma_j}_j\right]/(\mathbb{E}[R^{-\alpha\gamma_j}_j])^{\frac{2}{\alpha}}\leq 1$ by Jensen's inequality and thus $\widetilde{\lambda}^{pc}\leq \widetilde{\lambda}$. Since the lower bound in \eqref{Eqn:BoundSuccProbNoIntCan} are valid for any distribution of transmit powers, substituting $\mathbb{E}[(S^{pc}_k)^{-\frac{2}{\alpha}}]$, $\mathbb{E}[(S^{pc}_k)^{\frac{2}{\alpha}}]$  and $\widetilde{\lambda}^{pc}$ into \eqref{Eqn:BoundSuccProbNoIntCan} yields the lower bound in \eqref{Eqn:SuccProbPowerCon}. In order to have an explicit upper bound on $p^{pc}_k(\theta)$, we use the result of Theorem \ref{Thm:MomentSIR} instead of using the upper bond in \eqref{Eqn:SuccProbPowerCon}. The upper bound on $p^{pc}_k(\theta)$ can be found as follows
\begin{align*}
p^{pc}_k(\theta)& =\mathbb{P}\left[S^{pc}_k\geq\theta I^{pc}_k\right]=\mathbb{P}\left[S_k\geq (\theta \mathbb{E}[S^{\gamma_k}_k]I^{pc}_k)^{\frac{1}{1+\gamma_k}}\right]=\mathbb{E}\left\{F^c_{S_k}\left((\theta \mathbb{E}[S^{\gamma_k}_k]I^{pc}_k)^{\frac{1}{1+\gamma_k}}\right)\right\}\\
&\leq F^c_{S_k}\left((\theta \mathbb{E}[S^{\gamma_k}_k])^{\frac{1}{1+\gamma_k}}\mathbb{E}\left[(I^{pc}_k)^{\frac{1}{1+\gamma_k}}\right]\right).
\end{align*}
Then $\mathbb{E}\left[(I^{pc}_k)^{\frac{1}{1+\gamma_k}}\right]$ can be explicitly found by using the $\frac{1}{1+\gamma}$-moment of the SIR in \eqref{Eqn:NthMomSIR} with unit received signal power so that we have the upper bound shown in \eqref{Eqn:SuccProbPowerCon}. Finally, the result in \eqref{Eqn:SuccProbPowerContAlpha4} is obtained by substituting $S^{pc}_k$ and $\widetilde{\lambda}^{pc}$ into \eqref{Eqn:ExactSuccProbAlpha4} and the lower bound in \eqref{Eqn:LowBoundSuccProbPowerContAlpha4} is due to the erfc function with a positive argument that is convex.

\subsection{Proof of Theorem \ref{Thm:ErgodicLinkCapNoCan}}\label{App:ProofErgodicLinkCanNoCan}
Since the type-$k$ RXs do not cancel any interference, the type-$k$ ergodic link capacity in \eqref{Eqn:DefnSINR} can be rewritten as shown in the following:
\begin{align*}
c_k&=\frac{1}{\ln (2)}\mathbb{E}\left[\ln\left(1+\sinr_k\right)\right]=\frac{1}{\ln(2)}\int_{0}^{1}\mathbb{E}\left[\frac{\sinr_k}{1+y\sinr_k}\right]\dif y\\
&=\frac{1}{\ln(2)}\int_{0}^{1}\mathbb{E}\left[\frac{1}{1/\sinr_k+y}\right]\dif y
\end{align*}
Moreover, we know
\begin{align*}
\mathbb{E}\left[\frac{1}{1/\sinr_k+y}\right]=\int_{0}^{\infty} e^{-sy}\mathbb{E}\left[e^{-sI_k/S_k}\right]\dif s=\int_{0}^{\infty} e^{-sy}\mathbb{E}_{S_k}\left[\mathcal{L}_{I_k}\left(\frac{s}{S_k}\right)\right]\dif s
\end{align*}
and thus $c_k$ is given by
\begin{align*}
c_k& =\frac{1}{\ln(2)}\int_{0}^{\infty}\int_{0}^{1} e^{-sy}\mathbb{E}_{S_k}\left[\mathcal{L}_{I_k}\left(\frac{s}{S_k}\right)\right]\dif y\dif s\\
&=\frac{1}{\ln(2)}\int_{0^+}^{\infty} \frac{\left(1-e^{-s}\right)}{s}\mathbb{E}_{S_k}\left[\mathcal{L}_{I_k}\left(\frac{s/\mathbb{E}[S_k]}{S_k/\mathbb{E}[S_k]}\right)\right]\dif s\\
&= \frac{1}{\ln(2)}\int_{0^+}^{\infty}\int_{0}^{\infty} \left(1-e^{-s}\right)\mathcal{L}_{I_k}\left(\frac{s}{x\mathbb{E}[S_k]}\right)f_{\widehat{S}_k}(x)\dif x\dif s\\
&\stackrel{(a)}{=}\frac{1}{\ln(2)}\int_{0^+}^{\infty}\int_{0}^{\infty} \frac{\left(1-e^{-s}\right)}{se^{\pi\widetilde{\lambda}\Gamma(1-\frac{2}{\alpha})(s/x\mathbb{E}[S_k])^{\frac{2}{\alpha}}}}f_{\widehat{S}_k}(x)\dif x\dif s\\
&\stackrel{(b)}{=}\frac{1}{\ln(2)}\int_{0^+}^{\infty}\int_{0}^{\infty} \frac{\left(1-e^{-\vartheta x}\right)f_{\widehat{S}_k}(x)}{\vartheta e^{\pi\widetilde{\lambda}\Gamma(1-\frac{2}{\alpha})(\vartheta/\mathbb{E}[S_k])^{\frac{2}{\alpha}}}}\dif x\dif \vartheta=\frac{1}{\ln(2)}\bigintsss_{0^+}^{\infty} \frac{\left[1-\mathcal{L}_{\widehat{S}_k}(\vartheta)\right]}{\vartheta e^{\pi\widetilde{\lambda}\Gamma(1-\frac{2}{\alpha})(\vartheta/\mathbb{E}[S_k])^{\frac{2}{\alpha}}}}\dif \vartheta\\
&=\frac{1}{\ln(2)}\int_{0^+}^{\infty} \frac{1}{\vartheta }\left[1-\mathcal{L}_{\widehat{S}_k}(\vartheta)\right]\mathcal{L}_{I_k}\left(\frac{\vartheta}{\mathbb{E}[S_k]}\right)\dif \vartheta,
\end{align*}
where $(a)$ follows from the result of $\mathcal{L}_{I_k}(\cdot)$ in Theorem \ref{Thm:LaplaceFunCanKinterfer} and $(b)$ is obtained by letting $s=\theta x$ and carrying out the integral $\int_{0}^{\infty}(1-e^{-\theta x})f_{S_k}(x)\dif x$. Hence, the results in \eqref{Eqn:ErgodicLinkCapNoCan} are obtained. According to the result of $c_k$ in above, $c_{k,L}$ can also be found as shown in the following
\begin{align*}
c_{k,L}& =\frac{1}{\ln(2)}\int_{0}^{\infty}\int_{0}^{1} e^{-sy}\mathbb{E}_{S_k}\left[\mathcal{L}_{I_{k,L}}\left(\frac{s/\mathbb{E}[S_k]}{S_k/\mathbb{E}[S_k]}\right)\right]\dif y\dif s\\
&=\frac{1}{\ln(2)}\int_{0^+}^{\infty}\int_{0}^{\infty} \left(1-e^{-s}\right)\mathcal{L}_{I_{k,L}}\left(\frac{s}{x\mathbb{E}[S_k]}\right)f_{\widehat{S}_k}(x)\dif x\dif s\\
&\stackrel{(c)}{=}\frac{1}{\ln(2)}\bigintsss_{0^+}^{\infty} \frac{\left[1-\mathcal{L}_{\widehat{S}_k}(\vartheta)\right]}{\vartheta }\mathcal{L}_{I_{k,L}}\left(\frac{\vartheta}{\mathbb{E}[S_k]}\right)\dif \vartheta\\
&\stackrel{(d)}{=}\frac{1}{\ln(2)}\bigintss_{0^+}^{\infty} \frac{\left[1-\mathcal{L}_{\widehat{S}_k}(\vartheta)\right]}{\vartheta }\mathcal{L}_{I_k}\left(\frac{\vartheta}{\mathbb{E}[S_k]}\right) \mathcal{L}_{\ell_{D_L}\left(\vartheta/ \mathbb{E}[S_k]D_L^{\frac{\alpha}{2}},\frac{2}{\alpha}\right)}\left(\pi\widetilde{\lambda}\right)\dif \vartheta,
\end{align*}
where $(c)$ follows from the result of $c_k$ in above and $(d)$ follows from  the result of $\mathcal{L}_{I_{k,L}}(\cdot)$ in \eqref{Eqn:LaplaceCanKinterference}. Therefore, the results in \eqref{Eqn:ErgodicLinkCapKCan} are acquired.

\subsection{Proof of Theorem \ref{Thm:ErgodicRatePowerControl}}\label{App:ProofErgodicRatePowerControl}
Since $c_k$ in \eqref{Eqn:ErgodicLinkCapNoCan} is valid for any distribution of transmit power, $c^{pc}_k$ can be obtained by substituting $P_k$ in \eqref{Eqn:StochasticPowerControl} into $c_k$ in \eqref{Eqn:ErgodicLinkCapNoCan}. Thus, $\mathcal{L}_{I_k}(\vartheta/\mathbb{E}[S_k])$ in \eqref{Eqn:ErgodicLinkCapNoCan} with stochastic power control becomes $\mathcal{L}_{I^{pc}_k}(\vartheta/\mathbb{E}[S^{pc}_k])$ given by
\begin{align*}
\mathcal{L}_{I^{pc}_k}\left(\frac{\vartheta}{\mathbb{E}[S^{pc}_k]}\right)&= \exp\left(-\pi \Gamma\left(1-\frac{2}{\alpha}\right)\left(\frac{\vartheta}{\mathbb{E}[S^{pc}_k]}\right)^{\frac{2}{\alpha}}\widetilde{\lambda}^{pc}\right)\\
&=\exp\left(-\pi \Gamma\left(1-\frac{2}{\alpha}\right)\left(\frac{\vartheta\mathbb{E}[S^{\gamma_k}_k]}{\mathbb{E}[S^{1+\gamma_k}_k]}\right)^{\frac{2}{\alpha}}\widetilde{\lambda}^{pc}\right)
\end{align*}
because $S^{pc}_k=S^{1+\gamma_k}_k/\mathbb{E}[S^{\gamma_k}_k]$. The normalized type-$k$ received signal power with stochastic power control is $\widehat{S}^{pc}_k=S^{\gamma_k+1}_k/\mathbb{E}[S^{\gamma_k+1}_k]$ so that we have 
\begin{align*}
\mathcal{L}_{\widehat{S}^{pc}_k}(\vartheta)=\mathbb{E}\left[\exp\left(-\frac{\vartheta S^{\gamma_k+1}_k}{\mathbb{E}[S^{\gamma_k+1}_k]}\right)\right]=\mathcal{L}_{\widehat{S}^{\gamma_k+1}_k}\left(\frac{\vartheta(\mathbb{E}[S_k])^{\gamma_k+1}}{\mathbb{E}[S^{\gamma_k+1}_k]}\right).
\end{align*}
Substituting $\mathcal{L}_{I_k}\left(\frac{\vartheta}{\mathbb{E}[S_k]}\right)$ and $\mathcal{L}_{\widehat{S}_k}(\vartheta)$ found in above into $\eqref{Eqn:ErgodicLinkCapNoCan}$ leads to $c^{pc}_k$ in \eqref{Eqn:ErgodicRatePowerControl}. Also, the following lower bound 
\begin{align*}
1-\mathcal{L}_{\widehat{S}^{\gamma_k+1}_k}\left(\frac{\vartheta(\mathbb{E}[S_k])^{\gamma_k+1}}{\mathbb{E}\left[S^{\gamma_k+1}_k\right]}\right)\leq 1-e^{-\vartheta}.
\end{align*}
can be obtained by Jensen's inequality since $1-\mathcal{L}_{\widehat{S}^{\gamma_k+1}_k}(\cdot)$ is concave and the following upper bound
\begin{align*}
1-\mathcal{L}_{\widehat{S}^{\gamma_k+1}_k}\left(\frac{\vartheta(\mathbb{E}[S_k])^{\gamma_k+1}}{\mathbb{E}\left[S^{\gamma_k+1}_k\right]}\right)\geq \mathbb{E}\left[\frac{\vartheta S^{\gamma_k+1}_k}{\mathbb{E}\left[S^{\gamma_k+1}_k\right]+\vartheta S^{\gamma_k+1}_k}\right]\geq \frac{\vartheta}{1+\vartheta}
\end{align*}
is true due to $1-e^{-a x}\leq \frac{ax}{1+ax}$ for $a, x\in\mathbb{R}_+$ and Jensen's inequality. Hence, the bounds in \eqref{Eqn:BoundsErgodicLinkCapPowerCon} are valid. 

\bibliographystyle{ieeetran}
\bibliography{IEEEabrv,Ref_GenOutCapStoNets}

\begin{thebibliography}{10}
\providecommand{\url}[1]{#1}
\csname url@samestyle\endcsname
\providecommand{\newblock}{\relax}
\providecommand{\bibinfo}[2]{#2}
\providecommand{\BIBentrySTDinterwordspacing}{\spaceskip=0pt\relax}
\providecommand{\BIBentryALTinterwordstretchfactor}{4}
\providecommand{\BIBentryALTinterwordspacing}{\spaceskip=\fontdimen2\font plus
\BIBentryALTinterwordstretchfactor\fontdimen3\font minus
  \fontdimen4\font\relax}
\providecommand{\BIBforeignlanguage}[2]{{%
\expandafter\ifx\csname l@#1\endcsname\relax
\typeout{** WARNING: IEEEtran.bst: No hyphenation pattern has been}%
\typeout{** loaded for the language `#1'. Using the pattern for}%
\typeout{** the default language instead.}%
\else
\language=\csname l@#1\endcsname
\fi
#2}}
\providecommand{\BIBdecl}{\relax}
\BIBdecl

\bibitem{FBBBPM06}
F.~Baccelli, B.~Blaszczyszyn, and P.~M\"{u}hlethaler, ``An {A}loha protocol for
  multihop mobile wireless networks,'' \emph{{IEEE} Trans. Inf. Theory},
  vol.~52, no.~2, pp. 421--436, Feb. 2006.

\bibitem{SWXYJGAGDV05}
S.~P. Weber, X.~Yang, J.~G. Andrews, and G.~de~Veciana, ``Transmission capacity
  of wireless ad hoc networks with outage constraints,'' \emph{{IEEE} Trans.
  Inf. Theory}, vol.~51, no.~12, pp. 4091--4102, Dec. 2005.

\bibitem{MHRKG09}
M.~Haenggi and R.~K. Ganti, ``Interference in large wireless networks,''
  \emph{Foundations and Trends in Networking}, vol.~3, no.~2, pp. 127--248,
  2009.

\bibitem{MHJGAFBODMF10}
M.~Haenggi, J.~G. Andrews, F.~Baccelli, O.~Dousse, and M.~Franceschetti,
  ``Stochastic geometry and random graphs for the analysis and design of
  wireless networks,'' \emph{{IEEE} J. Sel. Areas Commun.}, vol.~27, pp.
  1029--1046, Sep. 2009.

\bibitem{CHLJGA12}
C.-H. Liu and J.~G. Andrews, ``Ergodic transmission capacity of wireless ad hoc
  networks with interference management,'' \emph{{IEEE} Trans. Wireless
  Commun.}, vol.~11, no.~6, pp. 2136--2147, Jun. 2012.

\bibitem{DSWKJM13}
S.~N. Chiu, D.~Stoyan, W.~S. Kendall, and J.~Mecke, \emph{Stochastic Geometry
  and Its Applications}, 3rd~ed.\hskip 1em plus 0.5em minus 0.4em\relax New
  York: John Wiley and Sons, Inc., 2013.

\bibitem{FBBBL101}
F.~Baccelli and B.~B{\l}aszczyszyn, ``Stochastic geometry and wireless
  networks: Volume {I Theory},'' \emph{Foundations and Trends in Networking},
  vol.~3, no. 3-4, pp. 249--449, 2010.

\bibitem{SWJGANJ07}
S.~Weber, J.~G. Andrews, and N.~Jindal, ``The effect of fading, channel
  inversion, and threshold scheduling on ad hoc networks,'' \emph{{IEEE} Trans.
  Inf. Theory}, vol.~53, no.~11, pp. 4127--4149, Nov. 2007.

\bibitem{CHL13}
C.-H. Liu, ``Distributed interferer-channel aware scheduling in large-scale
  wireless ad hoc networks,'' in \emph{IEEE Global Commun. Conf.}, Dec. 2013,
  pp. 213--215.

\bibitem{CHLYCT13}
C.-H. Liu and Y.-C. Tsai, ``Distributed dynamic scheduling in wireless ad hoc
  networks with spatial randomness,'' in \emph{Proc. IEEE Global Commun.
  Conf.}, Dec. 2013, pp. 177--182.

\bibitem{NJSPWJGA08}
N.~Jindal, S.~P. Weber, and J.~G. Andrews, ``Fractional power control for
  decentralized wireless networks,'' \emph{{IEEE} Trans. Wireless Commun.},
  vol.~7, no.~12, pp. 5482--5492, Dec. 2008.

\bibitem{XZMH12}
X.~Zhang and M.~Haenggi, ``Random power control in poisson networks,''
  \emph{{IEEE} Trans. Commun.}, vol.~60, no.~9, pp. 2602--2611, Sep. 2012.

\bibitem{CHLBRSC15}
C.-H. Liu, B.~Rong, and S.~Cui, ``Optimal discrete power control in
  {P}oisson-clustered ad hoc networks,'' \emph{{IEEE} Trans. Wireless Commun.},
  vol.~14, no.~1, pp. 138--151, Jan. 2015.

\bibitem{YGIBEZ13}
Y.~George, I.~Bergel, and E.~Zehavi, ``The ergodic rate density of aloha
  wireless ad-hoc networks,'' \emph{{IEEE} Trans. Wireless Commun.}, vol.~12,
  no.~12, pp. 6340--6351, Dec. 2013.

\bibitem{CHLJGA11}
C.-H. Liu and J.~G. Andrews, ``Multicast outage and transmission capacity in
  multihop wireless networks,'' \emph{{IEEE} Trans. Inf. Theory}, vol.~57,
  no.~7, pp. 4344--4358, Jul. 2011.

\bibitem{RVKTTSWRH11}
R.~Vaze, K.~T. Truong, S.~Weber, , and R.~W.~J. Heath, ``Two-way transmission
  capacity of wireless ad-hoc networks,'' \emph{{IEEE} Trans. Wireless
  Commun.}, vol.~10, no.~6, pp. 1966--1975, Jun. 2011.

\bibitem{HSRRSIALTTRJAG15}
H.~Shariatmadari, R.~Ratasuk, S.~Iraji, A.~Laya, T.~Taleb, R.~J\"{a}ntti, and
  A.~Ghosh, ``Machine-type communications: current status and future
  perspectives toward {5G} systems,'' \emph{{IEEE} Commun. Mag.}, vol.~53,
  no.~9, pp. 10--17, Sep. 2015.

\bibitem{SAOGAPMGTTJTJSMDYK15}
S.~Andreev, O.~Galinina, A.~Pyattaev, M.~Gerasimenko, T.~Tirronen, J.~Torsner,
  J.~Sachs, M.~Dohler, and Y.~Koucheryavy, ``Understanding the {IoT}
  connectivity landscape: a contemporary {M2M} radio technology roadmap,''
  \emph{{IEEE} Commun. Mag.}, vol.~53, no.~9, pp. 32--40, Sep. 2015.

\bibitem{SWJGAXYGVJ07}
S.~Weber, J.~G. Andrews, X.~Yang, and G.~de~Veciana, ``Transmission capacity of
  wireless ad hoc networks with successive interference cancellation,''
  \emph{{IEEE} Trans. Inf. Theory}, vol.~53, no.~8, pp. 2799--2814, Aug. 2007.

\bibitem{XZMH14}
X.~Zhang and M.~Haenggi, ``The performance of successive interference
  cancellation in random wireless networks,'' \emph{{IEEE} Trans. Inf. Theory},
  vol.~60, no.~10, pp. 6368--6388, Oct. 2014.

\bibitem{MH12}
M.~Haenggi, \emph{Stochastic Geometry for Wireless Networks}, 1st~ed.\hskip 1em
  plus 0.5em minus 0.4em\relax Cambridge University Press, 2012.

\bibitem{AMTSV04}
A.~M. Tulino and S.~Verd\'{u}, ``Random matrix theory and wireless
  communications,'' \emph{Foundations and Trends in Communications and
  Information Theory}, vol.~1, no.~1, pp. 1--182, 2004.

\bibitem{MAIAS72}
M.~Abramowitz and I.~A. Stegun, \emph{Handbook of Mathematical Functions: with
  Formulas, Graphs, and Mathematical Tables}, 9th~ed.\hskip 1em plus 0.5em
  minus 0.4em\relax Dover Publications, 1972.

\bibitem{NJSPWJGA11}
N.~Jindal, J.~G. Andrews, and S.~P. Weber, ``Multi-antenna communication in ad
  hoc networks: Achieving {MIMO} gains with {SIMO} transmission,'' \emph{{IEEE}
  Trans. Commun.}, vol.~59, no.~2, pp. 529--540, Feb. 2011.

\bibitem{PXCHLJGA13}
P.~Xia, C.-H. Liu, and J.~G. Andrews, ``Downlink coordinated multi-point with
  overhead modeling in heterogeneous cellular networks,'' \emph{{IEEE} Trans.
  Wireless Commun.}, vol.~12, no.~8, pp. 4025--4037, Jun. 2013.

\bibitem{SWJGAXYGDV07}
S.~P. Weber, J.~G. Andrews, X.~Yang, and G.~de~Veciana, ``Transmission capacity
  of wireless ad hoc networks with successive interference cancellation,''
  \emph{{IEEE} Trans. Inf. Theory}, vol.~53, no.~8, pp. 2799--2814, Aug. 2007.

\bibitem{DPB99}
D.~P. Bertsekas, \emph{Nonlinear Programming}, 2nd~ed.\hskip 1em plus 0.5em
  minus 0.4em\relax Belmont, MA: Athena Scientific, 1999.

\bibitem{MH05}
M.~Haenggi, ``On distances in uniformly random networks,'' \emph{{IEEE} Trans.
  Inf. Theory}, vol.~51, no.~10, pp. 3584 --3586, October 2005.

\bibitem{CHLLCW16}
C.-H. Liu and L.-C. Wang, ``Optimal cell load and throughput in green small
  cell networks with generalized cell association,'' \emph{{IEEE} J. Sel. Areas
  Commun.}, vol.~34, no.~5, pp. 1058--1072, May 2016.

\end{thebibliography}

\end{document}